\documentclass[journal,english]{IEEEtran}

\usepackage{epsfig}
\usepackage{times}
\usepackage{float}
\usepackage{afterpage}
\usepackage{amsmath}
\usepackage{amstext,cite}
\usepackage{amssymb,bm}
\usepackage{latexsym}
\usepackage{color}
\usepackage{graphicx}
\usepackage{amsmath}
\usepackage{amsthm}
\usepackage{graphicx}
\usepackage[center]{caption}
\usepackage{pstricks}
\usepackage{caption}
\usepackage{subcaption}
\usepackage{booktabs}
\usepackage{multicol}
\usepackage{lipsum}% dummy text
\usepackage[T1]{fontenc}
\usepackage{hyperref}
 \usepackage{aecompl}
 \usepackage{mathrsfs}

 \usepackage{arydshln}

\usepackage{soul}
\usepackage{cancel}

\usepackage{changes}
\definechangesauthor[color=red,name={Mingyue Ji}]{MJ}

\allowdisplaybreaks

\long\def\comment#1{}

\newcommand{\av}{{\mathbf a}}

\newcommand{\cv}{{\mathbf c}}

\newcommand{\uv}{{\mathbf u}}

\newcommand{\vv}{{\mathbf v}}
\newcommand{\xv}{{\mathbf x}}

\newcommand{\Ac}{{\mathcal A}}

\newcommand{\Gc}{{\mathcal G}}
\newcommand{\Hc}{{\mathcal H}}

\newcommand{\Oc}{{\mathcal O}}
\newcommand{\Pc}{{\mathcal P}}

\newcommand{\Rc}{{\mathcal R}}
\newcommand{\Sc}{{\mathcal S}}
\newcommand{\Tc}{{\mathcal T}}

\newcommand{\Zc}{{\mathcal Z}}

\newcommand{\asf}{{\mathsf a}}
\newcommand{\bsf}{{\mathsf b}}

\newcommand{\qsf}{{\mathsf q}}

\newcommand{\Ksf}{{\mathsf K}}
\newcommand{\Lsf}{{\mathsf L}}
\newcommand{\Msf}{{\mathsf M}}
\newcommand{\Nsf}{{\mathsf N}}

\newcommand{\Rsf}{{\mathsf R}}

\newcommand{\Tsf}{{\mathsf T}}

\renewcommand{\arg}{{\hbox{arg}}}

\newtheorem{thm}{Theorem}%[section]
\newtheorem{cor}{Corollary}
\newtheorem{lem}{Lemma}

\newtheorem{rem}{Remark}
\newtheorem{example}{Example}

\providecommand{\definitionname}{Definition}

\begin{document}

\title{Distributed Linearly Separable  Computation}  
\author{
Kai~Wan,~\IEEEmembership{Member,~IEEE,} 
Hua~Sun,~\IEEEmembership{Member,~IEEE,}
Mingyue~Ji,~\IEEEmembership{Member,~IEEE,}  
and~Giuseppe Caire,~\IEEEmembership{Fellow,~IEEE}

\thanks{
K.~Wan and G.~Caire are with the Electrical Engineering and Computer Science Department, Technische Universit\"at Berlin, 10587 Berlin, Germany (e-mail:  kai.wan@tu-berlin.de; caire@tu-berlin.de). The work of K.~Wan and G.~Caire was partially funded by the European Research Council under the ERC Advanced Grant N. 789190, CARENET.}
\thanks{
H.~Sun is with the Department of Electrical Engineering, University of North Texas, Denton, TX 76203, USA (email: hua.sun@unt.edu).  The work of Hua Sun was supported in part by funding from NSF grants CCF-2007108 and CCF-2045656.
}
\thanks{
M.~Ji is with the Electrical and Computer Engineering Department, University of Utah, Salt Lake City, UT 84112, USA (e-mail: mingyue.ji@utah.edu). The work of M.~Ji was supported in part by NSF Awards 1817154 and 1824558.}
}
\maketitle
%\IEEEpeerreviewmaketitle{}

\begin{abstract}
 This paper formulates a distributed computation problem, where a master asks $\Nsf$ distributed workers to compute a   linearly separable function. The task function can be expressed as $\Ksf_{\rm c}$ linear combinations of $\Ksf$ messages, where each message is a function of one dataset. Our objective is to find the optimal tradeoff between  the computation cost (number of uncoded datasets assigned to each worker)  and the communication cost (number of symbols the master must  download), such that   from the answers of any $\Nsf_{\rm r}$  out of $\Nsf$ workers the master can recover the task function with high probability, where the coefficients of the $\Ksf_{\rm c}$ linear combinations are uniformly i.i.d. over some large enough finite field. The formulated problem can be seen as a generalized version of some existing problems, such as   distributed gradient coding and  distributed linear transform.

 In this paper, we consider the specific case where  the   computation cost is minimum, and propose novel achievability schemes and converse bounds
for the optimal communication cost.  Achievability and converse bounds 
 coincide for some system parameters; when they do not match, we prove that the achievable distributed computing scheme is optimal under the constraint of a widely used `cyclic assignment' scheme on the datasets. Our results   also show that when $\Ksf=\Nsf$, with the same communication cost as the optimal  distributed gradient   coding scheme proposed by Tandon {\it et al}. from which the master recovers one linear combination of $\Ksf$ messages,  our proposed scheme can let the master recover any additional    $\Nsf_{\rm r}-1$ independent   linear combinations of messages with high probability.
\end{abstract}

\begin{IEEEkeywords}
 Distributed computation; linearly separable function; cyclic assignment
\end{IEEEkeywords}

\section{Introduction}
\label{sec:intro}
Enabling large-scale computations for a large dimension of data,
distributed computation systems  such as MapReduce \cite{dean2008mapreduce} and Spark \cite{zaharia2010spark} have received significant attention in recent years~\cite{largescale2012dean}.  
 The distributed computation system divides a computational task into several subtasks, which are then assigned to  some distributed workers. This  reduces significantly the computing time by exploiting parallel computing procedures and
thus enables handling of the computations over large-scale big data. However, while large scale distributed computing schemes
 have the potential for achieving unprecedented levels of accuracy and
providing dramatic insights into complex phenomena,  they also present some technical issues/bottlenecks.  First, due to the  presence of stragglers,  a subset of workers  may take
excessively long time or fail to return their computed sub-tasks, which leads to an undesirable and unpredictable  latency.  Second, 
data and computed results should  be  communicated among the master who wants to compute the task, and the workers. If the communication  bandwidth is limited, the communication cost becomes  another bottleneck of the distributed computation system.
In order to tackle these two bottlenecks, 
 coding techniques were introduced to the distributed computing algorithms~\cite{speedup2018Lee,distributedcomputing,unifiedDC2016li},
 with the purpose of increasing tolerance with respect to stragglers and reducing the master-workers communication cost.    
%  which bring  the tolerance on the stragglers and reduce  the communication cost for   distributed computing. 
More precisely, for the first bottleneck, using ideas similar to Minimum Distance Separable (MDS) codes,
the master can recover the task function from the answers of the fastest workers. 
 For the second bottleneck, inspired by concepts from coded caching networks~\cite{dvbt2fundamental,d2dcaching},     network coding techniques are used %among intermediate computed values 
to save significant communication cost exchanged in the network.
%{\red QUESTION OF KAI: SHOULD WE REFER TO SOME BACKGROUND OF LEARNING?}

In this paper,  a master aims to compute a    linearly separable function  $f$  (such as linear MapReduce, Fourier Transform, convolution, etc.) on $\Ksf$ datasets ($D_1,\ldots,D_{\Ksf}$),   which can be  written as 
 $$
 f(D_1,\ldots,D_{\Ksf})= g\big(f_1(D_1), \ldots, f_{\Ksf}(D_{\Ksf}) \big)= g(W_1,\ldots,W_{\Ksf}).
 $$
  $W_k = f_k(D_k)$ for all $k\in \{1, \ldots, \Ksf\}$ is the outcome of the  component function $f_k(\cdot)$ applied to dataset $D_k$, and it is represented as a string of $\Lsf$ symbols on an appropriate sufficiently large alphabet. 
    For example, $W_k$ can be  the intermediate value in linear MapReduce,  an input signal in Fourier Transform, etc.
  We consider 
the case where $g(\cdot)$ is a linear map defined by $\Ksf_{\rm c}$  linear combinations of the messages $W_1,\ldots,W_{\Ksf}$  with uniform i.i.d. coefficients over some   large enough finite field;  
 i.e.,  $g(W_1,\ldots,W_{\Ksf})$    can be seen as the matrix product ${\bf F W}$, where ${\bf F}$ is the coefficient matrix and ${\bf W}=[W_1;\ldots;W_{\Ksf}]$.\footnote{\label{foot:applications}   As   matrix multiplication is one of the key building blocks underlying many
data analytics, machine learning algorithms and  engineering problems, the considered model also has potential applications in those areas, where $f_1,\ldots,f_{\Ksf}$ represent the   pretreatment of the datasets.
 For example, each dataset $D_k$ where $k\in \{1, \ldots, \Ksf\}$  represents a raw dataset   and needs to be processed through some filters, where  $W_k$ represents the filtered dataset of $D_k$. For the sake of linear transforms (e.g., Wavelet Transform, Discrete Fourier Transform), we need to compute multiple linear combinations of  the filtered datasets, which can be expressed as  $g(W_1,\ldots,W_{\Ksf})$.   
  For another example, $D_1, \ldots, D_{\Ksf}$ are the $\Ksf$ ``input channels'' of a Convolutional    Neural Networks (CNN) stage. Each input channel $D_k$ where $k\in \{1,\ldots,\Ksf\}$ is filtered individually by a convolution operation yielding $W_k$. Then the convolutions are linearly mixed by the coefficients of $g(W_1,\ldots,W_{\Ksf})$ producing $\Ksf_{\rm c}$ new layers in the feature space. 
  Moreover,  if  ${\bf F}$ represents a MIMO precoding matrix, our  considered model can also be used in the MIMO systems.
  }
%An example of  the    linearly separable function $f$ can be  a  loss function where $\Ksf$ can be  can be the dimension of the data.
We consider the distributed computation scenario, where
 $ f(D_1,\ldots,D_{\Ksf})$ is computed in a distributed way by a group of $\Nsf$ workers. Each dataset is  assigned in an uncoded manner to a subset of workers and 
 the number of datasets assigned  to each worker cannot be larger than $\Msf$, which is referred to as the computation cost.\footnote{\label{foot:computation cost}We assume that each function $f_k(\cdot)$ is arbitrary such that in general it does not hold that computing less symbols for the result $W_k$ is less costly in terms of computation. Hence, each worker $n$ computes the whole $W_k=f_k(D_k)$ if $D_k$ is assigned to it. We also assume that the complexity of computing the messages from the datasets is much higher than computing the desired linear combinations of the messages. So we denote the computation cost by $\Msf$.}
   Each worker should compute and send   coded messages in terms of the datasets assigned  to it, such that  from the answers of any $\Nsf_{\rm r}$ 
 workers, the master can recover the task function with high probability.  Given $(\Ksf,\Nsf,\Nsf_{\rm r}, \Ksf_{\rm c}, \Msf)$, we aim to find the optimal distributed computing scheme with   {\it data assignment},   {\it computing}, and   {\it decoding} phases, which leads to the minimum communication cost (i.e., the number of downloaded symbols by the master, normalized by $\Lsf$). 
 
\begin{figure}
    \centering
    \begin{subfigure}[t]{0.5\textwidth}
        \centering
        \includegraphics[scale=0.20]{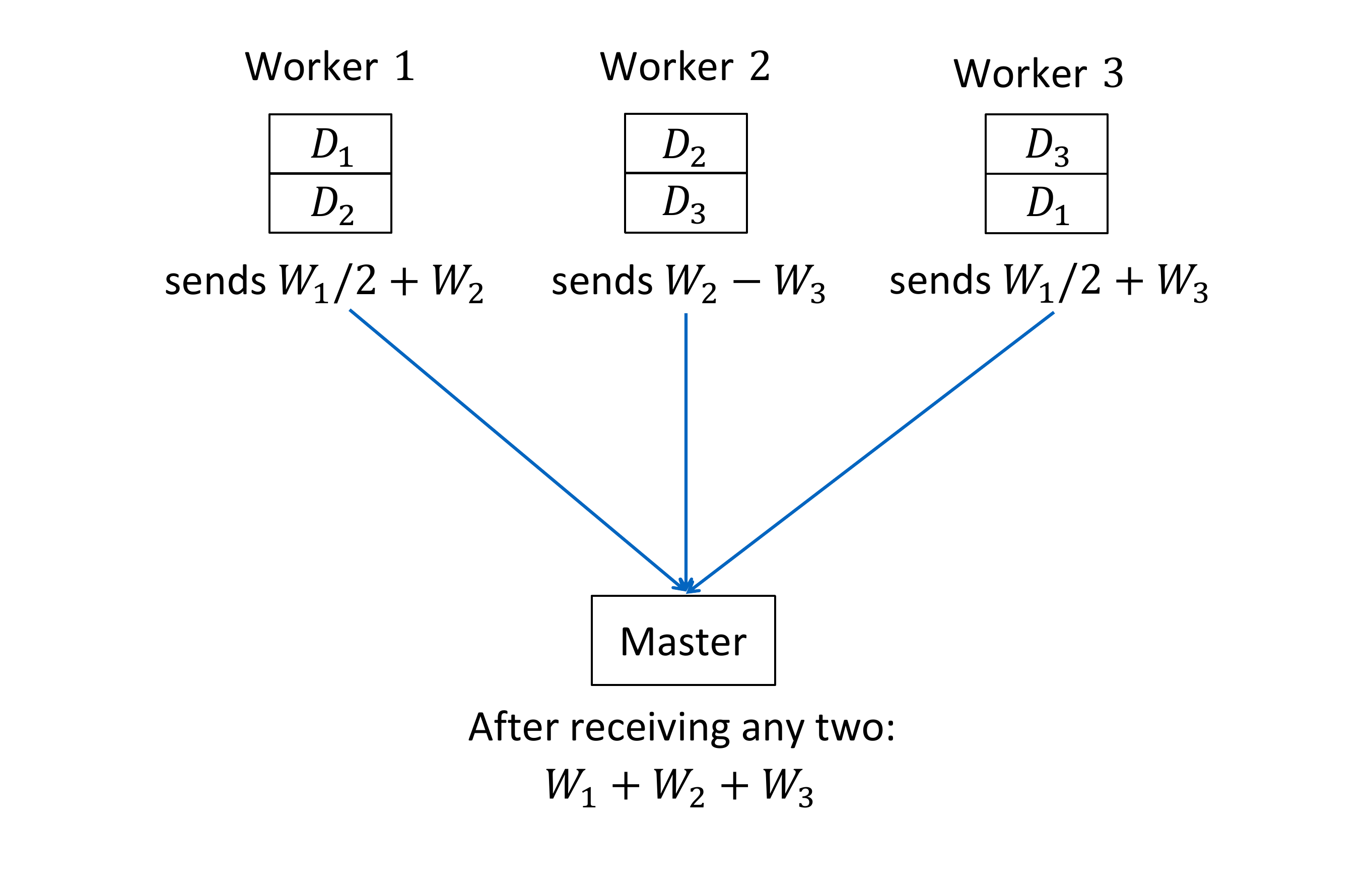}
        \caption{\small $\Ksf_{\rm c}=1$.}
        \label{fig:numerical 0a}
    \end{subfigure}%
    \\ 
    \begin{subfigure}[t]{0.5\textwidth}
        \centering
        \includegraphics[scale=0.20]{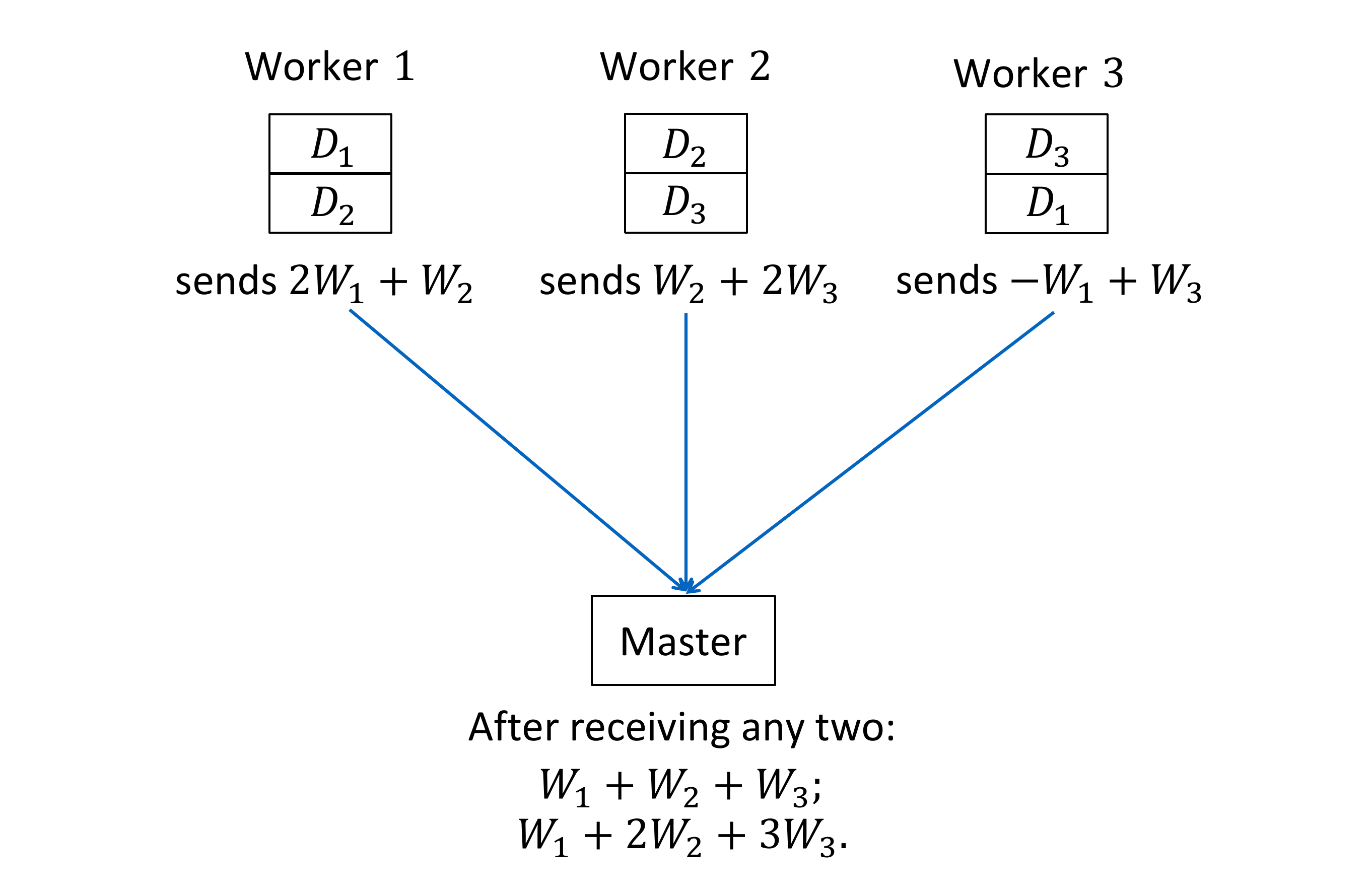}
        \caption{\small $\Ksf_{\rm c}=2$.}
        \label{fig:numerical 0b}
    \end{subfigure}
    \caption{\small Distributed linearly separable  computation with $\Ksf=\Nsf=3$ and $\Nsf_{\rm r}=2$. The number of datasets assigned to each worker is $\Msf=2$.}
    \label{fig:system model}
\end{figure}

 We illustrate two examples of the formulated distributed scenario in Fig.~\ref{fig:system model} where  $\Ksf_{\rm c}=1$ and  $\Ksf_{\rm c}=2$, respectively. 
 In both examples, we consider that $\Ksf=\Nsf=3, \Nsf_{\rm r}=2$,  and that the number of datasets assigned to each worker is $\Msf=2$.   Assume that the  characteristic of $\mathbb{F}_{\qsf}$ is larger than $3$. 
 \begin{itemize}
 \item  When $\Ksf_{\rm c}=1$, the considered problem (as shown in Fig.~\ref{fig:numerical 0a}) is equivalent to the distributed gradient coding problem in~\cite{gradiencoding}, which aims to compute the sum of gradients in   learning  tasks by distributed workers. The gradient coding proposed in~\cite{gradiencoding} assigns the datasets to the workers in a cyclic way, where   $D_1$ and $D_2$ are assigned to  worker $1$,   $D_2$ and $D_3$ are assigned to  worker $2$, and $D_3$ and $D_1$ are assigned to  worker $3$. Worker $1$ then %computes $W_1=f_1(D_1)$ and $W_2=f_2(D_2)$, and 
   computes and sends $\frac{W_1}{2}+W_2$. Worker $2$ sends $W_2-W_3$, and worker $3$ sends $\frac{W_1}{2}+W_3$. From any two sent coded messages, the master can recover the task function $W_1+W_2+W_3$. 
By the converse bound in~\cite{efficientgradientcoding}, it can be proved that    %if   the number of datasets assigned to each worker is $2$, 
     the gradient coding scheme~\cite{gradiencoding} is optimal under the constraint of   linear coding  in terms of communication cost. Note that in our paper, from a novel converse bound, we prove the   optimality of the gradient coding scheme~\cite{gradiencoding} when $\Ksf_{\rm c}=1$ by removing the constraint of linear coding.
   \item When $\Ksf_{\rm c}=2$,  besides  $W_1+W_2+W_3$ we let the master also request another   linear combination of the messages, e.g.,
   $W_1+2W_2+3W_3$.
    Here, we propose a novel distributed computing scheme (as shown in Fig.~\ref{fig:numerical 0a}), which can compute this additional sum but with the same   number of communicated symbols as the gradient coding scheme. With the same cyclic assignment, we let worker $1$ send $2W_1+W_2$, worker $2$ send $W_2 +2W_3$, worker $3$ send $-W_1 +W_3$. It can be checked that from any two sent coded messages, the master can recover both of the two requested sums. 
   Hence, with the same communication cost as the gradient coding scheme~\cite{gradiencoding}, the  proposed distributed computing scheme allows the master recover the  two requested linear combinations. 
 \end{itemize}

Since the seminal works on using coding techniques in distributed computing~\cite{speedup2018Lee,distributedcomputing,unifiedDC2016li}, different coded distributed computing schemes were proposed to compute various   tasks in machine  learning applications. The detailed comparison between the considered distributed linearly separable computation problem and each of the related existing works will be provided in Section~\ref{sub:connection to existing problem}. In short,
\begin{itemize}
\item  the distributed gradient coding problem considered in~\cite{gradiencoding,MDSGC2018raviv,improvedGC2017halbawi} is a special case of the considered problem in this paper with  $\Ksf_{\rm c}=1$ (i.e., the master requests one linear combination of the messages);
\item  the distributed linear transform problem considered in~\cite{shortdot2016dutta} is a special case of  the considered problem in this paper where    $\Lsf=1$ (i.e., each message contains one symbol) and each worker sends one symbol;
\item  in the distributed matrix-vector multiplication problem considered in~\cite{universally2019ramamoo,matrixvectorconvo2017das,finitematrixvector2018hadda}, the distributed matrix-matrix multiplication problem considered in~\cite{speedup2018Lee,highdimensional2017lee,sparsematrix2018wang,polyoptYu2017,yu2020stragglermitigation,dutta2020optimalRT,struresis2020ramamoo,jia2019matrixCSA}, and the  distributed multivariate polynomial computation problem considered in~\cite{lagrange2019yu},  coded assignments are allowed, i.e.,
linear combinations of all input datasets can be  assigned to each worker. 
Instead, 
in the considered problem the data assignment phase is uncoded, such that each worker can only compute functions of  the datasets which are assigned to it.
 \end{itemize}
 
\subsection*{Contributions}
In this paper, we formulate the distributed linearly separable computation problem and consider the case where $\Nsf$ divides $\Ksf$  and the computation cost is minimum, i.e., $\Msf=\frac{\Ksf}{\Nsf}(\Nsf-\Nsf_{\rm r}+1) $   by Lemma~\ref{lem:minimum computation cost}. Our main contributions on this case are as follows.
\begin{itemize}
\item We first propose an information theoretic converse bound on the minimum communication cost, inspired by  the converse bound for the coded caching problem with uncoded cache placement~\cite{indexcodingcaching2020,exactrateuncoded}.
\item With the cyclic assignment,   widely used in the existing works on   the distributed gradient coding problem such as~\cite{gradiencoding,MDSGC2018raviv,efficientgradientcoding},\footnote{\label{foot:cyc}   The main advantages of the cyclic assignment are that it can be used for any case where $\Nsf$ divides $\Ksf$  regardless of other system parameters, and its simplicity.  According to our knowledge, the other existing assignments, such as the repetition assignments in~\cite{gradiencoding,replicationcode2020}, can only be used for   limited number of cases. In addition, the cyclic assignment is independent of the task function; thus if the master has multiple tasks in different times, we need not assign the datasets in each time.
 } we propose a novel distributed computing scheme based on the linear space intersection and prove its decodability by  the Schwartz-Zippel lemma~\cite{Schwartz,Zippel,Demillo_Lipton}.\footnote{\label{foot:high proba} Note that the proposed computing is decodable with high probability; it will be explained in Remark~\ref{rem:non generic} that for some specific tasks, additional communication cost is needed.} 
\item Compared to the proposed converse bound, the achievable scheme is proved to be optimal when $\Nsf=\Ksf$, or $\Ksf_{\rm c} \in \left\{1,\ldots, \left\lceil \frac{\Ksf}{\binom{\Nsf}{\Nsf-\Nsf_{\rm r}+1}} \right\rceil  \right\}$, or $\Ksf_{\rm c} \in \left\{  \frac{\Ksf}{\Nsf}\Nsf_{\rm r} , \ldots, \Ksf \right\}$. In addition, the proposed achievable scheme is proved to be optimal under the constraint of the cyclic assignment for all system parameters. The optimality results are listed in Table~\ref{tab:optimality} at the top of the next page.
\item By the  derived optimality results, we obtain an interesting observation:   when $\Ksf =\Nsf$, for any $\Ksf_{\rm c} \in \{1,\ldots,\Nsf_{\rm r}\}$, the optimal communication cost is always $\Nsf_{\rm r}$. Thus by taking the same communicatoin cost as   the    optimal gradient coding scheme in~\cite{gradiencoding} for the distributed gradient coding problem (which is the case $\Ksf_{\rm c}=1$ of our problem), 
 with high probability our propose scheme can let the master recover  any additional   $\Nsf_{\rm r}-1$ linear combinations with  uniformly i.i.d. coefficients  over $\mathbb{F}_{\qsf}$.
\end{itemize}

Moreover, for the case where $\Nsf$ does not divide $\Ksf$,
the cyclic assignment cannot be directly used and 
we   propose    modified cyclic assignment and computing phases.

 \begin{table*}
 \centering
\protect\caption{Optimality results for the distributed linearly separable computation problem where $\Msf=\frac{\Ksf}{\Nsf} (\Nsf-\Nsf_{\rm r} +1)$ and $\Nsf$ divides $\Ksf $. }\label{tab:optimality}
\begin{tabular}{|c|c|}
\hline 
  Constraint of system parameters  &  Optimality  \tabularnewline
\hline 
%\hline 
%DIS, \cite{cachingincom}& $\Nsf\geq \Ksf$,  $\Msf=\Nsf t/\Ksf$ & order optimal within factor $\min\{\Hsf/\rsf,\left\lceil t\right\rceil +1\}$   \tabularnewline
\hline 
  $\Nsf=\Ksf $  &  optimal  \tabularnewline
\hline 
  $\Nsf\neq \Ksf$,    $\Ksf_{\rm c} \in \left\{1,\ldots, \left\lceil \frac{\Ksf}{\binom{\Nsf}{\Nsf-\Nsf_{\rm r}+1}} \right\rceil  \right\}$ & optimal \tabularnewline
\hline 
  $\Nsf\neq \Ksf$,    $\Ksf_{\rm c} \in \left\{  \frac{\Ksf}{\Nsf}\Nsf_{\rm r} , \ldots, \Ksf \right\}$ &   optimal         \tabularnewline
\hline 
  $\Nsf\neq \Ksf$,  $\Ksf_{\rm c} \in \left\{\left\lceil \frac{\Ksf}{\binom{\Nsf}{\Nsf-\Nsf_{\rm r}+1}} \right\rceil+1, \ldots,   \frac{\Ksf}{\Nsf}\Nsf_{\rm r} -1   \right\}$  &  optimal under   the cyclic assignment\tabularnewline
\hline  
\end{tabular}
\end{table*}

 %matrix inverse~\cite{codedinverse2017yang}
 
 %~\cite{improvedGC2017halbawi} improves decoding delay/complexity of GC by RS code.
 %~\cite{nearoptimalGC2018li} considers the tradeoff among the minimum recover threshold, the computation load, and the communication load.
 %~\cite{treeGC2019reisi} extend the GC strategy to a tree-topology  where the workers are located. The optimal communication load is characterized to tolerate the stragglers where a fix fraction of  children nodes per any parent node may straggle.
  %~\cite{MDSGC2018raviv} propose deterministic code construction based on MDS.

  %{\red MENTION CYCLIC ADVANTAGE}
\subsection*{Paper Organization}
The rest of this paper is organized as follows.
Section~\ref{sec:model} formulates the distributed linearly separable computation  problem and explains the differences from the existing distributed computation problem in the literature.
Section~\ref{sec:main} provides   the main results in this paper. 
Section~\ref{sec:achie}  describes the proposed achievable  distributed computing scheme.
Section~\ref{sec:extension} discusses the extensions of the proposed results.
Section~\ref{sec:conclusion} concludes the paper and some of the proofs are given in the Appendices.

\subsection*{Notation Convention}
%We use the following notation convention.
Calligraphic symbols denote sets, 
bold symbols denote vectors and matrices,
and sans-serif symbols denote system parameters.
We use $|\cdot|$ to represent the cardinality of a set or the length of a vector;
$[a:b]:=\left\{ a,a+1,\ldots,b\right\}$, $(a:b]:=\{a+1,a+2,\ldots,b \} $, $[a:b):=\{a,a+1,\ldots,b-1 \} $, $(a,b)=\{a+1,a+2,\ldots,b-1\}$ and $[n] := [1:n]$; 
$\oplus$ represents bit-wise XOR; $\mathbb{E}[\cdot]$ represents the expectation value of a random variable;  
$a!=a\times (a-1) \times \ldots \times 1$ represents the factorial of $a$;
$\mathbb{F}_{\qsf}$ represents a  finite field with order $\qsf$;         
$\mathbf{M}^{\text{T}}$  and $\mathbf{M}^{-1}$ represent the transpose  and the inverse of matrix $\mathbf{M}$, respectively;
 the matrix $[a;b]$ is written in a Matlab form, representing $[a,b]^{\text{T}}$;
$\text{rank}(\mathbf{M})$ represents the rank of matrix $\mathbf{M}$;
$\mathbf{I}_n$ represents the identity matrix with dimension $n \times n$;
${\bf 0}_{m \times n}$ represents the zero  matrix with dimension $m\times n$; 
$(\mathbf{M})_{m \times n}$ represents that the dimension of matrix $\mathbf{M}$ is $m \times n$;
$\mathbf{M}^{(\Sc)_{\rm r}}$ represents the sub-matrix of $\mathbf{M}$ which is composed of the rows  of $\mathbf{M}$ with indices in $\Sc$ (here $\rm r$ represents `rows'); 
$\mathbf{M}^{(\Sc)_{\rm c}}$ represents the sub-matrix of $\mathbf{M}$ which is composed of the columns  of $\mathbf{M}$ with indices in $\Sc$ (here $\rm c$ represents `columns'); 
 $\text{det}(\mathbf{M})$ represents the determinant matrix $\mathbf{M}$;
%$\mathbf{M}_{\Sc,\Vc}$ represents the sub-matrix of $\mathbf{M}$ by selecting from $\mathbf{M}$, the rows with indices in $\Sc$  and the columns with indices in $\Vc$.
 $\text{Mod} (b,a)$ represents the modulo operation on $b$ with  integer divisor $a$ and in this paper we let $\text{Mod}(b,a)\in \{1,\ldots,a \}$ (i.e., we let $ \text{Mod}(b,a)=a$ if $a$ divides $b$);
%the number of $k$-permutations of $n, n\geq k,$ is indicated as $P(n,k):=n\cdot(n-1)\ldots(n-k+1)$.
we let $\binom{x}{y}=0$ if $x<0$ or $y<0$ or $x<y$.
 In this paper, for each set  of integers  $\Sc$, we sort the elements in $\Sc$ in an increasing order and denote the $i^{\text{th}}$ smallest element by $\Sc(i)$, i.e., $\Sc(1)<\ldots<\Sc(|\Sc|)$.

The main network parameters and notations  are given in Table~\ref{tab:notations} at the top of the next page.
\begin{table*}
\centering
\protect\caption{Main notations}\label{tab:notations}
\begin{tabular}{|c|c|}
\hline 
\textbf{Notations} & \textbf{Semantics} \tabularnewline
\hline  
 $\Ksf$& number of datasets \tabularnewline
\hline  $\Nsf$ & number of workers\tabularnewline
\hline 
 $\Nsf_{\rm r}$& number of workers the master should wait for \tabularnewline
 \hline  $\Zc_n$ & set of datasets assigned to worker $n$ \tabularnewline
\hline  $\Msf$ & computation cost (i.e., number of datasets assigned to each worker)  \tabularnewline
  \hline  $X_n$ & transmission of worker $n$ \tabularnewline
    \hline  $T_n$ & number of symbols in $X_n$ \tabularnewline
  \hline  $X_{\Ac}$ &  $\{X_{n}:n\in \Ac\}$ \tabularnewline
  \hline 
 $\Rsf$&  communication cost \tabularnewline
 \hline 
 $\Rsf^{\star}$&  minimum communication cost over all   achievable computing schemes  \tabularnewline
  \hline 
 $\Rsf^{\star}_{\text{cyc}}$&  minimum communication cost over all   achievable computing schemes with the cyclic assignment  \tabularnewline
\hline  $D_n$ & the $n^{\text{th}}$ dataset \tabularnewline
\hline 
 $W_n=f_n(D_n)$ & the $n^{\text{th}}$ message \tabularnewline
 \hline 
 $\Lsf$& number of symbols of each message \tabularnewline
\hline   $g(W_1,\ldots,W_{\Nsf})={\bf F} [W_1;\ldots;W_{\Nsf}]$  &  task function (i.e., demanded linear combinations of messages)  \tabularnewline
\hline $\Ksf_{\rm c}$ &  number of  demanded linear combinations of messages (i.e., number of rows in ${\bf F} $)   \tabularnewline
\hline 
\end{tabular}
\end{table*}

\section{System Model}
\label{sec:model}

\subsection{Problem formulation}
\label{sub:problem formulation}
We formulate a $(\Ksf,\Nsf,\Nsf_{\rm r}, \Ksf_{\rm c}, \Msf)$ distributed linearly separable computation problem  over the canonical master-worker distributed system, as illustrated in Fig.~\ref{fig:system model}. The master wants to compute a function 
$$
f(D_1,   \ldots, D_{\Ksf})
$$
on  $\Ksf$ independent datasets $D_1,  \ldots, D_{\Ksf}$.   As the data sizes are large, we  distribute the computing task to a group of $\Nsf$ workers. 
 For distributed computation to be possible, we assume the function is \emph{separable} to some extent. As the simplest case, we assume the function is separable to each dataset, 
      \begin{subequations}
\begin{align}
   f(D_1, \ldots, D_{\Ksf}) &= g\big(f_1(D_1), \ldots, f_{\Ksf} (D_{\Ksf}) \big) \\
&=  g(W_1, \ldots, W_{\Ksf}),\label{eq:separated objective}
\end{align}
     \end{subequations} 
where we model $f_k(D_k)$,  $k \in [\Ksf]$ as the $k$-th message  $W_k$ and $f_k(\cdot)$ is an arbitrary function. We assume that the $\Ksf$ messages are independent and that each message is composed of $\Lsf$ uniformly i.i.d.   symbols  over a finite field $\mathbb{F}_{\qsf}$  for some large enough prime-power $\qsf$,  where $\Lsf$ is large enough such that any sub-message division is possible.\footnote{\label{foot:basis of log}In this paper, the basis of logarithm in the entropy terms  is   $\qsf$.} 
We consider the simplest case of the function $g(\cdot)$, the linear mapping. So we can rewrite the task function as
     \begin{subequations}
\begin{align}
g(W_1,   \ldots, W_{\Ksf})& = {\bf F}  \left[ \begin{array}{c}
W_1\\
\vdots \\
 W_{\Ksf}
\end{array} \right]=
\left[ \begin{array}{c}
F_1\\
\vdots \\
 F_{\Ksf_{\rm c}}
\end{array} \right],
\end{align}
\label{eq:objective matrix-matrix}
     \end{subequations}
where ${\bf F}$ is a matrix known by the master and the workers with dimension $\Ksf_{\rm c} \times \Ksf$, whose elements are  uniformly i.i.d.        over  $\mathbb{F}_{\qsf}$. In other words, $g(W_1,  \ldots, W_{\Ksf})$ contains 
 $\Ksf_{\rm c}$   linear combinations of   the $\Ksf$ messages, whose coefficients are uniformly i.i.d.    over $\mathbb{F}_{\qsf}$.  In this paper, we consider the case where $\Ksf_{\rm c} \leq \Ksf$.\footnote{\label{foot:trivial Kc>K} For the case where  $\Ksf_{\rm c}> \Ksf$, it is straightforward to use the same code for the case  where $ \Ksf_{\rm c}= \Ksf$, since all $\Ksf$ messages can be decoded individually. } 
Note that each component function $f_k$ where $k \in [\Ksf]$ is not restricted to be linear. 
We also assume that  $\frac{\Ksf}{\Nsf} $ is an integer.\footnote{\label{foot:extension on no division} The case $\Nsf$ does not divide $\Ksf$ will be specifically considered in Section~\ref{sub:N does not divide K} where we extend the proposed distributed computing scheme to the general case.}

A computing scheme for our problem contains three phases, {\it data assignment}, {\it computing}, and {\it decoding}. 
\paragraph*{Data assignment phase}
We assign each dataset $D_k$ where $k \in [\Ksf]$ to a subset of $\Nsf$ workers in an uncoded manner. The set of datasets assigned to worker $n \in [\Nsf]$ is denoted by $\Zc_n$, where $\Zc_n \subseteq [\Ksf]$. 
The   assignment  constraint is that 
\begin{align}
|\Zc_n| \leq \Msf, \  \forall  n \in [\Nsf], \label{eq:assignment constraint} 
\end{align}
where  $\Msf$ represents the computation cost as explained in Footnote~\ref{foot:computation cost}.
 The assignment function of worker $n$ is denoted by $\varphi_n$, where 
\begin{align}
& \Zc_n = \varphi_n({\bf F}) \subseteq [\Ksf], \\
&\varphi_n  :  [\mathbb{F}_{\qsf}]^{ \Ksf_{\rm c}   \Ksf } \to \Omega_{\Msf}(\Ksf),
\end{align}
and  $\Omega_{\Msf}(\Ksf)$  represents the set of all subsets of $[\Ksf]$ of size not larger than $\Msf$. In other words, the data assignment phase is uncoded.

\paragraph*{Computing phase}
Each worker $n \in[\Nsf]$ first computes the message $W_k = f_k (D_k)$ for each $k \in \Zc_n$. Then it computes 
\begin{align}
 X_n = \psi_n(\{W_k:  k \in \Zc_n\}, {\bf F} )
 \end{align}
  where the encoding function $\psi_n$ is such that
\begin{align} 
\psi_n &:  [\mathbb{F}_{\qsf}]^{ |\Zc_n| \Lsf} \times  [\mathbb{F}_{\qsf}]^{ \Ksf_{\rm c}   \Ksf } \to [\mathbb{F}_{\qsf}]^{ \Tsf_n },  
\label{eq: encoding function def}
\end{align}
and  $ \Tsf_n$ represents the length of $ X_n $. Finally, worker $n$ sends $X_n$ to the master. 

%We assume that the complexity of computing the messages is much higher than computing each function $\psi_n$. %such that the computation cost on $ X_n $ is negligible.  Hence, in this paper we denote the computation cost by $\Msf$.
 
\paragraph*{Decoding phase}
The master only waits for   the $\Nsf_{\rm r}$ fastest workers' answers to compute $ g(W_1, \ldots, W_{\Ksf})$. Hence, the computing scheme can tolerate   $\Nsf - \Nsf_{\rm r}$ stragglers.
Since the master does not know a priori which workers are stragglers, the computing scheme should be designed so that from the answers of any $\Nsf_{\rm r}$   workers, the master can recover $g(W_1,   \ldots, W_{\Ksf})$. More precisely, for any subset of workers $\Ac \subseteq [\Nsf]$ where $|\Ac|=\Nsf_{\rm r}$,  with the definition
\begin{align}
X_{\Ac}:=\{X_n: n\in \Ac \},
\end{align} 
 there exists a decoding function $\phi_{\Ac}$ such that
\begin{align}
&\hat{g}_{\Ac}= \phi_{\Ac}\big( X_{\Ac}, {\bf F} \big)   ,
\end{align}
 where  the decoding function $\phi_{\Ac}$ is such that
\begin{align}
& \phi_{\Ac} :  [\mathbb{F}_{\qsf}]^{\sum_{n \in \Ac} \Tsf_n } \times [\mathbb{F}_{\qsf}]^{\Ksf_{\rm c}   \Ksf}  \to [\mathbb{F}_{\qsf}]^{\Ksf_{\rm c}   \Lsf}.
\end{align}

The  worst-case   probability of
error is defined as
\begin{align}
 \varepsilon:= \max_{\Ac  \subseteq [\Nsf]: |\Ac|= \Nsf_{\rm r}} \Pr\{ \hat{g}_{\Ac} \neq g(W_1,   \ldots, W_{\Ksf}) \}. 
\end{align}

In addition, we denote the communication cost by,  %which measures the normalized number of symbols downloaded by the master  
\begin{align}
\Rsf  :=  \max_{\Ac  \subseteq [\Nsf]: |\Ac|= \Nsf_{\rm r}} \frac{ \sum_{n \in \Ac} \Tsf_n}{   \Lsf  }, \label{eq:communicaton rate}
\end{align}
representing %the ratio between the number of symbols in the objective function  and  
 the maximum normalized   number of symbols downloaded by the master from any $\Nsf_{\rm r}$ responding workers.
The communication cost $\Rsf$ is achievable if there exists a computing scheme with assignment, encoding, and decoding functions such that 
\begin{align}
 \lim_{\qsf \to \infty}  \varepsilon =0.
\end{align}
 The minimum  communication cost over all possible achievable computing schemes  is  denoted  by $\Rsf^{\star}$. 
 Since the elements of ${\bf F}$ are uniformly i.i.d. over larger enough field, ${\bf F}$ is   full-rank  with high probability. By the simple cut-set bound, we have 
 \begin{align}
 \Rsf^{\star} \geq \Ksf_{\rm c}.\label{eq:trivial lower bound}
 \end{align}

% It is easy to show that $\Rsf^{\star} \geq \Ksf_{\rm c}$.\footnote{\label{foot:trivial range of capacity}  
%Sine the elements of ${\bf F}$ are i.i.d. over larger enough field, ${\bf F}$ is  with high probability full-rank. The requested $\Ksf_{\rm c}$ linear combinations of the $\Ksf\geq \Ksf_{\rm c}$ %messages can be re-written as $\Ksf_{\rm c}$ independent transformed messages, each of which has $\Lsf$ i.i.d. symbols.  In order to recover these $\Ksf_{\rm c}$ independent transformed %messages, the master needs at least $\Ksf_{\rm c} \Lsf$ symbols. Hence, $\Rsf^{\star} \geq \Ksf_{\rm c} $.}
 
The following lemma   provides the minimum number of workers to whom each dataset should be assigned.
\begin{lem}
\label{lem:minimum computation cost}
 Each dataset must be assigned to  at least $\Nsf - \Nsf_{\rm r}+1$ workers.
 \hfill $\square$ 
\end{lem} 
\begin{IEEEproof}
Assume there exists one dataset (assumed to be $D_k$) assigned to only $\ell$ workers where  $\ell < \Nsf - \Nsf_{\rm r}+ 1$. It can be seen that there exist at least $\Nsf_{\rm r}$ workers which does not know  $D_k$. Hence,   the answers of these  $\Nsf_{\rm r}$ workers do not have any information of $W_k$,  and thus cannot reconstruct 
$g(W_1,   \ldots, W_{\Ksf})$  (recall that $g(W_1,   \ldots, W_{\Ksf})$ depends on $W_k$  with high probability).
\end{IEEEproof} 
 
In this paper, we consider the case where the computation cost is minimum, i.e., each dataset is assigned to $\Nsf - \Nsf_{\rm r}+1$ workers and   
$$
\Msf=|\Zc_1|=\cdots=|\Zc_{\Nsf}|= \frac{\Ksf}{\Nsf} (\Nsf-\Nsf_{\rm r} +1).
$$
 The objective of this paper is to characterize the minimum communication cost for the case where the computation cost is minimum.

 We then review the cyclic assignment, which was widely used in the  existing works on   the distributed gradient coding problem in~\cite{gradiencoding} (which is a special case of the consdered problem as explained in the next subsection), such as the gradient coding schemes in~\cite{gradiencoding,improvedGC2017halbawi,MDSGC2018raviv,efficientgradientcoding}.
 For each dataset $D_k$ where $k\in  [\Ksf]$, we assign $D_k$ to worker $j$, where $j \in \big\{\text{Mod}(k,\Nsf),  \text{Mod}(k-1,\Nsf),\ldots,  \text{Mod}(k-\Nsf+\Nsf_{\rm r}, \Nsf ) \big\}$.\footnote{By convention, we let $\text{Mod}(b,a) \in [1:a]$, and let $\text{Mod}(b,a) =a$ if $a$ divides $b$.} In other words, the set of datasets assigned  to  worker $n \in [\Nsf]$   is 
\begin{align}
\Zc_n &=  \underset{p \in \left[0:  \frac{\Ksf}{\Nsf} -1 \right]}{\cup}   \big\{\text{Mod}(n,\Nsf)+ p  \Nsf , \text{Mod}(n+1,\Nsf)+ p  \Nsf , \ldots, \nonumber\\& \text{Mod}(n+\Nsf-\Nsf_{\rm r},\Nsf)+ p  \Nsf  \big\}  \label{eq:cyclic assignment n divides k}
\end{align}
with cardinality $\frac{\Ksf}{\Nsf} (\Nsf-\Nsf_{\rm r} +1)$.
For example, if $\Ksf=\Nsf=4$ and $\Nsf_{\rm r}=3$, by the cyclic assignment  with $p=0$ in~\eqref{eq:cyclic assignment n divides k}, we assign 
\begin{align*}
&D_1, D_2, D_3 \ \text{to woker 1};\\
&D_2, D_3, D_4 \ \text{to woker 2};\\
&D_3, D_4, D_1 \ \text{to woker 3};\\
&D_4, D_1, D_2 \ \text{to woker 4}.
\end{align*}
  The minimum communication cost under  the cyclic assignment in~\eqref{eq:cyclic assignment n divides k} is  denoted   by $\Rsf^{\star}_{\rm cyc}$. 
 
  \subsection{Connection to existing problems}
\label{sub:connection to existing problem}
%The formulated distributed linearly separable computation  problem can be seen as the generalization of the distributed gradient coding problem in~\cite{gradiencoding} and the distributed linear transform problem in~\cite{shortdot2016dutta}.

\paragraph*{{\bf Distributed gradient coding}}
When $f_k(D_k)$, $k \in [\Ksf]$, represents the  partial gradient vector of the loss at the current estimate of the dataset $D_k$ and ${\bf F}= [1,\ldots,1]$, we have 
\begin{align}
f(D_1, \ldots, D_{\Ksf}) = f_1(D_1) + \cdots +f_{\Ksf}(D_{\Ksf}), 
\end{align}
representing the gradient of a  generic loss function. 
In this case, our problem reduces to  the distributed gradient coding problem in~\cite{gradiencoding}.
Hence, the distributed gradient coding problem in~\cite{gradiencoding} is a special case of the distributed linearly separable computation problem with   $\Ksf_{\rm c}=1$.  For the case where the   computation cost is minimum, based on the cyclic assignment in~\eqref{eq:cyclic assignment n divides k} and a random code construction, the authors in~\cite{gradiencoding} proposed a gradient coding scheme which lets each worker compute and send one linear combination of the messages related to its assigned datasets, while  the achieved  communication cost of this scheme is optimal under the constraint of linear coding~\cite{efficientgradientcoding}.  Instead of random code construction, a deterministic  code construction  was proposed in~\cite{MDSGC2018raviv}. The authors in~\cite{improvedGC2017halbawi} improved the decoding delay/complexity  by using     Reed–Solomon codes.

The authors in~\cite{efficientgradientcoding} characterized the optimal tradeoff between   the  computation cost  and communication cost for   the distributed gradient coding problem.   A distributed computing scheme achieving the same optimal computation-communication costs tradeoff as in~\cite{efficientgradientcoding} but with lower decoding complexity, was recently proposed in~\cite{communicationlowcom2020kadhe}. 

Some other extensions on the distributed gradient coding problem in~\cite{gradiencoding} were also considered in the literature. For instance, the authors in~\cite{treeGC2019reisi} extended the gradient coding strategy to a tree-topology  where the workers are located, and a fixed fraction of  children nodes per   parent node may be straggler. 
The case where the number of stragglers is not given in prior was considered in~\cite{nearoptimalGC2018li}. 
In~\cite{multimessageGC2020ozfa}, each worker sends multiple linear combinations such that the master does not always need to wait for the answers of $\Nsf_{\rm r}$ workers (i.e., from some `good' subset of workers with the cardinality less than $\Nsf_{\rm r}$, the master can recover the task function). 
It can be seen that these extended models are different from the considered problem in this paper.

\paragraph*{{\bf Distributed linear transform}}
The distributed linear transform problem in~\cite{shortdot2016dutta} aims to compute the linear transform ${\bf A} \xv$ where $\xv$   
is the input vector and ${\bf A} $ is a given matrix with dimension $\Ksf_{\rm c} \times \Ksf$. We should design a coding vector $\cv_n$ for each worker $n \in [\Nsf]$ (which then computes $\cv_n \xv$) such that from the computation results of any $\Nsf_{\rm r}$ workers we can reconstruct ${\bf A}\xv$. Meanwhile, in order to have low computation cost, each coding vector should be sparse and the number of its non-zero  elements should be no more than $\Msf$, where $\Msf$ should be minimized. %In other words, each worker can only access to up to $\Msf$ elements in $\xv$.
Hence, the distributed linear transform problem in~\cite{shortdot2016dutta}  can be seen a special case of the distributed linearly separable computation problem with       $  \Tsf_n =\Lsf=1$ for each $n \in [\Nsf]$ (recall that $\Tsf_n$ represents the number of symbols transmitted by worker $n$). In other words, in this paper we consider the case where the computation cost is minimum and search for the minimum communication cost, while the authors in~\cite{shortdot2016dutta} 
 considered the case where $\Lsf=1$ and the communication cost is minimum, and searched for the minimum computation cost.
 A computing scheme was proposed in~\cite{shortdot2016dutta} which needs $\Msf=\frac{\Ksf}{\Nsf} (\Nsf-\Nsf_{\rm r}+\Ksf_{\rm c})$.  
 The authors in~\cite{sparsification2017} further improved the distributed linear transform scheme in~\cite{shortdot2016dutta} by proposing a computing scheme to let each     worker  $n \in [\Nsf]$ only  access $\Msf^{\prime}_n$ elements in $\xv$, where   $ \Ksf  (\Nsf-\Nsf_{\rm r}+\Ksf_{\rm c})-\Nsf \Nsf_{\rm r} <\sum_{n\in [\Nsf]} \Msf^{\prime}_{n} < \Nsf \frac{\Ksf}{\Nsf} (\Nsf-\Nsf_{\rm r}+\Ksf_{\rm c}) $.
 
 The authors in~\cite{codedlineartrans2018wang} considered another distributed linear transform problem with a different   sparsity constraint compared to~\cite{shortdot2016dutta}. The distributed linear transform problem in~\cite{codedlineartrans2018wang} can be seen as a special case of the distributed linearly separable computation problem with       $  \Tsf_n =\Lsf=1$ and $\Ksf_{\rm c}=\Ksf$.

 \paragraph*{{\bf Distributed  matrix-vector and matrix-matrix multiplications}}
 Distributed computing techniques against stragglers were also used to compute matrix-vector multiplication as ${\bf A b}$~\cite{universally2019ramamoo,matrixvectorconvo2017das,finitematrixvector2018hadda} and matrix-matrix multiplication  as ${\bf A B}$~\cite{speedup2018Lee,highdimensional2017lee,sparsematrix2018wang,polyoptYu2017,yu2020stragglermitigation,dutta2020optimalRT,struresis2020ramamoo,jia2019matrixCSA}.
The general technique is to partition each input matrix  into sub-matrices and assign some linear combinations  of   all sub-matrices (from MDS codes, polynomial codes, etc.) 
to  the workers   
  without considering the sparsity of the coding vectors/matrices.   Thus, the assignment phase is coded.  
 %each worker can  fully access to all the inputs of the desired product.
 
% Instead, in our considered distributed linearly separable computation problem, to compute  ${\bf F}  [W_1; \ldots;W_{\Ksf}]$ in~\eqref{eq:objective matrix-matrix}, each worker can only access to a subset of the messages in $\{W_1, \ldots,W_{\Ksf} \}$.
 
  \paragraph*{{\bf Distributed multivariate polynomial computation}}
 Similar difference as above also appears between the considered distributed linearly separable computation problem and the  distributed multivariate polynomial computation problem in~\cite{lagrange2019yu}. It was shown  in~\cite{lagrange2019yu} that the gradient   descent can be computed distributedly by using a coding scheme based on the Lagrange polynomial.  However, the assignment phase of  the Lagrange distributed computing scheme  in~\cite{lagrange2019yu} is coded.

%where ${\bf W}$ is the collection of the messages, ${\bf W} = (W_1; W_2; \cdots; W_K)$ and ${\bf F}$ represents the $K_c$ (`$c$' stands for \emph{computing})  linear functions that we wish to compute. Let us emphasize that ${\bf F}_{K_c \times K}$ is assumed to be a constant and deterministic matrix (as $K_c$ and $K$ are finite constants) and $H(W_k)$ is allowed to grow (traditional Shannon theoretic setting).

%We adopt the convention that lower case letter represents the index over a range given by the upper case letter. For example, $k$ is the message index and takes value over $\{1, 2, \cdots, K\} \define [1:K]$. Similarly, $n \in [1:N]$ is the worker (server) index.

  In summary, compared to the distributed computing schemes with coded assignment phase, the main   challenge of designing computing schemes with uncoded assignment phase  is that besides the decodability constraint, 
we should additionally guarantee that in the transmitted linear combination(s) by each worker, the coefficients of the unassigned elements are $0$.
% even if any $\Nsf_{\rm r}$ coding vectors  (i.e., $\cv_n$ where $n \in \Ac$ and $\Ac$ is the set of responding workers) are linearly independent, we cannot guarantee that the master can reconstruct    ${\bf A} \xv$ from  $\cv_n \xv$ where $n \in \Ac$. 

 \section{Main Results}
\label{sec:main}
We first propose a converse bound on the   minimum communication cost in the following theorem, which will be proved in Appendix~\ref{sec:proof of converse} inspired by the converse bound for the coded caching problem with uncoded cache placement~\cite{indexcodingcaching2020,exactrateuncoded}.
\begin{thm}[Converse] 
\label{thm:converse}
 For the  $(\Ksf,\Nsf,\Nsf_{\rm r}, \Ksf_{\rm c}, \Msf)$ distributed linearly separable computation problem with  $\Msf= \frac{\Ksf}{\Nsf} (\Nsf-\Nsf_{\rm r} +1) $,
\begin{itemize}
\item when $\Ksf_{\rm c} \in \left[ \left\lceil \frac{\Ksf}{\binom{\Nsf}{\Nsf-\Nsf_{\rm r}+1}} \right\rceil  \right]$, we have 
\begin{subequations}
\begin{align}
\Rsf^{\star} \geq   \Nsf_{\rm r} \Ksf_{\rm c}. \label{eq:case 1 converse}
\end{align} 
\item when $\Ksf_{\rm c} \in  \left( \left\lceil \frac{\Ksf}{\binom{\Nsf}{\Nsf-\Nsf_{\rm r}+1}} \right\rceil :\Ksf \right] $, we have 
\begin{align}
\Rsf^{\star} \geq  \max\left\{  \Nsf_{\rm r} \left\lceil \frac{\Ksf}{\binom{\Nsf}{\Nsf-\Nsf_{\rm r}+1}} \right\rceil , \Ksf_{\rm c} \right\} . \label{eq:case 2 converse}
\end{align} 
\end{subequations}
\end{itemize}
\hfill $\square$ 
\end{thm}

For the case with  $\Ksf_{\rm c}=1$ and $\Msf= \frac{\Ksf}{\Nsf} (\Nsf-\Nsf_{\rm r} +1) $  which reduces to the   distributed gradient coding problem in~\cite{gradiencoding},
from Theorem~\ref{thm:converse}   and the gradient coding scheme in~\cite{gradiencoding} (each worker sends one linear combination of the assigned messages), we can directly prove the following corollary.
\begin{cor}
 \label{cor:direct optimality}
 For the  $(\Ksf,\Nsf,\Nsf_{\rm r}, \Ksf_{\rm c}, \Msf)$ distributed linearly separable computation problem  with  $\Msf= \frac{\Ksf}{\Nsf} (\Nsf-\Nsf_{\rm r} +1) $ and $\Ksf_{\rm c}=1$,  we have 
\begin{align}
\Rsf^{\star} =  \Nsf_{\rm r} . \label{eq:Kc 1 optimal}
\end{align} 
\hfill $\square$ 
 \end{cor}
 Note that the optimality of the gradient coding scheme in~\cite{gradiencoding} for the     distributed gradient coding problem was proved in~\cite{efficientgradientcoding}, but under the constraint that the encoding functions in~\eqref{eq: encoding function def} are linear.  In Corollary~\ref{cor:direct optimality}, we remove this constraint.
 
With the cyclic assignment in Section~\ref{sub:problem formulation}, we then propose a novel achievable distributed computing scheme  
%We provide the needed communication cost of the proposed scheme   in the following theorem,
 whose detailed proof could be found in Section~\ref{sec:achie}.  
 \begin{thm}[Proposed distributed computing scheme] 
\label{thm:achie}
 For the  $(\Ksf,\Nsf,\Nsf_{\rm r}, \Ksf_{\rm c}, \Msf)$ distributed linearly separable computation problem  with  $\Msf= \frac{\Ksf}{\Nsf} (\Nsf-\Nsf_{\rm r} +1) $, the communication cost $\Rsf_{{\rm ach}}$ is achievable, where 
\begin{itemize}
 \item when  $\Ksf_{\rm c} \in \left[ 1: \frac{\Ksf}{\Nsf} \right)$, 
 \begin{subequations}
\begin{align}
  \Rsf_{{\rm ach}}= \Nsf_{\rm r} \Ksf_{\rm c}; \label{eq:achie case 1}
\end{align} 
\item when  $ \Ksf_{\rm c}  \in \left[ \frac{\Ksf}{\Nsf}: \frac{\Ksf}{\Nsf}\Nsf_{\rm r} \right]  $,
\begin{align}
\Rsf_{{\rm ach}}= \frac{\Ksf}{\Nsf} \Nsf_{\rm r}  ; \label{eq:achie case 2}
\end{align}
\item when  $ \Ksf_{\rm c} \in \left(  \frac{\Ksf}{\Nsf}\Nsf_{\rm r}  : \Ksf \right]$,
\begin{align}
\Rsf_{{\rm ach}}=  \Ksf_{\rm c}. \label{eq:achie case 3}
\end{align}
\label{eq:achieve three cases}
\end{subequations}
\end{itemize}
\hfill $\square$ 
\end{thm}
  In Theorem~\ref{thm:achie}, we consider three regimes with respect to the value of $\Ksf_{\rm c}$ and the main ingredients are as follows.
\begin{enumerate}
\item  $\Ksf_{\rm c} \in \left[ 1: \frac{\Ksf}{\Nsf} \right)$. By some linear transformations on the request matrix ${\bf F}$, we treat  the considered problem as $\Ksf_{\rm c}$    sub-problems in each of which the master requests one linear combination of messages. Thus by using the coding scheme in Corollary~\ref{cor:direct optimality} for each sub-problem, we can let the master recover the general task function.
\item $   \Ksf_{\rm c}  \in \left[\frac{\Ksf}{\Nsf} :   \frac{\Ksf}{\Nsf}\Nsf_{\rm r} \right]$. This is the most  interesting case, where we propose a   computing scheme based on the linear space intersection (see Remark~\ref{rem:linear space explanation} for further explanations), with the communication cost  equal to 
the case where $\Ksf_{\rm c}=\frac{\Ksf}{\Nsf}$. We generate $ \frac{\Ksf}{\Nsf}\Nsf_{\rm r} - \Ksf_{\rm c}$ virtually requested linear combinations of messages such that the master totally recover $ \frac{\Ksf}{\Nsf}\Nsf_{\rm r} $ effective linear combinations of messages from   the responses of any $\Nsf_{\rm r}$ workers. Each worker transmits $\frac{\Ksf}{\Nsf} $ linear combinations of messages which lie in the intersection of the linear spaces of its known messages and the effective demanded linear combinations. From a highly non-trivial proof based on the Schwartz-Zippel lemma~\cite{Schwartz,Zippel,Demillo_Lipton}, where the main challenge is to prove that the multivariate polynomials are generally non-zero (see Appendix~\ref{sec:lemma nonzero n=k}), we  show that the responses of any $\Nsf_{\rm r}$ workers are linearly independent with high probability, and thus are able to reconstruct the effective demanded linear combinations.
\item $\Ksf_{\rm c} \in \left(  \frac{\Ksf}{\Nsf}\Nsf_{\rm r}:\Ksf \right] $. To recover $\Ksf_{\rm c}$ linear combinations of the $\Ksf$ messages, we propose a   computing scheme to let the master totally receive $\Ksf_{\rm c}$ coded messages with $\Lsf$ symbols each, i.e., $\Rsf^{\star} =\Ksf_{\rm c}$ is achieved.
\end{enumerate}

%{\blue
%Note that while proving the decodability of the proposed scheme in Theorem~\ref{thm:achie}, we use  the Schwartz-Zippel lemma~\cite{Schwartz,Zippel,Demillo_Lipton} in Appendix~\ref{sec:proof   %of SZlemma}. For  the non-zero multivariate polynomial condition, we   numerically verify   all cases that $\Nsf \leq 50$,  and conjecture that the condition holds for any case.
% }

\iffalse
\begin{rem}
\label{eq:assignment dependent on F}
Note that the proposed distributed computing scheme is with the cyclic assignment in~\cite{gradiencoding},  which is independent of the elements in the request matrix ${\bf F}$.
Instead, the distributed linear transformation scheme in~\cite{shortdot2016dutta} used an assignment   based on the   request matrix. Compared these two assignments, one advantage of the cyclic assignment   is its simplicity. Another advantage is that if the master wants to compute multiple computing tasks   on these $\Ksf$ datasets besides $f(D_1,\ldots,D_{\Ksf})$, the proposed scheme based on the cyclic assignment needs not to re-assign the datasets to the workers for each task, while the one  in~\cite{shortdot2016dutta} needs to do the re-assignment. 
\hfill $\square$ 
\end{rem} 
\fi
 
 \begin{rem}
\label{eq:real number}
Note that,  when the operations are on the field of real numbers, the proposed computing scheme in Theorem~\ref{thm:achie}  can   work with high probability if each element in ${\bf F}$ is uniformly i.i.d. over a large enough finite set of real numbers or over an interval of real numbers.
 \hfill $\square$ 
 \end{rem}

 By comparing the proposed converse bound in Theorem~\ref{thm:converse} and the achievable scheme in Theorem~\ref{thm:achie}, we can directly derive the  following optimality results.
 \begin{thm}[Optimality]
\label{thm:optimality} 
  For the  $(\Ksf,\Nsf,\Nsf_{\rm r}, \Ksf_{\rm c}, \Msf)$ distributed linearly separable computation problem with  $\Msf= \frac{\Ksf}{\Nsf} (\Nsf-\Nsf_{\rm r} +1) $,
  \begin{itemize}
  \item when $\Ksf=\Nsf$, we have 
  \begin{subequations}
  \begin{align}
  \Rsf^{\star}= \begin{cases} \Nsf_{\rm r} , & \text{ if }  \Ksf_{\rm c} \in [\Nsf_{\rm r}]; \\ \Ksf_{\rm c},  & \text{ if }  \Ksf_{\rm c} \in (\Nsf_{\rm r}: \Ksf] ; \end{cases} \label{eq:optimality N=K}
  \end{align}
    \item  when $\Ksf_{\rm c} \in \left[ \left\lceil \frac{\Ksf}{\binom{\Nsf}{\Nsf-\Nsf_{\rm r}+1}} \right\rceil  \right]$, we have 
\begin{align}
\Rsf^{\star} =  \Nsf_{\rm r} \Ksf_{\rm c}; \label{eq:optimality Kc small}
\end{align} 
  \item  when $\Ksf_{\rm c} \in \left[  \frac{\Ksf}{\Nsf}\Nsf_{\rm r}  : \Ksf \right]$, we have 
  \begin{align}
  \Rsf^{\star}= \Ksf_{\rm c}. \label{eq:optimality Kc large}
  \end{align}
  \end{subequations}
  \end{itemize}
  \hfill $\square$ 
 \end{thm}
 From Theorem~\ref{thm:optimality},  it can be seen that when $\Ksf =\Nsf$ and $\Ksf_{\rm c} \in [\Nsf_{\rm r}]$, the optimal communication cost is always $\Nsf_{\rm r}$ (i.e., each worker sends one linear combination of the messages from its assigned datasets). Thus %compared to  the distributed gradient coding problem in~\cite{efficientgradientcoding} with $\Ksf=\Nsf$, 
 we prove that with the same communication cost as the optimal gradient coding scheme in~\cite{gradiencoding} for the distributed gradient coding problem (from which the master recovers $W_1+\cdots W_{\Ksf}$), our propose scheme can let the master recover any additional $\Nsf_{\rm r}-1$ linear combinations of the $\Ksf$ messages whose coefficients are uniformly i.i.d.    over $\mathbb{F}_{\qsf}$ with high probability.
 
 \iffalse
From Theorem~\ref{thm:optimality}, we can also derive the minimum communication cost  for the case where $\Nsf_{\rm r} \in \{1,2,\Nsf\}$, whose proof could be found in Appendix~\ref{sec:direct optimality Nr 12N}.
\begin{cor}
 \label{cor:direct optimality Nr 12N}
 For the  $(\Ksf,\Nsf,\Nsf_{\rm r}, \Ksf_{\rm c}, \Msf)$ distributed linearly separable computation problem with  $\Msf= \frac{\Ksf}{\Nsf} (\Nsf-\Nsf_{\rm r} +1) $, we have 
 \begin{itemize}
 \item when $\Nsf_{\rm r}=1$, 
 \begin{subequations}
 \begin{align}
  \Rsf^{\star}= \Ksf_{\rm c};\label{eq:Nr=1}
  \end{align}
  \item when $\Nsf_{\rm r}=2$,
  \begin{align}
  \Rsf^{\star}= \begin{cases} 2 \Ksf_{\rm c} , & \text{ if }  \Ksf_{\rm c} \in \left[\frac{\Ksf}{\Nsf} \right]; \\ 2 \frac{\Ksf}{\Nsf},  & \text{ if }  \Ksf_{\rm c} \in \left(\frac{\Ksf}{\Nsf} :2\frac{\Ksf}{\Nsf}\right] ;   \\ \Ksf_{\rm c},  & \text{ if }  \Ksf_{\rm c} \in \left(2\frac{\Ksf}{\Nsf}: \Ksf \right] ;\end{cases} \label{eq:Nr=2}
 \end{align}
   \item when $\Nsf_{\rm r}=\Nsf$,
 \begin{align}
  \Rsf^{\star}= 
\begin{cases} \Nsf \Ksf_{\rm c} , & \text{ if }  \Ksf_{\rm c} \in \left[\frac{\Ksf}{\Nsf} \right]; \\ \Ksf,  & \text{ if }  \Ksf_{\rm c} \in \left(\frac{\Ksf}{\Nsf} : \Ksf \right] .\end{cases} \label{eq:Nr=N}   
  \end{align}
  \end{subequations}
 \end{itemize}
\hfill $\square$ 
 \end{cor}
Note that directly from Corollary~\ref{cor:direct optimality Nr 12N}, we have that the proposed scheme is optimal when $\Nsf \leq 3$. 
\fi
 
 In general, the minimum communication cost in the regime where  $\Ksf_{\rm c} \in \left(\left\lceil \frac{\Ksf}{\binom{\Nsf}{\Nsf-\Nsf_{\rm r}+1}} \right\rceil : \frac{\Ksf}{\Nsf}\Nsf_{\rm r}   \right)$ is still open. The following theorem claims that the proposed achievable scheme is optimal under the constraint of the cyclic assignment in~\cite{gradiencoding}, whose proof is in Appendix~\ref{sec:proof cyclic optimality}.
  \begin{thm}[Optimality under the cyclic assignment in~\cite{gradiencoding}] 
\label{thm:cyclic optimality}
 For the  $(\Ksf,\Nsf,\Nsf_{\rm r}, \Ksf_{\rm c}, \Msf)$ distributed linearly separable computation problem with   $\Msf= \frac{\Ksf}{\Nsf} (\Nsf-\Nsf_{\rm r} +1) $, the minimum communication cost  under  the cyclic assignment is
\begin{align}
\Rsf^{\star}_{\rm cyc}=\Rsf_{{\rm ach}}, \label{eq:cyclic optimality}
\end{align} 
  where $\Rsf_{{\rm ach}}$ is  given in~\eqref{eq:achieve three cases}. 
 \hfill $\square$ 
\end{thm}

\section{Achievable Distributed Computing Scheme}
\label{sec:achie}  
In this section, we introduce the proposed distributed computing scheme with the cyclic assignment in~\cite{gradiencoding}. As shown in Theorem~\ref{thm:achie}, we divide the range of $\Ksf_{\rm c} $ (which is $[\Ksf]$) into three regimes, and present the corresponding scheme in the order, $   \Ksf_{\rm c}  \in \left[\frac{\Ksf}{\Nsf} :   \frac{\Ksf}{\Nsf}\Nsf_{\rm r} \right]$,  $\Ksf_{\rm c} \in \left[1: \frac{\Ksf}{\Nsf} \right)$, and $\Ksf_{\rm c} \in \left(  \frac{\Ksf}{\Nsf}\Nsf_{\rm r}:\Ksf \right] $.

\subsection{$\Ksf_{\rm c}  \in \left[\frac{\Ksf}{\Nsf} :   \frac{\Ksf}{\Nsf}\Nsf_{\rm r} \right]$}
\label{sub:Kc in middle}
We first illustrate the main idea in the following example.
\begin{example}[$\Nsf = 3, \Ksf = 6, \Ksf_{\rm c} = 4, \Nsf_{\rm r} = 2$, $\Msf=4$]
\label{ex:regime 1}
\rm 
In this example, it can  be seen that $\Ksf_{\rm c}= \frac{\Ksf}{\Nsf} \Nsf_{\rm r}$. For the sake of simplicity, in the rest of this paper while illustrating the proposed schemes through examples, we assume that the field is a large enough prime field. It will be proved that in general this assumption is not necessary  in our proposed schemes where we only need the field size $\qsf$ is large enough. 
Assume that the task function is \begin{align*}
f(D_1,   \ldots, D_{6})&= 
 \left[ \begin{array}{c}
F_1\\
F_2\\
F_3\\
F_4\\
\end{array} \right]
= {\bf F}
\left[ \begin{array}{c}
W_1\\
W_2 \\
W_3 \\
W_4\\ 
W_5\\ 
W_6
\end{array} \right] \\
&  =\left[ \begin{array}{c}
1 ,1 , 1 ,1 ,1 ,1\\
1 ,2 , 3 ,4 ,5 ,6\\
1 ,0 , 2 ,3 ,5 ,4\\  
1 ,2 , 1 ,4 ,4 ,0\\   
\end{array} \right]
\left[ \begin{array}{c}
W_1\\
W_2 \\
W_3 \\
W_4\\ 
W_5\\ 
W_6
\end{array} \right].
\end{align*}
 
\paragraph*{Data assignment phase}
By the   cyclic assignment described in Section~\ref{sub:problem formulation}, we assign that 
\begin{align*}
\begin{array}{rl|c|c|c|}\cline{3-3}\cline{4-4}\cline{5-5}
&&\rule{0pt}{1.2em}\mbox{Worker 1} &\rule{0pt}{1.2em}\mbox{Worker 2} & \rule{0pt}{1.2em}\mbox{Worker 3} \\\cline{3-3}\cline{4-4}\cline{5-5}
&& D_1&D_2  &D_1\\
&& D_2& D_3 &D_3 \\ \cdashline{3-3}\cdashline{4-4}\cdashline{5-5}   
&& D_4&D_5 & D_4 \\
&& D_5&D_6 &  D_6\\ 
\cline{3-3}\cline{4-4}\cline{5-5}
\end{array}
\end{align*}

\paragraph*{Computing phase}
We first focus on worker $1$, who first computes $W_1$, $W_2$, $W_4$, and $W_5$ based on its assigned datasets. 
In other words, $W_{i}$ where $i\in \{3,6\}$ cannot be computed by worker $1$. We retrieve the $i^{\text{th}}$ column of ${\bf F}$ where $i\in \{3,6\}$, to obtain
\begin{align}
{\bf F}^{(\{3,6\})_{\rm c}}=\left[ \begin{array}{c}
  1 , 1\\
  3 , 6\\
  2  ,4\\  
  1 , 0\\   
\end{array} \right].
\end{align}
We then search for  a vector basis for the left-side null space of ${\bf F}^{(\{3,6\})_{\rm c}}$. Note that  ${\bf F}^{(\{3,6\})_{\rm c}}$ is a full-rank matrix with dimension $4 \times 2$. Hence, a vector basis for its left-side null space contains $4-2=2$   linearly independent vectors  with dimension $1 \times 4$, where the product of each vector and  ${\bf F}^{(\{3,6\})_{\rm c}}$ is ${\bf 0}_{1 \times 2}$ (i.e.,  the zero  matrix with dimension $1\times 2$).
A possible vector basis could be the set of vectors $(-6, 1, 0, 3)$ and $(0, -2, 3, 0)$. It can be seen that 
     \begin{subequations}
\begin{align}
& \negmedspace\negmedspace -6 F_1 +1 F_2 +0 F_3 +3 F_4 \negmedspace=\negmedspace  -2 W_1+2W_2 +10W_4+11 W_5,\\
& 0 F_1 -2 F_2 +3 F_3 +0 F_4= W_1-4 W_2+ W_4+5 W_5,
\end{align}
\label{eq:example 1 sent by worker 1} 
     \end{subequations}
both of which are independent of $W_3$ and $W_6$. Hence, the two linear combinations in~\eqref{eq:example 1 sent by worker 1}  could be computed and then sent   by worker $1$.
 
For worker $2$, who can compute  $W_2$, $W_3$, $W_5$, and $W_6$, we search for the   a vector basis for the left-side null space of ${\bf F}^{(\{1,4\})_{\rm c}}$. A possible vector basis could be the set of vectors $(0,-1,0,1)$ and $(-1,-2,3,0)$. Hence, we let worker $2$ compute and send
     \begin{subequations}
\begin{align}
& 0 F_1 -1 F_2 +0 F_3 + 1 F_4= -2W_3-W_5-6 W_6,\\
& -1  F_1 -2 F_2 + 3 F_3 +0 F_4 =-5 W_2- W_3 +4 W_5- W_6.
\end{align}
\label{eq:example 1 sent by worker 2} 
     \end{subequations}
 
 For worker $3$, who can compute  $W_1$, $W_3$, $W_4$, and $W_6$, we search for the   a vector basis for the left-side null space of ${\bf F}^{(\{2,5\})_{\rm c}}$. A possible vector basis could be the set of vectors $(-2, -2, 0, 3)$ and $(10, -5, 3, 0)$. Hence, we let worker $3$ compute and send
     \begin{subequations}
\begin{align}
& -2 F_1 -2 F_2 +0 F_3 + 3  F_4 =  -W_1  -5W_3+ 2W_4 -14W_6,\\
& 10 F_1 -5 F_2 +3 F_3 +0 F_4  = 8W_1 +  W_3 -W_4  -8 W_6.
\end{align}
\label{eq:example 1 sent by worker 3} 
     \end{subequations}

In summary, each worker sends two linear combinations of $(F_1,F_2,F_3,F_4)$.

\paragraph*{Decoding phase}
Assuming the set of responding workers is $\{1,2\}$. The master receives 
\begin{align}
 {\bf X}_{\{1,2\}}  = \left[ \begin{array}{c}
-6, 1, 0, 3 \\
0, -2, 3, 0\\
0,-1,0,1\\  
-1,-2,3,0 \\   
\end{array} \right]
\left[ \begin{array}{c}
F_1\\
F_2\\
F_3\\
F_4\\
\end{array} \right] := {\bf C}_{\{1,2\}} \left[ \begin{array}{c}
F_1\\
F_2\\
F_3\\
F_4\\
\end{array} \right].
\end{align}
Since matrix ${\bf C}_{\{1,2\}}$ is full-rank, the master can recover $[F_1;F_2;F_3;F_4]$ by computing ${\bf C}^{-1}_{\{1,2\}}  {\bf X}_{\{1,2\}}$.

Similarly, it can be checked that the four linear combinations sent from any two workers are linearly independent. Hence, by receiving the answers of any two workers, the master can recover  
 task function.

\paragraph*{Performance}
The needed communication cost is $\frac{2 \Lsf +2 \Lsf}{\Lsf}=4$, coinciding with the converse bound $\Rsf^{\star}\geq \Ksf_{\rm c}=4$.

  \hfill $\square$ 
 \end{example}
 
 \iffalse
\begin{rem}
\label{rem:existing cannot work}
Notice that the computing techniques  of the computing schemes with the cyclic assignment in~\cite{efficientgradientcoding} for  distributed gradient coding (i.e., for the case  $\Ksf_{\rm c}=1$) and in~\cite{shortdot2016dutta} for distributed linear transform (i.e., for the case $T_n=\Lsf=1$ and $\Msf=\frac{\Ksf}{\Nsf}(\Nsf-\Nsf_{\rm r}+\Ksf_{\rm c})$, $n\in [\Nsf]$), cannot work in this example.  

Obviously, for this example we have $\Nsf_{\rm r}=\Ksf_{\rm c}$, thus by the computing scheme   in~\cite{shortdot2016dutta}, we need to assign $\frac{\Ksf}{\Nsf}(\Nsf-\Nsf_{\rm r}+\Ksf_{\rm c})= \Ksf$ datasets  to each worker; i.e., all datasets should be assigned to each worker. 

Let us then focus on the 
the idea  of the   computing schemes in~\cite{efficientgradientcoding}. The transmissions of each worker $n$ can be expressed as 
\begin{align}
{\bf C}^{\prime}_{n}  {\bf F} [W_1;\ldots;W_{\Nsf}], \label{eq:Kc=1 cannot work}
\end{align} 
where ${\bf C}^{\prime}_{n} $   is a  Vandermonde matrix with   dimension of ${\bf C}^{\prime}_{n} $ is  $2 \times \Nsf_{\rm r}$. However,
in the linear combinations~\eqref{eq:Kc=1 cannot work}, 
the coefficients of the messages which cannot be computed by worker $n$  are not all zero.
  \hfill $\square$ 
\end{rem}
\fi
 
 We are now ready to generalize the proposed scheme in Example~\ref{ex:regime 1}. First we focus on $\Ksf_{\rm c}=\frac{\Ksf}{\Nsf} \Nsf_{\rm r}$. 
During the data assignment phase, we use the cyclic assignment  described in Section~\ref{sub:problem formulation}.

\paragraph*{Computing phase}
 Recall that by the cyclic assignment, the set of datasets assigned to worker $n \in [\Nsf]$ is 
\begin{align*}
 \Zc_n &=  \underset{p \in \left[0:  \frac{\Ksf}{\Nsf} -1 \right]}{\cup}   \big\{\text{Mod}(n,\Nsf)+ p  \Nsf , \text{Mod}(n+1,\Nsf)+ p  \Nsf , \ldots, \nonumber\\& \text{Mod}(n+\Nsf-\Nsf_{\rm r},\Nsf)+ p  \Nsf  \big\}
\end{align*}
 as  defined in~\eqref{eq:cyclic assignment n divides k}.
 We denote the set of datasets which are not assigned to worker $n$ by $\overline{  \Zc_n} := [\Ksf ] \setminus \Zc_n$.
We retrieve   columns   of ${\bf F}$ with indices in $\overline{\Zc_n}$  to obtain ${\bf F}^{(\overline{  \Zc_n})_{\rm c}}$. It can be seen that the dimension of ${\bf F}^{(\overline{  \Zc_n})_{\rm c}}$ is 
  $\Ksf_{\rm c} \times \frac{\Ksf}{\Nsf}(\Nsf_{\rm r}-1)= \frac{\Ksf}{\Nsf} \Nsf_{\rm r} \times \frac{\Ksf}{\Nsf}(\Nsf_{\rm r}-1)$, and the elements in  ${\bf F}^{(\overline{  \Zc_n})_{\rm c}}$  are uniformly i.i.d. over $\mathbb{F}_{\qsf}$.
 Hence,  a vector basis for the left-side null space  ${\bf F}^{(\overline{  \Zc_n})_{\rm c}}$  is the set of $\frac{\Ksf}{\Nsf }$  linearly independent vectors
with dimension $1 \times  \frac{\Ksf}{\Nsf} \Nsf_{\rm r}$, where the product of each vector and  ${\bf F}^{(\overline{  \Zc_n})_{\rm c}}$ is ${\bf 0}_{1 \times \frac{\Ksf}{\Nsf}(\Nsf_{\rm r}-1)}$.

We assume that a possible vector basis contains the vectors $\uv_{n,1},\ldots, \uv_{n, \frac{\Ksf}{\Nsf}}$.
For each $j \in \left[\frac{\Ksf}{\Nsf} \right]$, we focus on
\begin{align}
\uv_{n,j} {\bf F}   \left[ \begin{array}{c}
W_1\\
\vdots\\
W_{\Ksf}\\ 
\end{array} \right]. \label{eq:one term 0}
\end{align}  
Since $\uv_{n,j}  {\bf F}^{(\overline{  \Zc_n})_{\rm c}}={\bf 0}_{1 \times \frac{\Ksf}{\Nsf}(\Nsf_{\rm r}-1)}$, it can be seen that~\eqref{eq:one term 0} is a linear combination of $W_i$ where $i\in \Zc_n$, which could be computed by worker $n$.

 After computing $W_i=f_i(D_i)$ for each $i\in \Zc_n$, worker $n$ then computes 
\begin{align}
 {\bf X}_{\{n\}}  = \left[ \begin{array}{c}
\uv_{n,1} \\
\vdots\\
\uv_{n, \frac{\Ksf}{\Nsf}}\\    
\end{array} \right]
{\bf F}   \left[ \begin{array}{c}
W_1\\
\vdots\\
W_{\Ksf}\\ 
\end{array} \right] := {\bf C}_{\{n\}}
{\bf F}   \left[ \begin{array}{c}
W_1\\
\vdots\\
W_{\Ksf}\\ 
\end{array} \right],
\end{align}
which is then sent to the master. It can be seen that $ {\bf X}_{\{n\}}$ contains $\frac{\Ksf}{\Nsf}$ linear combinations of the messages in $\Zc_n$, each of which contains $\Lsf$ symbols. Hence, worker $n$ totally sends    $ \frac{\Ksf}{\Nsf} \Lsf$ symbols, i.e., 
\begin{align}
T_n= \frac{\Ksf}{\Nsf} \Lsf .\label{eq:Kc=Nr load of one worker}
\end{align}

\paragraph*{Decoding phase} 
 We provide the following lemma which will be proved in Appendix~\ref{sec:proof of SZlemma} based on the Schwartz-Zippel lemma~\cite{Schwartz,Zippel,Demillo_Lipton}.
\begin{lem}
\label{lem:SZlemma}
For any set $\Ac \subseteq [\Nsf]$ where $|\Ac|=\Nsf_{\rm r}$,   the vectors $\uv_{n,j}$ where $n\in \Ac$ and $j \in \left[ \frac{\Ksf}{\Nsf} \right]$  are linearly independent  (i.e., ${\bf C}_{\Ac}$ is full-rank)  with high probability.
\hfill $\square$ 
\end{lem} 

Assume that the set of responding workers is $\Ac=\left\{\Ac(1),\ldots,\Ac\left( \Nsf_{\rm r} \right) \right\}$ where $\Ac \subseteq [\Nsf]$ and $|\Ac|=\Nsf_{\rm r}$. Hence, the master receives 
\begin{align}
{\bf X}_{\Ac} &=    \left[ \begin{array}{c}
 {\bf X}_{\Ac(1)} \\
\vdots\\
 {\bf X}_{\Ac\left( \Nsf_{\rm r} \right)}\\    
\end{array} \right] = \left[ \begin{array}{c}
{\bf C}_{\Ac(1)} \\
\vdots\\  
{\bf C}_{\Ac\left( \Nsf_{\rm r} \right)}\\   
\end{array} \right] {\bf F}   \left[ \begin{array}{c}
W_1\\
\vdots\\
W_{\Ksf}\\ 
\end{array} \right] \nonumber\\& :=  {\bf C}_{\Ac} {\bf F}   \left[ \begin{array}{c}
W_1\\
\vdots\\
W_{\Ksf}\\ 
\end{array} \right].
\end{align}
By Lemma~\ref{lem:SZlemma}, matrix ${\bf C}_{\Ac}$ is full-rank. Hence, the master can recover the task function by taking
$$
{\bf C}_{\Ac}^{-1} {\bf X}_{\Ac} = {\bf F}   \left[ \begin{array}{c}
W_1\\
\vdots\\
W_{\Ksf}\\ 
\end{array} \right].
$$

\paragraph*{Performance} 
 From~\eqref{eq:Kc=Nr load of one worker}, the number of symbols sent by each worker is $\frac{\Ksf}{\Nsf} \Lsf$. Hence, the communication cost is 
 $ 
 \frac{\Ksf}{\Nsf} \Nsf_{\rm r}.
 $ 
 
\begin{rem}
\label{rem:linear space explanation}
The proposed scheme can be explained from the viewpoint on linear space. The request matrix ${\bf F}$ can be seen as a linear space composed of $\frac{\Ksf}{\Nsf}\Nsf_{\rm r}$ linearly independent vectors, each of which has the size $1\times \Ksf$. The assigned datasets to each worker $n\in [\Nsf]$, are $D_i$ where $i\in \Zc_n$. Thus all the linear combinations   which can be sent by worker $n$ are located at a linear space composed of the vectors $(0,\ldots,0,1,0,\ldots,0)$ where $1$ is at $i^{\text{th}}$ position for $i\in \Zc_n$.
The intersection of these two linear spaces contains $\frac{\Ksf}{\Nsf}$ linearly independent vectors. In other words, 
  the product of each of the $\frac{\Ksf}{\Nsf}$ vectors and $[W_1;\ldots;W_{\Ksf}]$ can be sent by worker $n$. In addition, considering any set of $\Nsf_{\rm r}$ workers, Lemma~\ref{lem:SZlemma} shows that the total  $\frac{\Ksf}{\Nsf} \Nsf_{\rm r}$ vectors are linearly independent, such that the master can recover the whole linear space generated by ${\bf F}$.
  \hfill$\square$
\end{rem}

 For each $\Ksf_{\rm c} \in \left[ \frac{\Ksf}{\Nsf}: \frac{\Ksf}{\Nsf} \Nsf_{\rm r}\right)$, the master generates a matrix ${\bf G}$ with dimension $\left(\frac{\Ksf}{\Nsf} \Nsf_{\rm r}-\Ksf_{\rm c} \right) \times \Ksf$, whose elements are uniformly i.i.d. over $\mathbb{F}_{\qsf}$. The master then requests ${\bf F}^{\prime} [W_1;\ldots;W_{\Ksf}]$, where ${\bf F}^{\prime}=[{\bf F}; {\bf G}]$. Hence, we can then use the above distributed computing scheme with $\Ksf_{\rm c}= \frac{\Ksf}{\Nsf} \Nsf_{\rm r}$  to  let the master recover ${\bf F}^{\prime} [W_1;\ldots;W_{\Ksf}]$, and the communication cost is also $ 
 \frac{\Ksf}{\Nsf} \Nsf_{\rm r},
 $ 
 which coincides with~\eqref{eq:achie case 2}.

  As stated in Footnote~\ref{foot:computation cost}, the computation complexity of each worker is mainly due to the computation on the messages from the assigned datasets.  Recall that $\Lsf$ is large enough.
For the proposed computing scheme in this case, the decoding complexity (i.e., the number of multiplications) of the master  is  $\Oc \left(\Ksf_{\rm c} \frac{\Ksf}{\Nsf}\Nsf_{\rm r} \Lsf\right)$.

  \subsection{$\Ksf_{\rm c} \in \left[1: \frac{\Ksf}{\Nsf} \right)$}
\label{sub:Kc first}
We also begin with an example to illustrate the main idea. 
\begin{example}[$\Nsf = 3, \Ksf = 9, \Ksf_{\rm c} = 2, \Nsf_{\rm r} = 2$, $\Msf=6$]
\label{ex:regime 2}
\rm 
Assume that the task function is \begin{align*}
f(D_1,   \ldots, D_{9})&= 
 \left[ \begin{array}{c}
F_1\\
F_2\\ 
\end{array} \right]
= {\bf F}
\left[ \begin{array}{c}
W_1\\
\vdots \\
 W_9
\end{array} \right]
  \nonumber\\& =\left[ \begin{array}{c}
1 ,1 , 1 ,1 ,1 ,1,1,1,1\\
1 ,2 , 3 ,4 ,5 ,6,7,8,9\\ 
\end{array} \right]
\left[ \begin{array}{c}
W_1\\
\vdots \\ 
W_9
\end{array} \right].
\end{align*}
 
By the   cyclic assignment described in Section~\ref{sub:problem formulation}, we assign that 
\begin{align*}
\begin{array}{rl|c|c|c|}\cline{3-3}\cline{4-4}\cline{5-5}
&&\rule{0pt}{1.2em}\mbox{Worker 1} & \rule{0pt}{1.2em}\mbox{Worker 2} &  \rule{0pt}{1.2em}\mbox{Worker 3} \\\cline{3-3}\cline{4-4}\cline{5-5}
&& D_1&  D_2  &  D_1\\
&& D_2&  D_3 &  D_3 \\ \cdashline{3-3}\cdashline{4-4}\cdashline{5-5}   
&& D_4&  D_5 &  D_4 \\
&& D_5&  D_6 &   D_6\\ \cdashline{3-3}\cdashline{4-4}\cdashline{5-5}   
&& D_7&  D_8  &  D_7\\
&& D_8&  D_9 & D_9 \\
\cline{3-3}\cline{4-4}\cline{5-5}
\end{array}
\end{align*}

Note that by the cyclic assignment, we can divide the $\Ksf=9$ datasets into $\Nsf=3$ groups, where in each group there are $\frac{\Ksf}{\Nsf}=3$ datasets. The first group contains $D_1, D_4, D_7$, which are assigned to workers $1$ and $3$. The coefficients of $(W_1, W_4, W_7)$ in $F_1$ are $(1,1,1)$ and in $F_2$ are $(1,4,7)$. We  define that 
\begin{subequations}
\begin{align}
&W^{\prime}_{1,1}=W_1+W_4+W_7,\\
&W^{\prime}_{2,1}=W_1+4 W_4+7 W_7 ,
\end{align}
\end{subequations}
which are computed by workers $1$ and $3$. 
Similarly,  the second group contains $D_2, D_5, D_8$, which are assigned to workers $1$ and $2$. The coefficients of $(W_2, W_5, W_8)$ in $F_1$ are $(1,1,1)$ and in $F_2$ are $(2,5,8)$. We  define that 
\begin{subequations}
\begin{align}
&W^{\prime}_{1,2}=W_2+W_5+W_8,\\
&W^{\prime}_{2,2}=2W_2+5 W_5+8 W_8 ,
\end{align}
\end{subequations}
which are computed by workers $1$ and $2$.  The third group contains $D_3, D_6, D_9$, which are assigned to workers $2$ and $3$. The coefficients of $(W_3, W_6, W_9)$ in $F_1$ are $(1,1,1)$ and in $F_2$ are $(3,6,9)$. We  define that 
\begin{subequations}
\begin{align}
&W^{\prime}_{1,3}=W_3+W_6+W_9,\\
&W^{\prime}_{2,3}=3W_3+6 W_6+9 W_9 ,
\end{align}
\end{subequations}
which are computed by workers $2$ and $3$. 

Now we treat this example as two separated sub-problems,
where each sub-problem is a $(\Ksf^{\prime},\Nsf^{\prime},\Nsf_{\rm r}^{\prime}, \Ksf_{\rm c}^{\prime}, \Msf^{\prime})=\left(3,3,2, 1, 2 \right)$ distributed linearly separable computation problem.
In the first sub-problem,   the   three messages  are $W^{\prime}_{1,1}$, $W^{\prime}_{1,2}$, and $W^{\prime}_{1,3}$, and the master aims to compute $W^{\prime}_{1,1}+W^{\prime}_{1,2}+W^{\prime}_{1,3}$. In the second sub-problem, the  three messages are $W^{\prime}_{2,1}$, $W^{\prime}_{2,2}$, and $W^{\prime}_{2,3}$, and the master aims to compute $W^{\prime}_{2,1}+W^{\prime}_{2,2}+W^{\prime}_{2,3}$.
Hence, each  sub-problem  can be solved by the proposed scheme in Section~\ref{sub:Kc in middle} with communication cost equal to $\frac{\Ksf^{\prime}}{\Nsf^{\prime}}\Nsf^{\prime}_{\rm r} =2$. The total communication cost is $ 4$.
  \hfill $\square$ 
 \end{example}

We are now ready to generalize Example~\ref{ex:regime 2}.
  For each integer $n \in [\Nsf]$,  we focus on the set of messages 
$
\left\{ W_{n+ p \Nsf }:p\in \left[0: \frac{\Ksf}{\Nsf} -1 \right] \right\}.
$ 
We define 
\begin{align}
W^{\prime}_{j,n}= \sum_{p\in \left[0: \frac{\Ksf}{\Nsf} -1 \right]}  f_{j,n+ p \Nsf } W_{n+ p \Nsf }, \ \forall j\in [\Ksf_{\rm c}],
\end{align}
where $f_{j,n+ p \Nsf }$ is the element located at the $j^{\text{th}}$ row and $(n+ p \Nsf)^{\text{th}}$ column of  matrix ${\bf F}$. Note that each message $W_{n+ pN }$ can be computed by workers in $[n: \text{Mod}(n-\Nsf+\Nsf_{\rm r})]$. Hence, $W^{\prime}_{j,n}$ can also be computed by   workers in $[n: \text{Mod}(n-\Nsf+\Nsf_{\rm r})]$.

We can re-write the task function as
     \begin{subequations}
\begin{align}
&f(D_1, \ldots, D_{\Ksf})=
 \left[ \begin{array}{c}
F_1\\
%F_2\\
\vdots\\
F_{\Ksf_{\rm c}}
\end{array} \right]  = \left[ \begin{array}{c}
W^{\prime}_{1,1}+ \cdots+ W^{\prime}_{1,\Nsf} \\
%W^{\prime}_{2,1}+ \cdots+ W^{\prime}_{2,\Nsf} \\
\vdots\\
W^{\prime}_{\Ksf_{\rm c},1}+ \cdots+ W^{\prime}_{\Ksf_{\rm c},\Nsf} \\
\end{array} \right].
\end{align}
     \end{subequations}
We then treat the problem as $\Ksf_{\rm c}$ separate sub-problems, where in the $j^{\text{th}}$ sub-problem, the master requests $W^{\prime}_{j,1}+ \cdots+ W^{\prime}_{j,\Nsf}$. Hence, each sub-problem is equivalent to the $(\Ksf^{\prime},\Nsf^{\prime},\Nsf_{\rm r}^{\prime}, \Ksf_{\rm c}^{\prime}, \Msf^{\prime})=\left(\Nsf,\Nsf,\Nsf_{\rm r}, 1, %\frac{\Ksf}{\Nsf}
 \Nsf-\Nsf_{\rm r}+1  \right)$ distributed linearly separable computation problem. 
%with $\Nsf^{\prime}=\Nsf$ workers, $\Ksf^{\prime}=\Ksf$ datasets, where the master requests $\Ksf^{\prime}_{\rm c}=1$ linear combination of the messages and wait for the fastest $\Nsf_{\rm r}$ workers. 
Each sub-problem can be solved by the proposed scheme in Section~\ref{sub:Kc in middle} with communication cost equal to $\frac{\Ksf^{\prime}}{\Nsf^{\prime}}\Nsf^{\prime}_{\rm r}=\Nsf_{\rm r}$. Hence, considering all the $\Ksf_{\rm c}$ sub-problems, the total communication cost is $\Ksf_{\rm c}\Nsf_{\rm r}$, which coincides with~\eqref{eq:achie case 1}.

  For the proposed computing scheme in this case, the decoding complexity  of the master  is  $\Oc \left(\Ksf_{\rm c} \Nsf_{\rm r}\Lsf \right)$.
 
 \subsection{$\Ksf_{\rm c} \in \left(  \frac{\Ksf}{\Nsf}\Nsf_{\rm r}:\Ksf \right] $}
\label{sub:Kc last}
 We still use an example to illustrate the main idea. 
\begin{example}[$\Nsf = 3, \Ksf =3, \Ksf_{\rm c} = 3, \Nsf_{\rm r} = 2$, $\Msf=2$]
\label{ex:regime 3}
\rm 
Assume that the task function is \begin{align*}
f(D_1,   \ldots, D_{3})&= 
 \left[ \begin{array}{c}
F_1\\
F_2\\ 
F_3\\
\end{array} \right]
= {\bf F}
\left[ \begin{array}{c}
W_1\\
 W_2\\
 W_3
\end{array} \right]
 \nonumber\\&  =\left[ \begin{array}{c}
1 ,1 , 1 \\
1 ,2 , 3 \\ 
1 ,4 , 9\\ 
\end{array} \right]
\left[ \begin{array}{c}
W_1\\
 W_2\\
 W_3
\end{array} \right].
\end{align*}
 
By the   cyclic assignment described in Section~\ref{sub:problem formulation}, we assign that 
\begin{align*}
\begin{array}{rl|c|c|c|}\cline{3-3}\cline{4-4}\cline{5-5}
&&\rule{0pt}{1.2em}\mbox{Worker 1}  &\rule{0pt}{1.2em}\mbox{Worker 2} &  \rule{0pt}{1.2em}\mbox{Worker 3} \\\cline{3-3}\cline{4-4}\cline{5-5}
&& D_1&  D_2    & D_1\\
&& D_2&  D_3   & D_3 \\
\cline{3-3}\cline{4-4}\cline{5-5}
\end{array}
\end{align*} 
 
For each message $W_k$ where $k\in [\Ksf]$, we divide $W_{k}$ into $2$ non-overlapping and equal-length sub-messages, denoted by $W_{k,1}$ and $W_{k,2}$. We then use a  $(3,2)$ MDS (Maximum Distance Separable) code to obtain $3$ MDS-coded packets:
$$
W_{k, \{ 1,2\} }= W_{k,1}, \ W_{k, \{ 1,3\} }= W_{k,2}, \ W_{k, \{ 2,3\} }= W_{k,1}+ W_{k,2}. 
$$
 
 Next we     treat this example  as $3$ sub-problems, where each sub-problem is a  $(\Ksf^{\prime},\Nsf^{\prime},\Nsf_{\rm r}^{\prime}, \Ksf_{\rm c}^{\prime}, \Msf^{\prime})=\left(3,3,2, 2, 2 \right)$  distributed linearly separable computation problem. In the first sub-problem, 
 the three messages are $W_{1, \{ 1,2\} }, W_{2, \{ 1,2\} }, W_{3, \{ 1,2\} }$, and the master requests 
 $$
{\bf F}^{(\{1,2\})_{\rm r}} \left[\negmedspace \begin{array}{c}
W_{1, \{ 1,2\} }\\
W_{2, \{ 1,2\} }\\
W_{3, \{ 1,2\} }
\end{array} \negmedspace\right]  \negmedspace\negmedspace=\negmedspace\negmedspace  \left[ \negmedspace\negmedspace \begin{array}{c}
W_{1, \{ 1,2\} }+ W_{2, \{ 1,2\} }+ W_{3, \{ 1,2\} }\\
W_{1, \{ 1,2\} }+2 W_{2, \{ 1,2\} }+ 3W_{3, \{ 1,2\} }
\end{array} \negmedspace\negmedspace\right].
 $$
  In the second sub-problem, 
 the three messages are $W_{1, \{ 1,3\} }, W_{2, \{ 1,3\} }, W_{3, \{ 1,3\} }$, and the master requests 
 $$
{\bf F}^{(\{1,3\})_{\rm r}} \left[\negmedspace \begin{array}{c}
W_{1, \{ 1,3\} }\\
W_{2, \{ 1,3\} }\\
W_{3, \{ 1,3\} }
\end{array} \negmedspace\right]  \negmedspace\negmedspace= \negmedspace\negmedspace \left[ \negmedspace\negmedspace \begin{array}{c}
W_{1, \{ 1,3\} }+ W_{2, \{ 1,3\} }+ W_{3, \{ 1,3\} }\\
W_{1, \{ 1,3\} }+4 W_{2, \{ 1,3\} }+ 9W_{3, \{ 1,3\} }
\end{array}\negmedspace\negmedspace \right].
 $$
  In the third sub-problem, 
 the three messages are $W_{1, \{ 2,3\} }, W_{2, \{ 2,3\} }, W_{3, \{ 2,3\} }$, and the master requests 
 $$
{\bf F}^{(\{2,3\})_{\rm r}} \left[ \negmedspace\begin{array}{c}
W_{1, \{ 2,3\} }\\
W_{2, \{ 2,3\} }\\
W_{3, \{ 2,3\} }
\end{array}\negmedspace \right] \negmedspace\negmedspace=\negmedspace\negmedspace \left[\negmedspace\negmedspace \begin{array}{c}
W_{1, \{ 2,3\} }+ 2W_{2, \{ 2,3\} }+3 W_{3, \{ 2,3\} }\\
W_{1, \{ 2,3\} }+4 W_{2, \{2,3\} }+ 9W_{3, \{ 2,3\} }
\end{array} \negmedspace\negmedspace\right].
 $$
 Each sub-problem can be solved by the proposed scheme in Section~\ref{sub:Kc in middle}, where each worker sends $ \frac{\Ksf^{\prime}}{\Nsf^{\prime}} =1$ linear combination of sub-messages with $\frac{\Lsf}{2}$ symbols. Hence,  each worker totally sends $ \frac{3\Lsf}{2}$ symbols, and thus the  
 communication cost equal to $\frac{3\Lsf \Nsf_{\rm r}}{2 \Lsf}=3$.

 Now we show that by solving the three sub-problems, the master can recover the task, i.e., 
 $F_1=W_{1}+W_{2}+W_{3}$, $F_2=W_{1}+2W_{2}+3W_{3}$, and $F_3=W_{1}+4W_{2}+9W_{3}$. 
 
 From the first and second sub-problems, the master can recover 
 \begin{subequations}
\begin{align}
&W_{1, \{ 1,2\} }+ W_{2, \{ 1,2\} }+ W_{3, \{ 1,2\} }= W_{1, 1 }+ W_{2, 1 }+ W_{3,1  }, \label{eq:W1,12F1}\\
&  W_{1, \{ 1,3\} }+ W_{2, \{ 1,3\} }+ W_{3, \{ 1,3\} }= W_{1, 2}+ W_{2, 2 }+ W_{3,2  }. \label{eq:W1,13F1}  
\end{align} 
\end{subequations}
 Hence, by concatenating~\eqref{eq:W1,12F1} and~\eqref{eq:W1,13F1},
 the master can recover $F_1$.
 
  From the first and third sub-problems, the master can recover 
   \begin{subequations}
\begin{align}
&W_{1, \{ 1,2\} }+ 2W_{2, \{ 1,2\} }+ 3 W_{3, \{ 1,2\} } = W_{1, 1 }+ 2 W_{2, 1 }+3  W_{3,1  }, \label{eq:W1,12F2}\\
 &  W_{1, \{ 2,3\} }+ 2 W_{2, \{ 2,3\} }+ 3W_{3, \{ 2,3\} }= (W_{1, 1}+W_{1, 2})+\nonumber\\& 2 (W_{2,1}+ W_{2, 2}) +3  (W_{3,1}+W_{3,2 }). \label{eq:W1,23F2}
\end{align} 
\end{subequations}
From~\eqref{eq:W1,12F2} and~\eqref{eq:W1,23F2}, the master can first recover $ W_{1, 2 }+ 2 W_{2, 2 }+3  W_{3,2 } $, which is then concatenated with~\eqref{eq:W1,12F2}. Hence,
 the master can recover $F_2$.
 
   From the second and third sub-problems, the master can recover 
   \begin{subequations}
\begin{align}
&W_{1, \{ 1,3\} }+ 4W_{2, \{ 1,3\} }+ 9 W_{3, \{ 1,3\} }= W_{1, 2 }+ 4 W_{2, 2 }+9  W_{3,2  }, \label{eq:W1,13F3}\\
&  W_{1, \{ 2,3\} }+ 4W_{2, \{ 2,3\} }+ 9W_{3, \{ 2,3\} }= (W_{1, 1}+W_{1, 2}) \nonumber\\& +4 (W_{2,1}+ W_{2, 2}) +9  (W_{3,1}+W_{3,2 }). \label{eq:W1,23F3}
\end{align} 
\end{subequations}
From~\eqref{eq:W1,13F3} and~\eqref{eq:W1,23F3}, the master can first recover $ W_{1, 1 }+ 4 W_{2, 1 }+9  W_{3,1 } $, which is then concatenated with~\eqref{eq:W1,13F3}. Hence,
 the master can recover $F_3$.
   \hfill $\square$ 
 \end{example}
 
 We are now ready to generalize Example~\ref{ex:regime 3}.
  We divide each message $W_k$ into $\binom{\Ksf_{\rm c}-1}{ \frac{\Ksf}{\Nsf} \Nsf_{\rm r}-1}$ equal-length and non-overlapped sub-messages,
$W_{k}=\left( W_{k,1}, \ldots, W_{k,\binom{\Ksf_{\rm c}-1}{ \frac{\Ksf}{\Nsf} \Nsf_{\rm r}-1}} \right)$,  
  which are then encoded by a $\left(\binom{\Ksf_{\rm c} }{ \frac{\Ksf}{\Nsf} \Nsf_{\rm r} }, \binom{\Ksf_{\rm c}-1}{ \frac{\Ksf}{\Nsf} \Nsf_{\rm r}-1} \right)$ MDS code. Each MDS-coded packet is denoted by $W_{k, \Sc}$ where $\Sc \subseteq [\Ksf_{\rm c}]$ where $|\Sc|= \frac{\Ksf}{\Nsf} \Nsf_{\rm r}$.
  Since $W_{k, \Sc}$ is a linear combination of $\left( W_{k,1}, \ldots, W_{k,\binom{\Ksf_{\rm c}-1}{ \frac{\Ksf}{\Nsf} \Nsf_{\rm r}-1}} \right)$, we define that
  \begin{align}
   W_{k, \Sc}= \vv_{\Sc} \left[ \begin{array}{c}
 W_{k,1}\\
\vdots\\
W_{k,\binom{\Ksf_{\rm c}-1}{ \frac{\Ksf}{\Nsf} \Nsf_{\rm r}-1}} 
\end{array} \right] , \ \forall \Sc \subseteq [\Ksf_{\rm c}]: |\Sc|= \frac{\Ksf}{\Nsf} \Nsf_{\rm r}, \label{eq:vv_S}
  \end{align}
  where $\vv_{\Sc}$ with  $\binom{\Ksf_{\rm c}-1}{ \frac{\Ksf}{\Nsf} \Nsf_{\rm r}-1}$ elements represents the   generation vector to generate the MDS-coded packet $W_{k, \Sc}$.
Note that each MDS-coded packet has $\frac{\Lsf}{\binom{\Ksf_{\rm c}-1}{ \frac{\Ksf}{\Nsf} \Nsf_{\rm r}-1}}$ symbols.%\footnote{\label{foot:large L} Here we assume that $\Lsf$ is large enough such that the above division is possible.}
  
  Next we     treat the problem  as $\binom{\Ksf_{\rm c}}{\frac{\Ksf}{\Nsf} \Nsf_{\rm r}}$ sub-problems, where each sub-problem is a  $(\Ksf^{\prime},\Nsf^{\prime},\Nsf_{\rm r}^{\prime}, \Ksf_{\rm c}^{\prime}, \Msf^{\prime})=\left(\Ksf,\Nsf, \Nsf_{\rm r} , \frac{\Ksf}{\Nsf}\Nsf_{\rm r},  \Msf\right)$  distributed linearly separable computation problem.
For each   
  $\Sc \subseteq [\Ksf_{\rm c}]$ where $|\Sc|= \frac{\Ksf}{\Nsf} \Nsf_{\rm r}$, there is a sub-problem. In this sub-problem 
the messages are $W_{1,\Sc}, \ldots, W_{\Ksf,\Sc}$, and the master requests 
$$  
  {\bf F}^{(\Sc)_{\rm r}} \left[ \begin{array}{c}
W_{1,\Sc}\\
\vdots\\
W_{\Ksf, \Sc }
\end{array} \right].
  $$
 Each sub-problem can be solved by the proposed scheme in Section~\ref{sub:Kc in middle}, where each worker sends $ \frac{\Ksf}{\Nsf}$ linear combination of sub-messages with $\frac{\Lsf}{\binom{\Ksf_{\rm c}-1}{ \frac{\Ksf}{\Nsf} \Nsf_{\rm r}-1}}$  symbols. Hence,  each worker totally sends 
$$
\binom{\Ksf_{\rm c}}{\frac{\Ksf}{\Nsf} \Nsf_{\rm r}}  \frac{\Ksf}{\Nsf} \frac{\Lsf}{\binom{\Ksf_{\rm c}-1}{ \frac{\Ksf}{\Nsf} \Nsf_{\rm r}-1}}  = \frac{\Lsf \Ksf_{\rm c}  }{  \Nsf_{\rm r}}   
$$ 
 symbols, and thus the  
 communication cost equal to $ \Nsf_{\rm r} \frac{\Lsf \Ksf_{\rm c}  }{  \Nsf_{\rm r}\Lsf }=\Ksf_{\rm c} $, which coincides with~\eqref{eq:achie case 3}.

 Now we show that by solving all the sub-problems, the master can recover the task, i.e.,  for each $j\in [\Ksf_{\rm c}]$ the master can recover
 \begin{subequations}
\begin{align}
 &F_{j}={\bf F}^{(\{j\})_{\rm r}} [W_1;\ldots; W_{\Ksf}]=f_{j,1} W_1+ \cdots+f_{j,\Ksf}W_{\Ksf}  \\
 &= f_{j,1} \left[ \begin{array}{c}
 W_{1,1}\\
\vdots\\
W_{1,\binom{\Ksf_{\rm c}-1}{ \frac{\Ksf}{\Nsf} \Nsf_{\rm r}-1}} 
\end{array} \right]+ \cdots+f_{j,\Ksf} \left[ \begin{array}{c}
 W_{\Ksf,1}\\
\vdots\\
W_{\Ksf,\binom{\Ksf_{\rm c}-1}{ \frac{\Ksf}{\Nsf} \Nsf_{\rm r}-1}} 
\end{array} \right], \label{eq:Fj task}
\end{align}
\end{subequations}
 where we define that ${\bf F}^{(\{j\})_{\rm r}}:= [f_{j,1}, \ldots, f_{j,\Ksf}]$.
 
 For each $\Sc\subseteq [\Ksf_{\rm c}]$ where $|\Sc|=\frac{\Ksf}{\Nsf} \Nsf_{\rm r}$ and $j \in \Sc$, in the corresponding sub-problem  the master has recovered 
 \begin{subequations}
\begin{align}
& {\bf F}^{(\{j\})_{\rm r}} \left[W_{1,\Sc};\ldots; W_{\Ksf, \Sc} \right]=   f_{j,1} W_{1,\Sc}+ \cdots+f_{j,\Ksf}W_{\Ksf, \Sc}\\
&= f_{j,1} \vv_{\Sc} \left[ \begin{array}{c}
 W_{1,1}\\
\vdots\\
W_{1,\binom{\Ksf_{\rm c}-1}{ \frac{\Ksf}{\Nsf} \Nsf_{\rm r}-1}} 
\end{array} \right] +\cdots +  f_{j,\Ksf} \vv_{\Sc} \left[ \begin{array}{c}
 W_{\Ksf,1}\\
\vdots\\
W_{\Ksf,\binom{\Ksf_{\rm c}-1}{ \frac{\Ksf}{\Nsf} \Nsf_{\rm r}-1}} 
\end{array} \right].
\end{align}
\end{subequations}
We assume that all the sets  $\Sc\subseteq [\Ksf_{\rm c}]$ where $|\Sc|=\frac{\Ksf}{\Nsf} \Nsf_{\rm r}$ and $j \in \Sc$, are $\Sc_1, \ldots, \Sc_{\binom{\Ksf_{\rm c}-1}{ \frac{\Ksf}{\Nsf} \Nsf_{\rm r}-1}}$. By considering all the sub-problems corresponding to the above sets, the master has recovered
\begin{align}
  & f_{j,1} \left[ \begin{array}{c}
 \vv_{\Sc_1}\\
\vdots\\
\vv_{\Sc_{\binom{\Ksf_{\rm c}-1}{ \frac{\Ksf}{\Nsf} \Nsf_{\rm r}-1}}} 
\end{array} \right] \left[ \begin{array}{c}
 W_{1,1}\\
\vdots\\
W_{1,\binom{\Ksf_{\rm c}-1}{ \frac{\Ksf}{\Nsf} \Nsf_{\rm r}-1}} 
\end{array} \right] +\cdots + \nonumber\\&   f_{j,\Ksf} \left[ \begin{array}{c}
 \vv_{\Sc_1}\\
\vdots\\
\vv_{\Sc_{\binom{\Ksf_{\rm c}-1}{ \frac{\Ksf}{\Nsf} \Nsf_{\rm r}-1}}} 
\end{array} \right] \left[ \begin{array}{c}
 W_{\Ksf,1}\\
\vdots\\
W_{\Ksf,\binom{\Ksf_{\rm c}-1}{ \frac{\Ksf}{\Nsf} \Nsf_{\rm r}-1}} 
\end{array} \right]:= {\bf H}_{j }.\label{eq:before inverse}
  \end{align}
Note that   $\left[ \begin{array}{c}
 \vv_{\Sc_1}\\
\vdots\\
\vv_{\Sc_{\binom{\Ksf_{\rm c}-1}{ \frac{\Ksf}{\Nsf} \Nsf_{\rm r}-1}}} 
\end{array} \right]$ is full-rank with size $\binom{\Ksf_{\rm c}-1}{ \frac{\Ksf}{\Nsf} \Nsf_{\rm r}-1} \times \binom{\Ksf_{\rm c}-1}{ \frac{\Ksf}{\Nsf} \Nsf_{\rm r}-1}$, and thus invertible. Hence, the master can recover $F_j$ in~\eqref{eq:Fj task} by taking $\left[ \begin{array}{c}
 \vv_{\Sc_1}\\
\vdots\\
\vv_{\Sc_{\binom{\Ksf_{\rm c}-1}{ \frac{\Ksf}{\Nsf} \Nsf_{\rm r}-1}} }
\end{array} \right]^{-1}   {\bf H}_{j }$.
 
 For the proposed computing scheme in this case, the decoding complexity  of the master  is  %$\Oc \left(\Ksf_{\rm c} \frac{\Ksf}{\Nsf}\Nsf_{\rm r}\Lsf + \Ksf_{\rm c} \binom{\Ksf_{\rm c}-1}{\frac{\Ksf}{\Nsf}\Nsf_{\rm r}-1} \Lsf\right)$
 $\Oc \left(  \Ksf_{\rm c} \binom{\Ksf_{\rm c}-1}{\frac{\Ksf}{\Nsf}\Nsf_{\rm r}-1} \Lsf\right)$.

\begin{rem}
\label{rem:non generic}
By using the   Schwartz-Zippel Lemma, we prove that the proposed scheme is decodable with high probability if the elements in the demand matrix ${\bf F}$  are uniformly i.i.d. over some large field. However, for some specific ${\bf F}$, the proposed scheme is not decodable (i.e., ${\bf C}_{\Ac}$ is not full-rank) and we may need more communication load. 

Let us focus on the $(\Ksf,\Nsf,\Nsf_{\rm r}, \Ksf_{\rm c}, \Msf)=(3,3,2,2,2)$ distributed linearly separable computation problem. In this example, there is only one possible assignment, which is as follows, 
\begin{eqnarray*}
\begin{array}{rl|c|c|c|}\cline{3-3}\cline{4-4}\cline{5-5}
&&\rule{0pt}{1.2em}\mbox{Worker 1} &\rule{0pt}{1.2em}\mbox{Worker 2}  & \rule{0pt}{1.2em}\mbox{Worker 3} \\\cline{3-3}\cline{4-4}\cline{5-5}
&& W_1&   W_2   & W_1 \\
&& W_2  &   W_3   & W_3 \\\cline{3-3}\cline{4-4}\cline{5-5}
\end{array}
\end{eqnarray*} 
Noting that in this case we have $\Nsf=\Ksf$ and $\Ksf_{\rm c}=\Nsf_{\rm r}$. From Theorem~\ref{thm:optimality}, the proposed scheme in Section~\ref{sub:Kc in middle} is decodable with high probability if the elements in the demand matrix ${\bf F}$ are uniformly i.i.d. over some large field, and achieves the optimal communication cost   $2$. 

In the following, we focus on a specific demand matrix
\begin{align}
{\bf F}^{\prime} = \left[ \begin{array}{c}
1,1,1\\
2,1,1\\
\end{array} \right] \left[ \begin{array}{c}
W_1\\
W_2\\
W_3\\
\end{array} \right] 
= \left[ \begin{array}{c}
W_1 + W_2 + W_3\\
2W_1 + W_2 + W_3\\
\end{array} \right] .
\end{align}
Note that the demand is equivalent to $(W_1, W_2 + W_3)$. If we use the proposed scheme in Section~\ref{sub:Kc in middle}, it can be seen that $C_{\{1\}}=[1,-1]$, $C_{\{2\}}=[2,-1]$, and $C_{\{3\}}=[1,-1]$. So we have $C_{\{1,3\}}= \left[ \begin{array}{c}
1,-1  \\
1,-1\\
\end{array} \right]$
is not full-rank, and thus the proposed scheme is not decodable. In the following, we will prove that the optimal communication cost for this demand matrix is $3$.  

{\bf [Converse]:} 
We now prove that the communication cost is no less than $3$. Note that
from $X_1$ and $X_3$, the master can recover $W_1$ and $W_2+W_3$. Hence, we have 
 \begin{subequations}
\begin{align}
 0 & = H(W_2+W_3| X_1, X_3) \\
&  \geq  H(W_2+W_3| X_1, X_3, W_1 , W_3)\\
& =  H(W_2+W_3| X_1,  W_1 , W_3) \label{eq:X3 from W1 W3}\\
&= H(W_2 | X_1,  W_1 , W_3) \\
& = H(W_2 | X_1,  W_1  )\label{eq:W3 indep from W12 X1},
\end{align}
 \end{subequations}
where~\eqref{eq:X3 from W1 W3} comes from that $X_3$ is a function of $(W_1, W_3)$ and~\eqref{eq:W3 indep from W12 X1} comes from that $W_3$ is independent of $(W_1, W_2, X_1)$.
 Since the master can recover $W_1$ from $(X_1,X_3)$,~\eqref{eq:W3 indep from W12 X1} shows that from $(X_1, X_3)$ the master can also recover $W_2$, i.e.,
 \begin{align}
 H(W_1,W_2|X_1,X_3)=0.\label{eq:W12 from X13}
 \end{align}
  Moreover, we have 
 \begin{subequations}
\begin{align}
 0 & = H(W_2+W_3| X_1, X_3) \\
&  \geq  H(W_2+W_3| X_1, X_3, W_1, W_2)\\
&= H(W_3 | X_1, X_3, W_1 , W_2 ) \\
& = H(W_3 | X_1,  X_3  )\label{eq:W12 from X1X3},
  \end{align}
 \end{subequations}
where~\eqref{eq:W12 from X1X3} comes from~\eqref{eq:W12 from X13}. Hence, we have 
\begin{align}
H(W_1,W_2,W_3| X_1, X_3)=0. \label{eq:all from X13}
\end{align}

Note that from   $X_1$ and $X_2$, the master can recover $W_1$ and $W_2+W_3$.  
Since  the master can recover $W_1$ from $(X_1,X_2)$,~\eqref{eq:W3 indep from W12 X1} shows that from $(X_1, X_2)$ the master can also recover $W_2$, i.e.,
 \begin{align}
 H(W_1,W_2|X_1,X_2)=0.\label{eq:W12 from X12}
 \end{align}
Moreover, we have  
  \begin{subequations}
\begin{align}
 0 & = H(W_2+W_3| X_1, X_2) \\
 &  \geq  H(W_2+W_3| X_1, X_2, W_1, W_2)\\
&= H(W_3 | X_1, X_2, W_1 , W_2 ) \\
 &= H(W_3 | X_1, X_2),  \label{eq:W3 from X12}
  \end{align}
 \end{subequations}
where~\eqref{eq:W3 from X12} comes from~\eqref{eq:W12 from X12}.  From~\eqref{eq:W12 from X12} and~\eqref{eq:W3 from X12},  we have 
\begin{align}
H(W_1,W_2,W_3| X_1, X_2)=0. \label{eq:all from X12}
\end{align}

Similarly, we also have 
\begin{align}
H(W_1,W_2,W_3| X_2, X_3)=0. \label{eq:all from X23}
\end{align}

From~\eqref{eq:all from X13},~\eqref{eq:all from X12}, and~\eqref{eq:all from X23},   it can be seen that for any set of workers $\Ac \subseteq [3]$ where $|\Ac| = 2$,   we have (recall that $X_{\Ac}:= \{X_n:n\in \Ac\}$)
\begin{align}
H(X_{\Ac}) \geq 3\Lsf,
\end{align}
  Hence, we have the communication cost is no less than $3$.

{\bf [Achievability]:} 
We can use the proposed scheme in Example~\ref{ex:regime 3} to let the master recover $3$ linearly independent linear combinations of $(W_1,W_2,W_3)$, such that the master can recover 
each message and then recover $(W_1, W_2+W_3)$. The needed communication cost is $3$ as shown in Example~\ref{ex:regime 3}, which coincides with the above converse bound.

From the above proof, we can also see that for the $(\Ksf,\Nsf,\Nsf_{\rm r}, \Ksf_{\rm c}, \Msf)=(3,3,2,2,2)$ distributed linearly separable computation problem,  
\begin{itemize}
\item if the demand matrix is full-rank and it contains  a sub-matrix   with dimension $2 \times 2$ which is not full-rank, the optimal communication cost is $3$;
\item  otherwise, the optimal communication cost is $2$.
\end{itemize}
 It is one of our on-going works to study the specific demand matrices for more general case.
\hfill $\square$ 
 \end{rem}

 \section{Extensions}
 \label{sec:extension} 
 In this section, we will discuss about the extension of the proposed scheme in Section~\ref{sec:achie}. In Section~\ref{sub:N does not divide K}, we propose an extended scheme for the general values of $\Ksf$ and $\Nsf$ (i.e., $\Nsf$ does not necessarily divide $\Ksf$).  In Section~\ref{sub:non cyclic assignment}, we provide an example to show that the cyclic assignment is sub-optimal.
 
 \subsection{General values of $\Ksf$ and $\Nsf$}
 \label{sub:N does not divide K}
 We assume that $\Ksf = \asf \Nsf + \bsf$, where $\asf$ is a non-negative integer and $\bsf \in [\Nsf-1]$. Since we still consider the minimum computation cost and each dataset should be assigned to at least $\Nsf-\Nsf_{\rm r}+1$ workers, thus now the minimum computation cost is 
\begin{align}
 \left\lceil  \frac{\Ksf}{\Nsf}(\Nsf-\Nsf_{\rm r}+1) \right\rceil  
= \asf (\Nsf-\Nsf_{\rm r}+1) +  \left\lceil  \frac{\bsf}{\Nsf}(\Nsf-\Nsf_{\rm r}+1) \right\rceil . \label{eq:total memory}
\end{align}
It will be explained later that in order to enable the extension of the cyclic assignment to the general values of $\Ksf$ and $\Nsf$, we consider the computation cost
\begin{align}
\Msf_1 :=\asf (\Nsf-\Nsf_{\rm r}+1) +  \left\lceil  \frac{ \Nsf-\Nsf_{\rm r}+1 }{\left\lfloor  \frac{\Nsf}{\bsf}  \right\rfloor }  \right\rceil , \label{eq:M1}
\end{align}
 which may  be slightly larger than the minimum computation cost in~\eqref{eq:total memory}.
 
 We generalize the proposed scheme in Section~\ref{sec:achie}  by introducing $\Nsf-\bsf$ virtual datasets, to obtain the following theorem, which is the generalized version of Theorem~\ref{thm:achie}.
 \begin{thm}
 \label{thm:achie N does not divide K}
 For the  $(\Ksf,\Nsf,\Nsf_{\rm r}, \Ksf_{\rm c}, \Msf)$ distributed linearly separable computation problem  with $\Ksf = \asf \Nsf +\bsf$ and $\Msf= \Msf_1  $ where $\asf$ is a non-negative integer and $\bsf \in [\Nsf-1]$,  the communication cost $\Rsf^{\prime}_{{\rm ach}}$ is achievable, where 
\begin{itemize}
 \item when  $\Ksf_{\rm c} \in \left[  \left\lfloor \frac{\Ksf}{\Nsf} \right\rfloor   \right]$, 
 \begin{subequations}
\begin{align}
  \Rsf^{\prime}_{{\rm ach}}= \Nsf_{\rm r} \Ksf_{\rm c}; \label{eq:general achie case 1}
\end{align} 
\item when  $ \Ksf_{\rm c}  \in \left[ \left\lceil  \frac{\Ksf}{\Nsf} \right\rceil   : \left\lceil \frac{\Ksf}{\Nsf}  \right\rceil   \Nsf_{\rm r} \right]  $,
\begin{align}
\Rsf^{\prime}_{{\rm ach}}=\left\lceil \frac{\Ksf}{\Nsf}  \right\rceil  \Nsf_{\rm r}  ; \label{eq:general achie case 2}
\end{align}
\item when  $ \Ksf_{\rm c} \in \left(  \left\lceil \frac{\Ksf}{\Nsf}  \right\rceil   \Nsf_{\rm r}  : \Ksf \right]$,
\begin{align}
\Rsf^{\prime}_{{\rm ach}}= \Rsf^{\star}= \Ksf_{\rm c}, \label{eq:general achie case 3}
\end{align}
 \end{subequations}
where $\Rsf^{\star}$ represents the optimal communication cost for this case.
 \end{itemize}
 \hfill $\square$ 
 \end{thm}
\begin{IEEEproof} 
%Since the case where $\Nsf$ divides $\Ksf$ has been considered in Section~\ref{sec:achie}, in the following we assume $\bsf \in [ \Nsf-1]$.
We first extend the cyclic assignment in Section~\ref{sub:problem formulation} to the general case  
  by dividing the $\Ksf$ datasets into two groups, $[\asf\Nsf]$ and $[\asf\Nsf+1:\Ksf]$, respectively. 
\begin{itemize}
\item For each dataset $D_k$ where $k\in  [\asf \Nsf]$, we assign $D_k$ to worker $j$, where $j \in \big\{\text{Mod}(k,\Nsf),  \text{Mod}(k-1,\Nsf),\ldots,  \text{Mod}(k-\Nsf+\Nsf_{\rm r}, \Nsf ) \big\}$. Hence, the assignment on the datasets in the first group  is the same as   the cyclic assignment in Section~\ref{sub:problem formulation}. 
The number of datasets in the first group assigned to each worker is 
\begin{align}
\asf (\Nsf-\Nsf_{\rm r}+1). \label{eq:memory second group}
\end{align} 
\item For the second group, we introduce $\Nsf-\bsf$ virtual datasets and thus there are totally $\Nsf$ effective (real or virtual) datasets. We then use the cyclic assignment in Section~\ref{sub:problem formulation} to assign the $\Nsf$ effective datasets to the workers, such that the number of effective datasets assigned to each worker is $\Nsf-\Nsf_{\rm r}+1$. 
To satisfy the assignment constraint (i.e., $|\Zc_n|\leq \Msf$ for each $n \in [\Nsf]$), it can be seen from~\eqref{eq:M1} and~\eqref{eq:memory second group} that the number of real datasets in the second group assigned to each worker should be no more than $\left\lceil  \frac{ \Nsf-\Nsf_{\rm r}+1 }{\left\lfloor  \frac{\Nsf}{\bsf}  \right\rfloor }  \right\rceil.$
Hence, our objective is to choose $\bsf$ datasets from  $\Nsf$ effective datasets as the real datasets, such that by the cyclic assignment on these $\Nsf$ effective datasets the number of real datasets assigned to each worker is  no more than $\left\lceil  \frac{ \Nsf-\Nsf_{\rm r}+1 }{\left\lfloor  \frac{\Nsf}{\bsf}  \right\rfloor }  \right\rceil.$ 
We will propose an allocation algorithm in Appendix~\ref{sec:choosing method} which can generally attain the above objective. Here we provide an example to illustrate the idea, where $\Ksf=\bsf=3$, $\asf=0$, $\Nsf=6$, and $\Nsf_{\rm r}=4$. We have  totally $6$ effective datasets denoted by, $E_1,\ldots, E_6$. By the cyclic assignment, the number of effective datasets assigned to each worker is $\Nsf-\Nsf_{\rm r}+1=3$. Thus 
we assign that 
\begin{align*}
\begin{array}{rl|c|c|c|}\cline{3-3}\cline{4-4}\cline{5-5} 
&&\rule{0pt}{1.2em}\mbox{Worker 1} &\rule{0pt}{1.2em}\mbox{Worker 2}  & \rule{0pt}{1.2em}\mbox{Worker 3} \\\cline{3-3}\cline{4-4}\cline{5-5}\
&& {\magenta E_1}&  E_2  &  {\magenta E_3}\\
&& E_2&  {\magenta E_3}   & E_4   \\
&& {\magenta E_3} &  E_4   & {\magenta E_5}  \\ 
\cline{3-3}\cline{4-4}\cline{5-5}
&&\rule{0pt}{1.2em}\mbox{Worker 4} &\rule{0pt}{1.2em}\mbox{Worker 5}  & \rule{0pt}{1.2em}\mbox{Worker 6} \\\cline{3-3}\cline{4-4}\cline{5-5}
&&   E_4&  {\magenta E_5}& E_6\\
&&  {\magenta E_5}  & E_6  & {\magenta E_1}\\
&&  E_6 & {\magenta E_1} & E_2\\ 
\cline{3-3}\cline{4-4}\cline{5-5}
\end{array}
\end{align*}
By choosing $E_1$, $E_3$, and $E_5$ as the real datasets, it can be seen that the number of real datasets assigned to each worker is no more than $   \left\lceil  \frac{ \Nsf-\Nsf_{\rm r}+1 }{\left\lfloor  \frac{\Nsf}{\bsf}  \right\rfloor }  \right\rceil =2$.
\end{itemize}  
  After the data assignment phase, each worker then computes the message for each assigned real dataset. The virtual message which comes from each virtual dataset, is set to be a vector of $\Lsf$ zeros. We then directly use the computing phase of the  proposed scheme in Section~\ref{sec:achie} for  the  $(\Ksf^{\prime},\Nsf^{\prime},\Nsf_{\rm r}^{\prime}, \Ksf_{\rm c}^{\prime}, \Msf^{\prime})=\left((\asf+1)\Nsf,\Nsf, \Nsf_{\rm r}, \Ksf_{\rm c}, (\asf+1)(\Nsf-\Nsf_{\rm r}+1) \right)$ distributed linearly separable computation problem, to achieve the communication cost in Theorem~\ref{thm:achie N does not divide K}. 
  \end{IEEEproof}

 \subsection{Improvement on the cyclic assignment}
 \label{sub:non cyclic assignment} 
In the following, we will provide an example which shows the sub-optimality of the cyclic assignment. 

\begin{example}[$\Ksf=12$, $\Nsf=4$, $\Nsf_{\rm r}=3$, $\Ksf_{\rm c}=3$, $\Msf=6$]
\label{ex:improvement}
\rm 
Consider the example where $\Ksf=12$, $\Nsf=4$, $\Nsf_{\rm r}=3$, $\Ksf_{\rm c}=3$, and we assign  $\Msf=\frac{\Ksf}{\Nsf} (\Nsf-\Nsf_{\rm r}+1)=6$ datasets to each worker. Each dataset is assigned to $\Nsf-\Nsf_{\rm r}+1=2$ workers. By the proposed   scheme with the cyclic assignment for the case where $\Ksf_{\rm c}=\frac{\Ksf}{\Nsf}$ in Theorem~\ref{thm:achie},  the needed communication cost is $\frac{\Ksf}{\Nsf} \Nsf_{\rm r}=9$, which is optimal under the constraint of   the cyclic assignment. 
However, by the proposed converse bound in Theorem~\ref{thm:converse}, the minimum communication cost is upper bounded by $6$. We will introduce a novel distributed computing scheme to 
achieve the minimum communication cost. As a result, we   show the sub-optimality of the cyclic assignment.

\paragraph*{Data assignment phase}
Inspired by the placement phase of the coded caching scheme in~\cite{dvbt2fundamental}, we assign that
\begin{align*}
\begin{array}{rl|c|c|c|c|}\cline{3-3}\cline{4-4}\cline{5-5}\cline{6-6}
&&\rule{0pt}{1.2em}\mbox{Worker 1}  &\rule{0pt}{1.2em}\mbox{Worker 2}  & \rule{0pt}{1.2em}\mbox{Worker 3}  &\rule{0pt}{1.2em}\mbox{Worker 4} \\\cline{3-3}\cline{4-4}\cline{5-5}\cline{6-6}
&& D_1&  D_1 &  D_3 &  D_{5}\\
&& D_2&   D_2 &  D_4 &  D_{6} \\ 
&& D_3&  D_7 &  D_7 &   D_{9} \\
&& D_4&   D_8 &  D_8 &  D_{10}\\
&& D_5&  D_9 &  D_{11} &   D_{11}\\
&& D_6&   D_{10} &  D_{12} &   D_{12}\\
\cline{3-3}\cline{4-4}\cline{5-5}\cline{6-6}
\end{array}
\end{align*}
More precisely, we partition the $12$ datasets into $\binom{4}{2}=6$ groups, 
each of which is denoted by $\Hc_{\Tc}$ where $\Tc \subseteq [4]$ where $|\Tc|=2$ and contains $2$ datasets. In this example, we let
\begin{align*}
&\Hc_{\{1,2\}}=\{1,2\}, \ \Hc_{\{1,3\}}=\{3,4\}, \ \Hc_{\{1,4\}}=\{5,6\}, \\
&\Hc_{\{2,3\}}=\{7,8\}, \ \Hc_{\{2,4\}}=\{9,10\}, \ \Hc_{\{3,4\}}=\{11,12\}.
\end{align*}
  For each set $\Tc \subseteq [4]$ where $|\Tc|=2$, we assign   dataset $D_k$ where $k\in \Hc_{\Tc}$ to workers in $\Tc$. Hence, each dataset is assigned to $2$ workers, and the number of datasets assigned to each worker is $2\binom{4-1}{2-1}=6$ (e.g., the datasets in groups $\Hc_{\{1,2\}}, \Hc_{\{1,3\}},\Hc_{\{1,4\}}$ are assigned to worker $k$), satisfying the assignment constraint.

\paragraph*{Computing phase}
We assume that the task function is 
\begin{align*}
&f(D_1,   \ldots, D_{\Ksf})= \left[ \begin{array}{c}
F_1\\
F_2\\
F_3\\
\end{array} \right]  
=  {\bf F} 
\left[ \begin{array}{c}
W_1\\
\vdots\\
W_{12}\\
\end{array} \right] \\&=
\left[ \begin{array}{c}
1 ,1 , 1 ,1 ,1 ,1, 1 ,1 , 1 ,1 ,1 ,1\\
1 ,2 , 3 ,4 ,5 ,6,7,8,9,10,11,12\\
1 ,0 , 3,2,8,4,1,2,9,4,5,10\\     
\end{array} \right]
\left[ \begin{array}{c}
W_1\\
\vdots\\
W_{12}\\
\end{array} \right] .
\end{align*}
Note that  the following proposed scheme works for any request  with high probability, where the elements ${\bf F}$ are uniformly i.i.d. 

We now focus on  each group  $\Hc_{\Tc}$ where $\Tc \subseteq [6]$ and $|\Tc|=2$. When $\Tc=\{1,2\}$, we have $\Hc_{\{1,2\}}=\{1,2\}$.  We retrieve the sub-matrix 
$$
{\bf F}^{(\{1,2\})_{\rm c}}= \left[ \begin{array}{c}
1 ,1 \\
1 ,2 \\
1 ,0 \\     
\end{array} \right],
$$
i.e., columns with indices in $\Hc_{\{1,2\}}=\{1, 2 \}$ of ${\bf F}$. Since the dimension of ${\bf F}^{(\{1,2\})_{\rm c}}$ is  $3 \times 2$, the left-side null-space of ${\bf F}^{(\{1,2\})_{\rm c}}$ contains one   vector. Now we choose the vector $(-2,1,1)$, where  $(-2,1,1) {\bf F}^{(\{1,2\})_{\rm c}}=(0,0)$. 
Hence, in the product $(-2,1,1) [F_1;F_2;F_3]$, the coefficients of $W_1$ and $W_2$ are $0$. We define that 
     \begin{subequations}
\begin{align}
& U_{\Tc}=U_{\{1,2\}}:=(-2,1,1) [F_1;F_2;F_3]= -2 F_1+ 1 F_2 +1 F_3 \\
&=  {\bf 0 W_1+ 0 W_2} + 4 W_3 + 4 W_4+ 11 W_5 + 8 W_6+ 6 W_7 \nonumber\\& + 8 W_8 +16 W_9 + 12 W_{10} +14 W_{11}+ 20 W_{12}. \label{eq:U12}
\end{align}
      \end{subequations}

Similarly, when $\Tc=\{1,3\}$,  we have $\Hc_{\{1,3\}}=\{3,4\}$.  By choosing the vector $(-6,1,1)$ as the left-side null-space of ${\bf F}^{(\{3,4\})_{\rm c}}$, and define that 
     \begin{subequations}
\begin{align}
& U_{\{1,3\}}:=(-6,1,1) [F_1;F_2;F_3]= -6 F_1+ 1 F_2 +1 F_3 \\
&=  -4 W_1  -4 W_2 + {\bf 0 W_3 + 0 W_4} + 7 W_5 + 4 W_6+ 2 W_7 \nonumber\\& + 4 W_8 +12 W_9 + 8 W_{10} +10 W_{11}+ 16 W_{12}.  \label{eq:U13}
\end{align}
      \end{subequations}
 
When $\Tc=\{1,4\}$,  we have $\Hc_{\{1,4\}}=\{5,6\}$. By choosing the vector $(-28,4,1)$ as the left-side null-space of ${\bf F}^{(\{5,6\})_{\rm c}}$, and define that 
      \begin{subequations}
\begin{align}
& U_{\{1,4\}}:=(-28,4,1) [F_1;F_2;F_3]= -28 F_1+ 4 F_2 +1 F_3 \\
&=  -23 W_1  -20 W_2 - 13 W_3 - 10 W_4+ {\bf 0 W_5 + 0 W_6}+ 1 W_7 \nonumber\\& + 6 W_8 +17 W_9 + 16 W_{10} +21 W_{11}+ 30 W_{12}.   \label{eq:U14}
\end{align}
      \end{subequations}
 
 When $\Tc=\{2,3\}$,  we have $\Hc_{\{2,3\}}=\{7,8\}$. By choosing the vector $(6,-1,1)$ as the left-side null-space of ${\bf F}^{(\{7,8\})_{\rm c}}$, and define that 
       \begin{subequations}
\begin{align}
& U_{\{2,3\}}:=(6,-1,1) [F_1;F_2;F_3]= 6 F_1-1 F_2 +1 F_3 \\
&=  6 W_1  +4 W_2 + 6 W_3 +4 W_4+ 9 W_5 + 4 W_6\nonumber\\& + {\bf 0 W_7+ 0 W_8} +6 W_9 + 0 W_{10} +0 W_{11}+ 4 W_{12}.    \label{eq:U23}
\end{align}
      \end{subequations}
      
   When $\Tc=\{2,4\}$,  we have $\Hc_{\{2,4\}}=\{9,10\}$. By choosing the vector $(-54,5,1)$ as the left-side null-space of ${\bf F}^{(\{9,10\})_{\rm c}}$, and define that    
        \begin{subequations}
\begin{align}
& U_{\{2,4\}}:=(-54,5,1)[F_1;F_2;F_3]= -54 F_1 +5 F_2 +1 F_3 \\
&= -48 W_1  -44 W_2 -36 W_3 -32 W_4 -21 W_5 -20 W_6 \nonumber\\& -18 W_7 -12 W_8 + {\bf 0 W_9 + 0 W_{10}} +6 W_{11}+ 16 W_{12}.     \label{eq:U24}
\end{align}
      \end{subequations}
 
 When $\Tc=\{3,4\}$,  we have $\Hc_{\{3,4\}}=\{11,12\}$. By choosing the vector $(50,-5,1)$ as the left-side null-space of ${\bf F}^{(\{11,12\})_{\rm c}}$, and define that     
        \begin{subequations}
\begin{align}
& U_{\{3,4\}}:=(50,-5,1)[F_1;F_2;F_3]= 50 F_1 -5 F_2 +1 F_3 \\
&=  46 W_1  +40 W_2 +38 W_3 +32 W_4 +33 W_5 +24 W_6 \nonumber\\& +16 W_7 +12 W_8 + 14 W_9 + 4 W_{10} + {\bf 0 W_{11}+ 0 W_{12}}.      \label{eq:U34}
\end{align}
      \end{subequations}

{\bf Our main strategy is that for any set of two workers $\Sc \subseteq [4]$ where $|\Sc|=\Nsf-\Nsf_{\rm r}+1=2$,
from the transmitted coded messages by the workers in $\Sc$, the master can recover  $U_{[4]\setminus \Sc}$.}
\begin{itemize}
\item Assume that   the straggler is worker $4$. From workers $1$ and $2$, the master can recover $U_{\{3,4\}}$; from workers $1$ and $3$, the master can recover $U_{\{2,4\}} $; from workers $2$ and $3$, the master can recover $U_{\{1,4\}} $. In addition, it can be seen that $U_{\{1,4\}} $, $U_{\{2,4\}} $, and $U_{\{3,4\}} $ are linearly independent. Hence, the master can recover $F_1$, $F_2$, and $F_3$.
\item Assume that   the straggler is worker $3$. The master can recover $U_{\{1,3\}} $, $U_{\{2,3\}} $, and $U_{\{3,4\}} $, which are linearly independent, such that it can recover $F_1$, $F_2$, and $F_3$. 
\item Assume that   the straggler is worker $2$. The master can recover $U_{\{1,2\}} $, $U_{\{2,3\}} $, and $U_{\{2,4\}} $, which are linearly independent, such that it can recover $F_1$, $F_2$, and $F_3$. 
\item Assume that   the straggler is worker $1$. The master can recover $U_{\{1,2\}} $, $U_{\{1,3\}} $, and $U_{\{1,4\}} $, which are linearly independent, such that it can recover $F_1$, $F_2$, and $F_3$. 
\end{itemize}
In the following, we provide a code construction such that the above strategy can be achieved.

When $\Sc=\{1,2\}$,  workers $1$ and $2$ should send cooperatively 
\begin{align*}
U_{\{3,4\}} &=  46 W_1  +40 W_2 +38 W_3 +32 W_4 +33 W_5 +24 W_6 \nonumber\\& +16 W_7 +12 W_8 + 14 W_9 + 4 W_{10} + {\bf 0 W_{11}+ 0 W_{12}}.
\end{align*} 
Between workers $1$ and $2$, it can be seen that $W_{3}$, $W_4$, $W_5$, and $W_6$ can only be computed by worker $1$, while  $W_{7}$, $W_8$, $W_9$, and $W_{10}$ can only be computed by worker $2$. In addition,   both workers $1$ and $2$ can compute $W_{1}$ and $W_2$.
Hence, we let worker $1$  send 
$$
A_{1,\{3,4\}}=x_5 W_{1}+x_6 W_{2}+38 W_{3} +32 W_4+33 W_5+24 W_6,
$$
and let worker $2$ send
$$
A_{2,\{3,4\}}=x_{11} W_{1}+x_{12} W_{2}+16 W_{7} +12 W_8+ 14 W_9+4 W_{10},
$$
where   $A_{1,\{3,4\}}+A_{2,\{3,4\}}=U_{\{3,4\}}$. Note that
$x_5$, $x_6$, $x_{11}$, and $x_{12}$ are the coefficients which we can design. 
 Hence, we have 
\begin{align}
&x_5+x_{11}=46; \label{eq:x511}\\
&x_6+x_{12}=40. \label{eq:x612}
\end{align}

Similarly, by considering all sets $\Sc \subseteq [4]$ where $|\Sc|= 2$, the transmissions of worker $1$ can be expressed as
\begin{align}
A_{1,\{2,3\}}&= 6 W_{1}+ 4 W_{2}+6 W_{3}+4 W_{4} \nonumber\\& +x_1 W_{5}+x_2 W_{6},\\
 A_{1,\{2,4\}}&= -48 W_{1}- 44 W_{2}+ x_3 W_{3}+x_4 W_{4} \nonumber\\& -21 W_{5} -20 W_{6},\\
 A_{1,\{3,4\}}&= x_5 W_{1}+x_6 W_{2}+38 W_{3} +32 W_4 \nonumber\\& +33 W_5+24 W_6.
\end{align}
The transmissions of worker $2$ can be expressed as
\begin{align}
 A_{2,\{1,4\}}&= -23 W_{1} -20 W_{2}+ x_7 W_{7}+ x_8 W_{8} \nonumber\\& + 17 W_{9}+16 W_{10},\\
 A_{2,\{1,3\}}&= -4 W_{1} -4 W_{2}+2 W_{7}+4 W_{8} + x_9 W_{9} \nonumber\\&+ x_{10} W_{10},\\
 A_{2,\{3,4\}}&= x_{11} W_{1} +x_{12} W_{2}+16 W_{7}+12 W_{8}  \nonumber\\&+ 14 W_{9}+ 4 W_{10}.
\end{align}
The transmissions of worker $3$ can be expressed as
\begin{align}
 A_{3,\{1,2\}}&= 4 W_{3} +4 W_{4}+ 6 W_{7}+ 8 W_{8} \nonumber\\&+ x_{13} W_{11}+x_{14} W_{12},\\
A_{3,\{1,4\}}&= -13 W_{3} -10 W_{4}+ x_{15} W_{7}+ x_{16} W_{8} \nonumber\\& + 21 W_{11}+ 30 W_{12},\\
 A_{3,\{2,4\}}&= x_{17} W_{3} +x_{18} W_{4} -18  W_{7}-12 W_{8} \nonumber\\& + 6 W_{11}+16 W_{12}.
\end{align}
The transmissions of worker $4$ can be expressed as
\begin{align}
A_{4,\{1,2\}}&= 11 W_{5} +8 W_{6}+ 16 W_{9}+ 12 W_{10}  \nonumber\\&+ x_{19} W_{11}+x_{20} W_{12},\\
 A_{4,\{1,3\}}&= 7 W_{5} +4 W_{6}+ x_{21} W_{9}+ x_{22} W_{10} \nonumber\\& + 10 W_{11}+ 16 W_{12},\\
 A_{4,\{2,3\}}&= x_{23} W_{5} +x_{24} W_{6}+ 6 W_{9}+ 0 W_{10} \nonumber\\&  + 0 W_{11}+ 4 W_{12}.
\end{align}
The coefficients of $(x_1,\ldots, x_{12})$ should satisfy~\eqref{eq:x511},~\eqref{eq:x612}, and
\begin{align}
&x_{1}+x_{23}=9; \label{eq:x123}\\
&x_{2}+x_{24}=4; \label{eq:x224}\\
&x_{3}+x_{17}=-36; \label{eq:x317}\\
&x_{4}+x_{18}=-32; \label{eq:x418}\\
&x_{7}+x_{15}=1; \label{eq:x715}\\
&x_{8}+x_{16}=6; \label{eq:x816}\\
&x_{9}+x_{21}=12; \label{eq:x921}\\
&x_{10}+x_{22}=8; \label{eq:x1022}\\
&x_{13}+x_{19}=14; \label{eq:x1319}\\
&x_{14}+x_{20}=20. \label{eq:x1420}
\end{align}

Finally, we will introduce how to choose  $(x_1,\ldots, x_{12})$ such that the above constraints are satisfied.
Meanwhile, 
the rank of the transmissions of each worker is $2$ (i.e., among the three sent sums by each worker, one sum can be obtained by the linear combinations of the other two sums), such that we can let each worker send only two linear combinations of messages and the   needed communication cost is $2\Nsf_{\rm r}=6$,
         which coincides with  the proposed converse bound in Theorem~\ref{thm:converse}.  

We let $A_{1,\{2,3\}}+A_{1,\{2,4\}}=A_{1,\{3,4\}}$. Hence, we have 
$$
x_{1}=54, \ x_{2}=44, \ x_{3}=32, \ x_{4}=28, \ x_{5}= -42, \ x_{6}=-40.
$$

With $x_{5}= -42$ and $   x_{6}=-40$, from~\eqref{eq:x511} and~\eqref{eq:x612}  we can see that 
$$
x_{11}=88, \ x_{12}=80.
$$
Since we fix $x_{11}=88$ and  $x_{12}=80$, if the rank of the transmissions of worker  $2$ is $2$, we should have 
$$
x_{7}=-11, \ x_{8}=-29/2, \ x_{9}= -89/10, \ x_{10}=-7.
$$

With $x_{3}= 32$ and $   x_{4}=28$, from~\eqref{eq:x317} and~\eqref{eq:x418}  we can see that 
$$
x_{17}=-68, \ x_{18}=-60.
$$
Since we fix $x_{17}=-68$ and  $x_{18}=-60$, if the rank of the transmissions of worker  $3$ is $2$, we should have 
$$
x_{13}=6, \ x_{14}=192/25, \ x_{15}=12, \ x_{16}=41/2.
$$

With $x_{1}= 54$ and $   x_{2}=44$, from~\eqref{eq:x123} and~\eqref{eq:x224}  we can see that 
$$
x_{23}=-45, \ x_{24}=-40.
$$
Since we fix $x_{23}=-45$ and  $x_{24}=-40$, if the rank of the transmissions of worker  $4$ is $2$, we should have 
$$
x_{19}=8, \ x_{20}= 308/25, \ x_{21}=418/20, \ x_{22}=15.
$$

With the above choice of $(x_1,\ldots, x_{12})$, we can find that \\
$x_{7}+x_{15}= -11+ 12=1$, satisfying~\eqref{eq:x715};\\
 $x_{8}+x_{16}= -29/2+ 41/2=6$, satisfying~\eqref{eq:x816};\\
  $x_{9}+x_{21}=  -89/10+ 418/20=12$, satisfying~\eqref{eq:x921};\\
    $x_{10}+x_{22}=  -7+ 15=8$, satisfying~\eqref{eq:x1022};\\
     $x_{13}+x_{19}= 6+ 8=14$, satisfying~\eqref{eq:x1319};\\
          $x_{14}+x_{20}= 192/25+ 308/25=20$, satisfying~\eqref{eq:x1420}.
          
          In conclusion  the above choice of $(x_1,\ldots, x_{12})$ satisfies all constraints in~\eqref{eq:x511},~\eqref{eq:x612},~\eqref{eq:x123}-\eqref{eq:x1420},
          while the rank of the transmissions of each worker is $2$.  
          
   Note that the above assignment based on coded caching can only be used for very limited number of cases in our problem, i.e., when $\binom{\Nsf}{\Nsf-\Nsf_{\rm r}+1}$ divides $\Ksf$. In addition, it is   part of on-going works to generalize the above computing phase under the coded caching assignment to the general case where $\binom{\Nsf}{\Nsf-\Nsf_{\rm r}+1}$ divides $\Ksf$.
         \hfill $\square$ 
 \end{example}

 \section{Conclusions}
\label{sec:conclusion}
In this paper, we introduced a distributed linearly separable computation problem and studied the optimal communication cost when the computation cost is minimum. 
We proposed a converse bound inspired by coded caching converse bounds and an achievable distributed computing scheme based on linear space intersection. The proposed scheme was proved to be optimal under some system parameters. In addition, it was also proved to be optimal under the constraint of the cyclic assignment on the datasets. 

Further works include the extension of the proposed scheme to the case where the computation cost is increased, the design of the distributed computing scheme with some improved assignment    rather  than the cyclic assignment, and novel achievable schemes  on specific demand matrices for   general case. 
 
  \appendices

\section{Proof of Theorem~\ref{thm:converse}}
\label{sec:proof of converse}
Recall that  the computation cost is minimum, and thus each dataset is assigned to $\Nsf-\Nsf_{\rm r}+1$ workers. 
For each set  $\Sc \subseteq [\Nsf]$ where $|\Sc|=\Nsf-\Nsf_{\rm r}+1$,  we define $\Gc_{\Sc}$ as the set of datasets uniquely assigned to all workers in $\Sc$. For example, in Example~\ref{ex:regime 1}, $\Gc_{\{1,2\}}=\{2,5\}$, $\Gc_{\{1,3\}}=\{1,4\}$, and $\Gc_{\{2,3\}}=\{3,6\}$.

Let us focus one worker $n \in [\Nsf]$.
Since the number of datasets assigned to each worker is $\frac{\Ksf}{\Nsf} (\Nsf-\Nsf_{\rm r}+1)$, we have
 \begin{align}
\sum_{\Sc\subseteq [\Nsf]:|\Sc|= \Nsf-\Nsf_{\rm r}+1, n \in \Sc} |\Gc_{\Sc}|= \frac{\Ksf}{\Nsf} (\Nsf-\Nsf_{\rm r}+1).\label{eq:storage constraint}
\end{align}
From~\eqref{eq:storage constraint}, it can be seen that  
\begin{subequations}
\begin{align}
\max_{\Sc\subseteq [\Nsf]:|\Sc|= \Nsf- \Nsf_{\rm r}+1, n \in \Sc} |\Gc_\Sc| &\geq \left\lceil \frac{\Ksf  (\Nsf- \Nsf_{\rm r}+1) }{\Nsf \binom{\Nsf-1}{\Nsf-\Nsf_{\rm r}}} \right\rceil\\ &=  \left\lceil \frac{ \Ksf    }{\binom{\Nsf }{\Nsf-\Nsf_{\rm r}+1}} \right\rceil .\label{eq:max constraint}
\end{align} 
\end{subequations}
In addition, with a slight abuse of notation we define that 
\begin{align}
\Sc_{\rm max}=\underset{\Sc \subseteq [\Nsf]:|\Sc|= \Nsf-\Nsf_{\rm r}+1, n \in \Sc}{ \arg\max}   |\Gc_\Sc|
\end{align}
Consider now the set of responding workers   $\Sc_1= \{n\} \cup ([\Nsf] \setminus \Sc_{\rm max})$.  Note that among the workers in $\Sc_1$, each dataset $D_k$ where $k \in \Gc_{\Sc_{\rm max}}$ is only assigned to worker $n$. In addition, since the elements in  ${\bf F}$ are uniformly i.i.d. over a large enough field, 
matrix ${\bf F}^{(\Gc_{\Sc_{\rm max}})_{\rm c}}$ (representing the sub-matrix  containing    the columns with indices in $\Gc_{\Sc_{\rm max}}$ of  ${\bf F}$)  has the  rank equal to 
$
\min \left\{\Ksf_{\rm c}, |\Gc_{\Sc_{\rm max}}|  \right\}
$
 with high probability. 
In addition, each message has $\Lsf$ uniformly i.i.d. symbols. Hence, we have 
\begin{align}
T_n \geq H(X_n) \geq \min \left\{\Ksf_{\rm c}, |\Gc_{\Sc_{\rm max}}|  \right\} \Lsf.\label{eq:converse of Xn}
\end{align}
%Similarly,~\eqref{eq:converse of Xn} holds for each worker $n\in [\Nsf]$.

Now we consider each   $\Ac \subseteq [\Nsf]$ where $|\Ac|=\Nsf_{\rm r}$ as the set of responding worker.
From the definition of the communication cost in~\eqref{eq:communicaton rate}, we have 
    \begin{subequations}
\begin{align}
\Rsf & \geq  \frac{ \sum_{n_1 \in \Ac} T_{n_1}}{ \Lsf} \\
&\geq  \frac{  \Nsf_{\rm r}  \min \left\{\Ksf_{\rm c}, |\Gc_{\Sc_{\rm max}}|  \right\} \Lsf }{ \Lsf} \label{eq:come from converse Xn}\\
&\geq     \Nsf_{\rm r}  \min \left\{\Ksf_{\rm c}, \left\lceil \frac{ \Ksf    }{\binom{\Nsf }{\Nsf-\Nsf_{\rm r}+1}} \right\rceil  \right\}  , \label{eq:come from max constraint}
\end{align}
    \end{subequations}
where~\eqref{eq:come from converse Xn} comes from~\eqref{eq:converse of Xn} and~\eqref{eq:come from max constraint} comes from~\eqref{eq:max constraint}.  
By  the definition of the minimum communication cost   and   the fact that $\Rsf^{\star} \geq \Ksf_{\rm c} $, from~\eqref{eq:come from max constraint} we prove Theorem~\ref{thm:converse}.

\section{Proof of Theorem~\ref{thm:cyclic optimality}}
\label{sec:proof cyclic optimality} 
We fix  an integer $n \in [\Nsf]$.
By the cyclic assignment described in Section~\ref{sub:problem formulation},   each dataset $D_{n+p \Nsf}$ where $p \in \left[0: \frac{\Ksf}{\Nsf}-1\right]$ is assigned to  $\Nsf-\Nsf_{\rm r}+1$ workers. The set of these $\Nsf-\Nsf_{\rm r}+1$ workers is 
 $$
 \Sc_1=\big\{n,  \text{Mod}(n-1,\Nsf),\ldots,  \text{Mod}(n-\Nsf+\Nsf_{\rm r}, \Nsf ) \big\}.
 $$
 Now we assume the set of the responding workers is $\Rc_1=\{n\} \cup ([\Nsf]\setminus \Sc_1)$.  It can be seen that among the workers in $\Rc_1$, each dataset $D_k$ where $k \in \left\{n+p \Nsf: p \in \left[0: \frac{\Ksf}{\Nsf}-1\right] \right\}$ is only assigned to worker $n$.  
 In addition, since the elements in  ${\bf F}$ are uniformly i.i.d. over a large enough field, matrix ${\bf F}^{(\left\{n+p \Nsf: p \in \left[0: \frac{\Ksf}{\Nsf}-1\right] \right\})_{\rm c}}$
   has the  rank equal to 
$
\min \left\{\Ksf_{\rm c}, \frac{\Ksf}{\Nsf}  \right\} 
$
 with high probability.
In addition, each message has $\Lsf$ uniformly i.i.d. symbols. Hence, we have 
\begin{align}
T_n \geq H(X_n) \geq \min \left\{\Ksf_{\rm c}, \frac{\Ksf}{\Nsf}  \right\} \Lsf.\label{eq:converse of Xi}
\end{align}

Now we consider each   $\Ac \subseteq [\Nsf]$ where $|\Ac|=\Nsf_{\rm r}$ as the set of responding worker. We have 
    \begin{subequations}
\begin{align}
\Rsf & \geq  \frac{ \sum_{n_1 \in \Ac} T_{n_1}}{ \Lsf} \\
&\geq  \frac{  \Nsf_{\rm r}  \min \left\{\Ksf_{\rm c}, \frac{\Ksf}{\Nsf}  \right\} \Lsf}{ \Lsf} \label{eq:come from converse Xi}, 
\end{align}
    \end{subequations}
where~\eqref{eq:come from converse Xi} comes from~\eqref{eq:converse of Xi}. Hence, when $\Ksf_{\rm c}\leq \frac{\Ksf}{\Nsf}$, we have $\Rsf \geq  \Nsf_{\rm r} \Ksf_{\rm c}$; when $\Ksf_{\rm c}\geq \frac{\Ksf}{\Nsf}$, we have $\Rsf \geq  \Nsf_{\rm r}  \frac{\Ksf}{\Nsf}$. Together with $\Rsf \geq \Ksf_{\rm c}$, we obtain the converse bound in Theorem~\ref{thm:cyclic optimality}.

\section{Proof of Lemma~\ref{lem:SZlemma}}
\label{sec:proof of SZlemma}
 We  first focus one $\Ac\subseteq [\Nsf]$ where $|\Ac|= \Nsf_{\rm r}$. We assume that $\Ac=\{\Ac(1), \ldots, \Ac(\Nsf_{\rm r}) \}$ where $\Ac(1)  < \cdots < \Ac(\Nsf_{\rm r})$.

%sort   the elements in $\Ac$ in an increasing order and denote the $i$ th smallest element by $\Ac(i)$, i.e., $\Ac(1)  < \cdots < \Ac(\Nsf_{\rm r})$.

Recall that $\Ksf_{\rm c}=\frac{\Ksf}{\Nsf} \Nsf_{\rm r}$ and that the task function is (recall that $(\mathbf{M})_{m \times n}$ indicates that the dimension of matrix $\mathbf{M}$ is $m \times n$)
$$
({\bf F})_{\frac{\Ksf}{\Nsf} \Nsf_{\rm r} \times \Ksf}  
 ([W_1;\ldots;W_{\Ksf}])_{\Ksf \times \Lsf},
$$  
where each element in  ${\bf F}$ is  uniformly  i.i.d. over large enough finite field $\mathbb{F}_{\qsf}$.  
By the construction of our proposed achievable scheme, each worker $\Ac(i)$ where $i\in [\Nsf_{\rm r}]$ sends 
\begin{align}
{\bf C}_{\{ \Ac(i)\}}
{\bf F}   \left[ \begin{array}{c}
W_1\\
\vdots\\
W_{\Ksf}\\ 
\end{array} \right]=
\left[ \begin{array}{c}
\uv_{\Ac(i),1} \\
\vdots\\
\uv_{\Ac(i), \frac{\Ksf}{\Nsf}}\\    
\end{array} \right]
{\bf F}   \left[ \begin{array}{c}
W_1\\
\vdots\\
W_{\Ksf}\\ 
\end{array} \right], \label{eq:transformed Cn}
\end{align}
where   $\uv_{\Ac(i),j}  {\bf F}^{\left(\overline{  \Zc_{\Ac(i)}} \right)_{\rm c}}={\bf 0}_{1 \times \frac{\Ksf}{\Nsf}(\Nsf_{\rm r}-1)}$ for each $j \in \left[ \frac{\Ksf}{\Nsf}\right]$, and $\overline{  \Zc_{\Ac(i)}} \subseteq [\Ksf]$ represents the set of  datasets which are not assigned to worker $\Ac(i)$.
To simplify the notations, we let 
\begin{align}
\overline{ {\bf F}_{\Ac(i)} }:={\bf F}^{\left(\overline{  \Zc_{\Ac(i)} }  \right)_{\rm c}},
\end{align}
with dimension $\Ksf_{\rm c} \times \frac{\Ksf}{\Nsf}(\Nsf_{\rm r}-1)=\frac{\Ksf}{\Nsf} \Nsf_{\rm r}  \times \frac{\Ksf}{\Nsf}(\Nsf_{\rm r}-1) $.
% Recall that the dimension of $ \overline{ {\bf F_k}}$ defined in~\eqref{eq:bar F_k} is $\frac{K}{N} N_r \times \frac{K}{N}  (N_r-1)$. Hence, the left null-space of $ \overline{ {\bf F_k}}$  contains $\frac{K}{N}$ linearly independent vectors with dimension  $1 \times \frac{K}{N} N_r$.
 By some linear transformations on the rows of ${\bf C}_{\{\Ac(i)\}}$ (we will prove very soon that this transformation exists with high probability),    we have~\eqref{eq:CAi} at the top of the next page.
 \begin{figure*}
  \begin{subequations}
\begin{align}
&\left( {\bf C}_{\{\Ac(i)\}} \right)_{\frac{\Ksf}{\Nsf} \times \frac{\Ksf}{\Nsf} \Nsf_{\rm r}} =
\left[ \begin{array}{cccc}
c_{\Ac(i),1,1} & c_{\Ac(i),1,2} & \cdots & c_{\Ac(i),1,\frac{\Ksf}{\Nsf} \Nsf_{\rm r} } \\
\vdots & \vdots &\ddots & \vdots \\
c_{\Ac(i),\frac{\Ksf}{\Nsf},1} & c_{\Ac(i),\frac{\Ksf}{\Nsf},2} & \cdots & c_{\Ac(i),\frac{\Ksf}{\Nsf},\frac{\Ksf}{\Nsf} \Nsf_{\rm r} } \\ 
\end{array} \right] \\
&=\left[ \begin{array}{cccccccccc}
c_{\Ac(i),1,1} & \cdots &  c_{\Ac(i),1, \frac{\Ksf}{\Nsf}(i-1)} &   1 & 0 & \cdots &0 & c_{\Ac(i),1,\frac{\Ksf}{\Nsf} i + 1 } & \cdots & c_{\Ac(i),1,\frac{\Ksf}{\Nsf} \Nsf_{\rm r}  }  \\
c_{\Ac(i),2,1} & \cdots &    c_{\Ac(i),2,\frac{\Ksf}{\Nsf}(i-1)} &  0 & 1 & \cdots &0 & c_{\Ac(i),2,\frac{\Ksf}{\Nsf} i + 1 } & \cdots & c_{\Ac(i),2,\frac{\Ksf}{\Nsf} \Nsf_{\rm r}  } \\
 \vdots         & \ddots &                            \vdots & \vdots&  \vdots& \ddots& \vdots& \vdots&   \ddots&  \vdots \\
c_{\Ac(i),\frac{\Ksf}{\Nsf},1} & \cdots &    c_{\Ac(i),\frac{\Ksf}{\Nsf},\frac{\Ksf}{\Nsf}(i-1)} &  0 & 0 & \cdots &1  & c_{\Ac(i),\frac{\Ksf}{\Nsf},\frac{\Ksf}{\Nsf} i + 1 } & \cdots & c_{\Ac(i),\frac{\Ksf}{\Nsf},\frac{\Ksf}{\Nsf} \Nsf_{\rm r}  } \\
\end{array} \right]. 
\end{align}
\label{eq:CAi}
 \end{subequations}
  \end{figure*}
In other words, we let 
\begin{align}
\left[ \begin{array}{ccc}
c_{\Ac(i),1,\frac{\Ksf}{\Nsf}(i-1)+1} &  \cdots & c_{\Ac(i),1,\frac{\Ksf}{\Nsf} i } \\
\vdots & \ddots & \vdots \\
c_{\Ac(i),\frac{\Ksf}{\Nsf},\frac{\Ksf}{\Nsf}(i-1)+1 }   & \cdots & c_{\Ac(i),\frac{\Ksf}{\Nsf},\frac{\Ksf}{\Nsf} i} \\ 
\end{array} \right] ={\bf I}_{\frac{\Ksf}{\Nsf}}
\end{align} 
 where ${\bf I}_{\frac{\Ksf}{\Nsf}}$ represents the identity matrix with dimension $\frac{\Ksf}{\Nsf} \times \frac{\Ksf}{\Nsf}$. 
 
 \iffalse
 The sub-matrix ${\bf C}^{\prime}_{\{n\}} $   can be  further expressed as 
$$
{\bf C}^{\prime}_{\{n\}}= 
\left[ \begin{array}{c}
\av_{n,1}\\
\vdots\\
\av_{n,\frac{\Ksf}{\Nsf}}\\ 
\end{array} \right] 
$$
whose $j^{\text{th}}$ row is  $\av_{n,j}=\left[a_{n,j,1}, \ldots, a_{n, j,\frac{\Ksf}{\Nsf}  (\Nsf_{\rm r}-1)} \right]$ for each $j\in \left[\frac{\Ksf}{\Nsf}\right]$.
\fi

Recall that $\mathbf{M}^{(\Sc)_{\rm r}}$ represents the sub-matrix of $\mathbf{M}$ which is composed of the rows  of $\mathbf{M}$ with indices in $\Sc$. From 
\begin{align}
{\bf C}_{\{\Ac(i)\}} \overline{ {\bf F}_{\Ac(i)}}=  %\left[ {\bf I}_{\frac{\Ksf}{\Nsf}}  ,  {\bf C}^{\prime}_{\{n\}}   \right] \overline{ {\bf F}_n} =
 {\bf 0}_{\frac{\Ksf}{\Nsf} \times \frac{\Ksf}{\Nsf}  (\Nsf_{\rm r}-1)},
\end{align} 
we  have 
\begin{align} 
& {\bf C}_{\{\Ac(i)\}}^{\left( \left[ \frac{\Ksf}{\Nsf} \Nsf_{\rm r} \right] \setminus \left[\frac{\Ksf}{\Nsf}(i-1)+1  : \frac{\Ksf}{\Nsf} i \right] \right)_{\rm c}}
 \ \overline{ {\bf F}_{\Ac(i)}}^{\left( \left[ \frac{\Ksf}{\Nsf} \Nsf_{\rm r} \right] \setminus \left[\frac{\Ksf}{\Nsf}(i-1)+1  : \frac{\Ksf}{\Nsf} i \right] \right)_{\rm r}} \nonumber\\
 &= -\overline{ {\bf F}_{\Ac(i)}}^{\left(\left[\frac{\Ksf}{\Nsf}(i-1)+1  : \frac{\Ksf}{\Nsf} i \right] \right)_{\rm r}}:= \left[ \begin{array}{c}
\overline{ {\bf f}_{\Ac(i),\frac{\Ksf}{\Nsf}(i-1)+1  }}\\
\vdots\\
\overline{ {\bf f}_{\Ac(i),\frac{\Ksf}{\Nsf} i }}\\ 
\end{array} \right],
\end{align}
where each vector $\overline{ {\bf f}_{\Ac(i),j}}$, $j\in \left[\frac{\Ksf}{\Nsf}(i-1)+1  : \frac{\Ksf}{\Nsf} i \right]$, is with dimension $1 \times \frac{\Ksf}{\Nsf}(\Nsf_{\rm r}-1)$.

 \iffalse
In other words, we have
\begin{align} 
\av_{n,j}    \overline{ {\bf F}_n}^{\left( \left[ \frac{\Ksf}{\Nsf}+1 : \frac{\Ksf}{\Nsf} \Nsf_{\rm r} \right] \right)_{\rm r}}=\overline{ {\bf f}_{n,j}}, \ \forall j\in \left[\frac{\Ksf}{\Nsf} \right].
\end{align}
\fi

By the Cramer's rule, it can be seen that 
\begin{align}
c_{\Ac(i),j,m} = \frac{\text{det}({\bf Y}_{\Ac(i),j,m})}{\text{det}\left(\overline{ {\bf F}_{\Ac(i)}}^{\left(  \left[ \frac{\Ksf}{\Nsf} \Nsf_{\rm r} \right] \setminus \left[\frac{\Ksf}{\Nsf}(i-1)+1  : \frac{\Ksf}{\Nsf} i \right] \right)_{\rm r}} \right)},    \label{eq:expression of anjm}
\end{align}
$\forall m \in \left[ \frac{\Ksf}{\Nsf} \Nsf_{\rm r} \right] \setminus \left[\frac{\Ksf}{\Nsf}(i-1)+1  : \frac{\Ksf}{\Nsf} i \right].$
Assuming $m$ is the $s^{\text{th}}$ smallest value in $\left[ \frac{\Ksf}{\Nsf} \Nsf_{\rm r} \right] \setminus \left[\frac{\Ksf}{\Nsf}(i-1)+1  : \frac{\Ksf}{\Nsf} i \right]$, we define
  ${\bf Y}_{\Ac(i),j,m}$ as the matrix formed by replacing the $s^{\text{th}}$ row of $\overline{ {\bf F}_{\Ac(i)}}^{\left( \left[ \frac{\Ksf}{\Nsf} \Nsf_{\rm r} \right] \setminus \left[\frac{\Ksf}{\Nsf}(i-1)+1  : \frac{\Ksf}{\Nsf} i \right] \right)_{\rm r}}$ by $\overline{ {\bf f}_{\Ac(i),j}}$.
  
In addition, $ 
\text{det}\left(\overline{ {\bf F}_{\Ac(i)}}^{\left(  \left[ \frac{\Ksf}{\Nsf} \Nsf_{\rm r} \right] \setminus \left[\frac{\Ksf}{\Nsf}(i-1)+1  : \frac{\Ksf}{\Nsf} i \right] \right)_{\rm r}} \right)$ is the determinant of a $\frac{\Ksf}{\Nsf}  (\Nsf_{\rm r}-1) \times \frac{\Ksf}{\Nsf}  (\Nsf_{\rm r}-1)$ matrix, which can be viewed as a multivariate polynomial whose variables are the elements in ${\bf F} $. Since the elements in ${\bf F} $ are uniformly i.i.d. over $\mathbb{F}_{\qsf}$, it is with high probability that the multivariate polynomial $\text{det}\left(\overline{ {\bf F}_{\Ac(i)}}^{\left(  \left[ \frac{\Ksf}{\Nsf} \Nsf_{\rm r} \right] \setminus \left[\frac{\Ksf}{\Nsf}(i-1)+1  : \frac{\Ksf}{\Nsf} i \right] \right)_{\rm r}} \right)$ is a non-zero multivariate polynomial  (i.e., a multivariate polynomial whose coefficients are not all $0$)  of degree $\frac{\Ksf}{\Nsf}  (\Nsf_{\rm r}-1) $. Hence, by the Schwartz-Zippel Lemma~\cite{Schwartz,Zippel,Demillo_Lipton}, we have
    \begin{subequations}
 \begin{align}
& \Pr \{ c_{\Ac(i),j,m} \ \text{exsits}   \} \nonumber\\& =  \Pr \left\{\text{det}\left(\overline{ {\bf F}_{\Ac(i)}}^{\left(  \left[ \frac{\Ksf}{\Nsf} \Nsf_{\rm r} \right] \setminus \left[\frac{\Ksf}{\Nsf}(i-1)+1  : \frac{\Ksf}{\Nsf} i \right] \right)_{\rm r}} \right)     \ \text{is non-zero}   \right\}\\& \geq 1- \frac{\Ksf (\Nsf_{\rm r}-1)  }{\Nsf \qsf}.  \label{eq:a_njm exist}
 \end{align}
       \end{subequations}
 Note that the above probability~\eqref{eq:a_njm exist} is over all possible realizations of ${\bf F}$ whose elements are uniformly i.i.d. over $\mathbb{F}_{\qsf}$.
 
 By the probability union bound, we have
      \begin{subequations}
 \begin{align}
&  \Pr \left\{ c_{\Ac(i),j,m} \ \text{exsits}, \  \forall i \in [\Nsf_{\rm r}], j \in \left[ \frac{\Ksf}{\Nsf} \right], \right. \nonumber\\& \left. m\in  \left[ \frac{\Ksf}{\Nsf} \Nsf_{\rm r} \right] \setminus \left[\frac{\Ksf}{\Nsf}(i-1)+1  : \frac{\Ksf}{\Nsf} i \right]  \right\} \nonumber\\ & \geq  1- \frac{\Ksf (\Nsf_{\rm r}-1)  }{\Nsf \qsf}  \Nsf \frac{\Ksf}{\Nsf} \frac{\Ksf}{\Nsf}  (\Nsf_{\rm r}-1)\\
 &=1- \frac{\Ksf^3 (\Nsf_{\rm r}-1)^2  }{\Nsf^2 \qsf}  \label{eq:first proba}\\
 & \stackrel{\qsf \to \infty}{\longrightarrow} 1.  
 \end{align}
      \end{subequations}
Hence, we prove that   the coding matrix of  each worker $\Ac(i)$ where $i\in [\Nsf_{\rm r}]$, ${\bf C}_{\Ac(i)}$ in~\eqref{eq:transformed Cn}, exists  with high probability.

In the following, we will prove that matrix 
\begin{align}
{\bf C}_{\Ac}:=\left[ \begin{array}{c}
{\bf C}_{\Ac(1)} \\
\vdots\\
{\bf C}_{\Ac(\Nsf_{\rm r})} \\ 
\end{array} \right]
\end{align}
is full-rank  with high probability.

 \iffalse
Now let us focus on each set $\Ac=\{\Ac(1),\ldots,\Ac(\Nsf_{\rm r})\}$, where
$\Ac \subseteq [\Nsf]$ and $|\Ac|=\Nsf_{\rm r}$. 
  Note that the coding matrix of each worker $n \in [\Nsf]$ is ${\bf C}_{\{n\}}$ with dimension  $\frac{\Ksf}{\Nsf}  \times \frac{\Ksf}{\Nsf}   \Nsf_{\rm r} $. We will prove that the coding matrix of the workers in $\Ac$, 
\begin{align}
{\bf C}_{\Ac}:=\left[ \begin{array}{c}
{\bf C}_{\Ac(1)} \\
\vdots\\
{\bf C}_{\Ac(\Nsf_{\rm r})} \\ 
\end{array} \right],
\end{align}
is full-rank  with high probability, such that Lemma~\ref{lem:SZlemma} is proved. 
%from the transmissions of the workers  in $\Sc$ the master can recover $F_1, \ldots, F_{\frac{K}{N} N_r}$.
\fi

Note that ${\bf C}_{\Ac}$ is a matrix with dimension $\frac{\Ksf}{\Nsf} \Nsf_{\rm r} \times \frac{\Ksf}{\Nsf} \Nsf_{\rm r}$. We expand the determinant of  ${\bf C}_{\Ac}$ as follows,
\begin{align}
 \text{det}({\bf C}_{\Ac})= \sum_{i \in \left[\left(\frac{\Ksf}{\Nsf} \Nsf_{\rm r}\right) ! \right]} \frac{P_i}{Q_i},
\end{align}
  %$ \text{det}({\bf C}_{\Ac})$ 
  which contains $\left(\frac{\Ksf}{\Nsf} \Nsf_{\rm r} \right) !$ terms.  Each term can be expressed as $\frac{P_i}{Q_i}$, where $P_i$ and $Q_i$ are multivariate polynomials whose variables are the elements in ${\bf F }$.
From~\eqref{eq:expression of anjm}, it can be seen that  each element in ${\bf C}_{\Ac}$ is the ratio of two multivariate polynomials whose variables are the elements in ${\bf F }$ with degree $\frac{\Ksf}{\Nsf}  (\Nsf_{\rm r}-1)$. In addition, each term in $ \text{det}({\bf C}_{\Ac})$
is a multivariate polynomial whose variables are   the elements in ${\bf C}_{\Ac}$ with degree $\frac{\Ksf}{\Nsf}  \Nsf_{\rm r}$.
  Hence,  $P_i$ and $Q_i$ are multivariate polynomials whose variables are the elements in ${\bf F}$ with degree $\left( \frac{\Ksf}{\Nsf}  \right)^2  \Nsf_{\rm r} (\Nsf_{\rm r}-1)$.

  We then let 
\begin{align}
P_{\Ac} :=   \text{det}({\bf C}_{\Ac}) \ \prod_{i  \in \left[\left(\frac{\Ksf}{\Nsf} \Nsf_{\rm r}\right) ! \right] }Q_i. \label{eq:PA}
\end{align} 
 If ${\bf C}_{\Ac}$ exists and $P_{\Ac} \neq 0$, we have $ \text{det}({\bf C}_{\Ac}) \neq 0$ and thus ${\bf C}_{\Ac}$ is full-rank.

To apply the Schwartz-Zippel lemma~\cite{Schwartz,Zippel,Demillo_Lipton}, we need to guarantee that $P_{\Ac}$ is   a non-zero multivariate polynomial. To this end, we only need one specific realization of  ${\bf F} $ so that $P_{\Ac} \neq 0$ (or alternatively $\text{det}({\bf C}_{\Ac}) \neq 0$ and $Q_i \neq 0$ at the same time). 
We construct such specific ${\bf F} $  in Appendix~\ref{sec:lemma nonzero n=k} such that   the following lemma can be proved.
\begin{lem}
\label{lem:nonzero general}
	For the  $(\Ksf,\Nsf,\Nsf_{\rm r}, \Ksf_{\rm c}, \Msf)=\left(\Ksf,\Nsf,\Nsf_{\rm r}, \frac{\Ksf}{\Nsf}\Nsf_{\rm r}, \frac{\Ksf}{\Nsf}(\Nsf-\Nsf_{\rm r}+1) \right)$ distributed linearly separable computation problem,   $P_{\Ac}$ in~\eqref{eq:PA} is   a non-zero multivariate polynomial.
	\hfill $\square$ 
\end{lem}

 \iffalse
\begin{lem}
\label{lem:nonzero n=k}
If $P_{\Ac}$ is a non-zero multivariate polynomial for the  $(\Ksf,\Nsf,\Nsf_{\rm r}, \Ksf_{\rm c}, \Msf)=\left(\Nsf,\Nsf,\Nsf_{\rm r}, \Nsf-\Nsf_{\rm r}+1 \right)$ distributed linearly separable computation problem, 
then for any  \\
$(\Ksf,\Nsf,\Nsf_{\rm r}, \Ksf_{\rm c}, \Msf)=\left(\Ksf,\Nsf,\Nsf_{\rm r}, \frac{\Ksf}{\Nsf}\Nsf_{\rm r}, \frac{\Ksf}{\Nsf}(\Nsf-\Nsf_{\rm r}+1) \right)$ distributed linearly separable computation problem
where $\Nsf$ divides  $\Ksf$,  $P_{\Ac}$ is also a non-zero multivariate polynomial.
\hfill $\square$ 
\end{lem}
 \fi

 %By random  generation of  ${\bf F}$, we have tested all cases that $\Nsf=\Ksf \leq 50$. Then from Lemma~\ref{lem:nonzero n=k}  we have that  for all cases where $\Nsf\leq 50$,  $P_{\Ac}$ is also a non-zero polynomial.  We conjecture that for any case,  $P_{\Ac}$ is not a zero polynomial.
 
% By Lemmas~\ref{lem:nonzero n=k one case} and~\ref{lem:nonzero n=k}, it can be seen that for any    $(\Ksf,\Nsf,\Nsf_{\rm r}, \Ksf_{\rm c}, \Msf)=\left(\Ksf,\Nsf,\Nsf_{\rm r}, \frac{\Ksf}{\Nsf}\Nsf_{\rm r}, \frac{\Ksf}{\Nsf}(\Nsf-\Nsf_{\rm r}+1) \right)$ distributed linearly separable computation problem where $\Nsf$ divides  $\Ksf$,  $P_{\Ac}$ is also a non-zero polynomial.
 Recall that  $P_i$ and $Q_i$ are multivariate polynomials with degree $\left( \frac{\Ksf}{\Nsf}  \right)^2  \Nsf_{\rm r} (\Nsf_{\rm r}-1)$. Thus  the degree of $P_{\Ac}$ is less than 
$ \left( \frac{\Ksf}{\Nsf}  \Nsf_{\rm r} \right)^2 $. Hence, by the Schwartz-Zippel lemma~\cite{Schwartz,Zippel,Demillo_Lipton} we have
% \begin{align}
 %&\Pr \left\{P_{\Ac} \neq 0 \Big| c_{\Ac(i),j,m} \ \text{exsits}, \  \forall i \in [\Nsf_{\rm r}], j \in \left[ \frac{\Ksf}{\Nsf} \right], m\in  \left[ \frac{\Ksf}{\Nsf} \Nsf_{\rm r} \right] \setminus \left[\frac{\Ksf}{\Nsf}(i-1)+1  : \frac{\Ksf}{\Nsf} i \right]  \right\}  \nonumber \\ & \geq  1- \frac{ \left(\frac{\Ksf}{\Nsf} \Nsf_{\rm r} \right) !   \left( \frac{\Ksf}{\Nsf}  \Nsf_{\rm r} \right)^2 }{\qsf}.\label{eq:second proba}
% \end{align}
 \begin{align}
 &\Pr \left\{P_{\Ac} \neq 0 \right\} \geq  1- \frac{ \left(\frac{\Ksf}{\Nsf} \Nsf_{\rm r} \right) !   \left( \frac{\Ksf}{\Nsf}  \Nsf_{\rm r} \right)^2 }{\qsf}.\label{eq:second proba}
\end{align}

Hence, from~\eqref{eq:first proba} and~\eqref{eq:second proba}, we have 
      \begin{subequations}
\begin{align}
&\Pr \{ {\bf C}_{\Ac} \text{ is full-rank}\} \nonumber\\& \geq 1- \Pr \{ {\bf C}_{\Ac} \text{ does not exist}\}-\Pr \{ P_{\Ac} = 0 \} \\
& \geq 1- \frac{\Ksf^3 (\Nsf_{\rm r}-1)^2  }{\Nsf^2 \qsf} - \frac{ \left(\frac{\Ksf}{\Nsf} \Nsf_{\rm r} \right) !   \left( \frac{\Ksf}{\Nsf}  \Nsf_{\rm r} \right)^2 }{\qsf}.
\end{align}
       \end{subequations}

Finally, by considering all $\Ac \subseteq [\Nsf] $ where $|\Ac|=\Nsf_{\rm r}$, we have 
      \begin{subequations}
\begin{align}
& \Pr \{{\bf C}_{\Ac} \text{ is full-rank}, \ \forall \Ac \subseteq [\Nsf]: |\Ac|=\Nsf_{\rm r} \} \\
& \geq 1 - \sum_{\Ac \subseteq [\Nsf]: |\Ac|=\Nsf_{\rm r} }  \Pr \{ {\bf C}_{\Ac} \text{ is not full-rank} \}\\
&\geq  1 - \binom{\Nsf}{\Nsf_{\rm r}} \left( \frac{\Ksf^3 (\Nsf_{\rm r}-1)^2  }{\Nsf^2 \qsf} + \frac{ \left(\frac{\Ksf}{\Nsf} \Nsf_{\rm r} \right) !   \left( \frac{\Ksf}{\Nsf}  \Nsf_{\rm r} \right)^2 }{\qsf} \right)\\
 & \stackrel{\qsf \to \infty}{\longrightarrow} 1. 
 \end{align}
      \end{subequations}
      Hence, we prove Lemma~\ref{lem:SZlemma}.
 \iffalse
In conclusion, from~\eqref{eq:first proba} and~\eqref{eq:second proba}, we have 
      \begin{subequations}
\begin{align}
& \Pr \{P_{\Ac} \neq 0, \ \forall \Ac \subseteq [\Nsf]: |\Ac|=\Nsf_{\rm r} \} \\
& \geq 
1- \Pr \left\{ a_{n,j,m}  \ \text{does not exsit}, \  \text{for some } n \in [\Nsf], j \in \left[ \frac{\Ksf}{\Nsf} \right], m\in \left[\frac{\Ksf}{\Nsf}  (\Nsf_{\rm r}-1) \right]  \right\} \nonumber\\&
-\Pr \left\{ P_{\Ac} = 0  , \  \text{for some } \Ac \subseteq [\Nsf]: |\Ac|=\Nsf_{\rm r}   \Big| a_{n,j,m}  \ \text{exsits}, \  \forall n \in [\Nsf], j \in \left[ \frac{\Ksf}{\Nsf}\right], m\in \left[\frac{\Ksf}{\Nsf} \Nsf_{\rm r} \right]\right\}\\
&\geq 1- \Pr \left\{ a_{n,j,m}  \ \text{does not exsit}, \  \text{for some } n \in [\Nsf], j \in \left[ \frac{\Ksf}{\Nsf} \right], m\in \left[\frac{\Ksf}{\Nsf}  (\Nsf_{\rm r}-1) \right]  \right\} \nonumber \\&
- \sum_{\Ac \subseteq [\Nsf]: |\Ac|=\Nsf_{\rm r}  } \Pr \left\{ P_{\Ac} = 0     \Big| a_{n,j,m}  \ \text{exsits}, \  \forall n \in [\Nsf], j \in \left[ \frac{\Ksf}{\Nsf}\right], m\in \left[\frac{\Ksf}{\Nsf} \Nsf_{\rm r} \right]\right\} \\
 &\geq 1- \frac{ \Ksf^3 (\Nsf_{\rm r}-1)^2  }{\Nsf^2 \qsf} -\frac{ \left(\frac{\Ksf}{\Nsf} \Nsf_{\rm r}\right) !   \left( \frac{\Ksf}{\Nsf}  \Nsf_{\rm r} \right)^2 }{\qsf} \binom{\Nsf}{\Nsf_{\rm r}}\\
 & \stackrel{\qsf \to \infty}{\longrightarrow} 1. 
\end{align}
      \end{subequations}
      Hence, we prove Lemma~\ref{lem:SZlemma}.
 \fi

\section{Proofs of Lemma~\ref{lem:nonzero general}}
\label{sec:lemma nonzero n=k}
  \subsection{$\Nsf=\Ksf$}
\label{sub:lem:nonzero n=k part1}
  We first consider the case where $\Nsf=\Ksf$.   We aim to construct one demand matrix ${\bf F} $ where $\text{det}({\bf C}_{\Ac} )  \neq 0$, such that we can prove Lemma~\ref{lem:nonzero general} for this case.
   
  Note that  when $\Nsf=\Ksf$, we have that $\Ksf_{\rm c}=\frac{\Ksf}{\Nsf}\Nsf_{\rm r}=\Nsf_{\rm r}$ and that  the dimension of  ${\bf F} $ is $\Nsf_{\rm r} \times \Nsf$. 
We construct an ${\bf F}$ such that for each $i\in [\Nsf_{\rm r}]$ and $j \in \overline{  \Zc_{\Ac(i)} } $, the element located at   the $i^{\text{th}}$ row and the $j^{\text{th}}$ column is $0$. Recall that the number of datasets which are not assigned to each worker is $|\overline{  \Zc_{\Ac(i)} } |=\Nsf_{\rm r}-1$ and that by the cyclic assignment, the elements in $\overline{  \Zc_{\Ac(i)} } $ are adjacent; thus the $i^{\text{th}}$ row of ${\bf F}$ can be expressed as follows,
\begin{align}
{\bf F}^{(\{i\})_{\rm r}}= [ *, *, \cdots,*, 0, 0, \cdots, 0, *,*,\cdots ,*], \label{eq:ith row} 
\end{align}
where the number of  adjacent `$0$' in~\eqref{eq:ith row} is $\Nsf_{\rm r}-1$ and each `$*$' represents a    symbol uniformly i.i.d. over $\mathbf{F}_{\qsf}$.

To prove that $\Pc(\Ac) $ in~\eqref{eq:PA} is non-zero, we need to prove 
\begin{enumerate}
\item $ 
\text{det}\left(\overline{ {\bf F}_{\Ac(i)}}^{\left(  \left[ \frac{\Ksf}{\Nsf} \Nsf_{\rm r} \right] \setminus \left[\frac{\Ksf}{\Nsf}(i-1)+1  : \frac{\Ksf}{\Nsf} i \right] \right)_{\rm r}} \right) \neq 0$ for each $i \in [\Nsf_{\rm r}]$, such that ${\bf C}_{\Ac}$ exists (see~\eqref{eq:expression of anjm}); thus $\prod_{i  \in \left[\left(\frac{\Ksf}{\Nsf} \Nsf_{\rm r}\right) ! \right] }Q_i \neq 0$.
\item  $ \text{det}({\bf C}_{\Ac}) \neq 0$. %, such that the proposed scheme is decodable. 
\end{enumerate}

First, we  prove that  ${\bf C}_{\Ac}$ exists. We focus on worker $\Ac(i)$ where   $i \in [\Nsf_{\rm r}]$.
 Matrix $\overline{ {\bf F}_{\Ac(i)}}^{\left(  \left[ \frac{\Ksf}{\Nsf} \Nsf_{\rm r} \right] \setminus \left[\frac{\Ksf}{\Nsf}(i-1)+1  : \frac{\Ksf}{\Nsf} i \right] \right)_{\rm r}}$ is with dimension $(\Nsf_{\rm r}-1) \times (\Nsf_{\rm r}-1)$.
Each row of  $\overline{ {\bf F}_{\Ac(i)}}^{\left(  \left[ \frac{\Ksf}{\Nsf} \Nsf_{\rm r} \right] \setminus \left[\frac{\Ksf}{\Nsf}(i-1)+1  : \frac{\Ksf}{\Nsf} i \right] \right)_{\rm r}}$ corresponds to one worker in $\Ac \setminus \{\Ac (i)\}$. There are three cases:
\begin{itemize}
\item if this worker is $\text{Mod}(\Ac(i)+j,\Nsf)$ where $j \in [\Nsf_{\rm r}-2]$, the corresponding row is 
$$
[*,\cdots,*,0,\cdots,0],
$$
where the number of `$*$' is $j$ and the number of `$0$' is $\Nsf_{\rm r}-1-j$;
\item if this worker is $\text{Mod}(\Ac(i)-j,\Nsf)$ where $j \in [\Nsf_{\rm r}-2]$, the corresponding row is 
$$
[0,\cdots,0,*,\cdots,*],
$$
where the number of `$0$' is $j$ and the number of `$*$' is $\Nsf_{\rm r}-1-j$;
\item otherwise, the corresponding row is 
$$
[*,\cdots,*].
$$
\end{itemize}

By the above observation, it can be seen that   each column of $\overline{ {\bf F}_{\Ac(i)}}^{\left(  \left[ \frac{\Ksf}{\Nsf} \Nsf_{\rm r} \right] \setminus \left[\frac{\Ksf}{\Nsf}(i-1)+1  : \frac{\Ksf}{\Nsf} i \right] \right)_{\rm r}}$ contains at most $(\Nsf_{\rm r}-2)$ `$0$', and that there does not exist two columns with $(\Nsf_{\rm r}-2)$ `$0$' where these two columns have the same form (i.e., the positions of `$0$' are the same). 
 Hence, with some row permutation on rows, we can let the elements located at the right-diagonal of 
 $\overline{ {\bf F}_{\Ac(i)}}^{\left(  \left[ \frac{\Ksf}{\Nsf} \Nsf_{\rm r} \right] \setminus \left[\frac{\Ksf}{\Nsf}(i-1)+1  : \frac{\Ksf}{\Nsf} i \right] \right)_{\rm r}}$ 
 are all `$*$'. 
 In other words, 
 $ 
\text{det}\left(\overline{ {\bf F}_{\Ac(i)}}^{\left(  \left[ \frac{\Ksf}{\Nsf} \Nsf_{\rm r} \right] \setminus \left[\frac{\Ksf}{\Nsf}(i-1)+1  : \frac{\Ksf}{\Nsf} i \right] \right)_{\rm r}} \right) $ is a non-zero multivariate polynomial where each `$*$' in $\overline{ {\bf F}_{\Ac(i)}}^{\left(  \left[ \frac{\Ksf}{\Nsf} \Nsf_{\rm r} \right] \setminus \left[\frac{\Ksf}{\Nsf}(i-1)+1  : \frac{\Ksf}{\Nsf} i \right] \right)_{\rm r}} $ is a  variable uniformly i.i.d. over $\mathbb{F}_{\qsf}$. By the Schwartz-Zippel lemma~\cite{Schwartz,Zippel,Demillo_Lipton}, it can be seen that 
\begin{align}
\Pr \left\{ \text{det}\left(\overline{ {\bf F}_{\Ac(i)}}^{\left(  \left[ \frac{\Ksf}{\Nsf} \Nsf_{\rm r} \right] \setminus \left[\frac{\Ksf}{\Nsf}(i-1)+1  : \frac{\Ksf}{\Nsf} i \right] \right)_{\rm r}} \right) \neq 0 \right\}  \stackrel{\qsf \to \infty}{\longrightarrow} 1.
\end{align}

By the probability union bound, we have 
\begin{align}
\Pr \left\{ \text{det}\left(\overline{ {\bf F}_{\Ac(i)}}^{\left(  \left[ \frac{\Ksf}{\Nsf} \Nsf_{\rm r} \right] \setminus \left[\frac{\Ksf}{\Nsf}(i-1)+1  : \frac{\Ksf}{\Nsf} i \right] \right)_{\rm r}} \right) \neq 0, \ \forall i \in [\Nsf_{\rm r}] \right\}  \stackrel{\qsf \to \infty}{\longrightarrow} 1.
\end{align}
 Hence, there must exist some ${\bf F}$ such that $\text{det}\left(\overline{ {\bf F}_{\Ac(i)}}^{\left(  \left[ \frac{\Ksf}{\Nsf} \Nsf_{\rm r} \right] \setminus \left[\frac{\Ksf}{\Nsf}(i-1)+1  : \frac{\Ksf}{\Nsf} i \right] \right)_{\rm r}} \right) \neq 0$ for each $i \in [\Nsf_{\rm r}]$; thus we finish the proof on the existence of   ${\bf C}_{\Ac}$.

Next, we prove the proposed scheme is decodable.
Obviously, 
$$
{\bf F}^{(\{i\})_{\rm r}} \left[ \begin{array}{c}
W_1 \\
\vdots\\
W_{\Nsf} \\ 
\end{array} \right]
$$
 can be sent by worker $\Ac(i)$. With $\Nsf=\Ksf$, each worker sends $\frac{\Ksf}{\Nsf}=1$ linear combination of messages.  
 By the construction, we can see that for each $i\in [\Nsf_{\rm r}]$, the coding matrix is 
\begin{align}
{\bf C}_{\Ac(i)}= [0,\cdots,0,1,0,\cdots,0],
\end{align} 
 where $1$ is located at the $i^{\text{th}}$ column and the dimension of ${\bf C}_{\Ac(i)}$ is   $1 \times \Nsf_{\rm r}$.
Hence, it can be seen that 
 \begin{align}
 {\bf C}_{\Ac}= \left[ \begin{array}{c}
{\bf C}_{\Ac(1)} \\
\vdots\\
{\bf C}_{\Ac(\Nsf_{\rm r})} \\ 
\end{array} \right]
 \end{align}
 is an identity matrix and is thus full-rank, i.e., $ \text{det}({\bf C}_{\Ac}) \neq 0$.

\subsection{$\Nsf$ divides $\Ksf$}
\label{sub:lem:nonzero n=k part2}
 %{\red DO WE NEED ALWAYS REPEAT  $(\Ksf,\Nsf,\Nsf_{\rm r}, \Ksf_{\rm c}, \Msf)=\big(\asf \Nsf,\Nsf,\Nsf_{\rm r}, \asf \Nsf_{\rm r}, \asf(\Nsf-\Nsf_{\rm r}+1) \big)$???}
 %Assume that  $P_{\Ac}$ in~\eqref{eq:PA} is a non-zero polynomial for the  $(\Ksf,\Nsf,\Nsf_{\rm r}, \Ksf_{\rm c}, \Msf)=\left(\Nsf,\Nsf,\Nsf_{\rm r}, \Nsf_{\rm r}, \Nsf-\Nsf_{\rm r}+1 \right)$ distributed linearly separable computation problem. 
 Let us then focus on   the  $(\Ksf,\Nsf,\Nsf_{\rm r}, \Ksf_{\rm c}, \Msf)=\big(\asf \Nsf,\Nsf,\Nsf_{\rm r}, \asf \Nsf_{\rm r}, \asf(\Nsf-\Nsf_{\rm r}+1) \big)$ distributed linearly separable computation problem, where $\asf$ is a  positive integer.
Similarly, we also aim to construct one demand matrix ${\bf F} $ where $\text{det}({\bf C}_{\Ac} )  \neq 0$.
 
 More precisely, we let  (recall that ${\bf 0}_{m \times n}$ represents the zero  matrix with dimension $m\times n$; 
$(\mathbf{M})_{m \times n}$ represents the dimension of matrix $\mathbf{M}$ is $m \times n$)
 \begin{align}
 {\bf F}  = \left[\begin{array}{c:c:c:c}
 ({\bf F}_1)_{\Nsf_{\rm r} \times \Nsf}  & {\bf 0}_{\Nsf_{\rm r} \times \Nsf}  & \cdots & {\bf 0}_{\Nsf_{\rm r} \times \Nsf}   \\ \hdashline
{\bf 0}_{\Nsf_{\rm r} \times \Nsf} &  ({\bf F}_2)_{\Nsf_{\rm r} \times \Nsf}   & \cdots & {\bf 0}_{\Nsf_{\rm r} \times \Nsf}   \\ \hdashline 
 \vdots   & \vdots  &  \vdots& \vdots \\ \hdashline
 {\bf 0}_{\Nsf_{\rm r} \times \Nsf} &   {\bf 0}_{\Nsf_{\rm r} \times \Nsf}    & \cdots &  ({\bf F}_{\asf})_{\Nsf_{\rm r} \times \Nsf} 
 \end{array}
\right], \label{eq:constructed F for N divides K}
 \end{align}
 where each element in ${\bf F}_i, i\in [\asf]$, is   uniformly i.i.d.        over  $\mathbb{F}_{\qsf}$. 
    In the above construction, 
the  $(\Ksf,\Nsf,\Nsf_{\rm r}, \Ksf_{\rm c}, \Msf)=\big(\asf \Nsf,\Nsf,\Nsf_{\rm r}, \asf \Nsf_{\rm r}, \asf(\Nsf-\Nsf_{\rm r}+1) \big)$ distributed linearly separable computation problem is divided into $\asf$ independent  $(\Ksf,\Nsf,\Nsf_{\rm r}, \Ksf_{\rm c}, \Msf)=\left(\Nsf,\Nsf,\Nsf_{\rm r},  \Nsf_{\rm r}, \Nsf-\Nsf_{\rm r}+1 \right)$ distributed linearly separable computation sub-problems.
In each sub-problem, assuming that the coding matrix of the workers in $\Ac$ is ${\bf C}^{\prime}_{\Ac}$, from Appendix~\ref{sub:lem:nonzero n=k part1}, we have ${\bf C}^{\prime}_{\Ac} \neq 0$ with high probability. Hence,  in the  
   $(\Ksf,\Nsf,\Nsf_{\rm r}, \Ksf_{\rm c}, \Msf)=\big(\asf \Nsf,\Nsf,\Nsf_{\rm r}, \asf \Nsf_{\rm r}, \asf(\Nsf-\Nsf_{\rm r}+1) \big)$ distributed linearly separable computation problem with the constructed  ${\bf F}$ in~\eqref{eq:constructed F for N divides K}, we also have that $ \text{det}({\bf C}_{\Ac} ) \neq 0$ with high probability. 

  %  $ \text{det}({\bf C}_{\Ac}) \neq 0$ with high probability for the  $(\Ksf,\Nsf,\Nsf_{\rm r}, \Ksf_{\rm c}, \Msf)=\left(\Nsf,\Nsf,\Nsf_{\rm r},  \Nsf_{\rm r}, \Nsf-\Nsf_{\rm r}+1 \right)$ distributed linearly separable computation problem. Hence, we also have that in the constructed  $(\Ksf,\Nsf,\Nsf_{\rm r}, \Ksf_{\rm c}, \Msf)=\big(\asf \Nsf,\Nsf,\Nsf_{\rm r}, \asf \Nsf_{\rm r}, \asf(\Nsf-\Nsf_{\rm r}+1) \big)$ distributed linearly separable computation problem,  $ \text{det}({\bf C}_{\Ac} ) \neq 0$ with high probability. 

\section{An Allocation Algorithm for the Cyclic Assignment in the General Case}
\label{sec:choosing method}
Recall that  our objective is to choose $\bsf$ datasets from  $\Nsf$ effective datasets as the real datasets, such that by the cyclic assignment on these $\Nsf$ effective datasets the number of real datasets assigned to each worker is no more than $\left\lceil  \frac{ \Nsf-\Nsf_{\rm r}+1 }{\left\lfloor  \frac{\Nsf}{\bsf}  \right\rfloor }  \right\rceil.$ 
By the cyclic assignment, each effective dataset (denoted by $E_k$ where $k\in [\Nsf]$) is assigned to workers in $\big\{\text{Mod}(k,\Nsf),  \text{Mod}(k-1,\Nsf),\ldots,  \text{Mod}(k-\Nsf+\Nsf_{\rm r}, \Nsf ) \big\}$. The set of effective datasets assigned  to  worker $n \in [\Nsf]$   is  $\big\{\text{Mod}(n,\Nsf)  , \text{Mod}(n+1,\Nsf) , \ldots, \text{Mod}(n+\Nsf-\Nsf_{\rm r},\Nsf) \big\}$.
We propose an algorithm based on the following integer decomposition.

We  decompose the integer  $ \Nsf-\bsf$ into $\bsf$ parts, $ \Nsf-\bsf=p_1+\cdots+p_{\bsf}$, where $p_1\leq \cdots\leq p_{\bsf}$ and $p_i$ is either 
 $\left\lceil \frac{\Nsf-\bsf}{\bsf} \right\rceil $ or $\left\lfloor \frac{\Nsf-\bsf}{\bsf} \right\rfloor$ for each $i \in [\bsf]$. 
More precisely, by  defining $\alpha= \bsf\left\lceil \frac{\Nsf-\bsf}{\bsf} \right\rceil- (\Nsf-\bsf)$,  we let 
      \begin{subequations}
\begin{align}
&p_1 =\cdots=p_{\alpha} =\left\lfloor \frac{\Nsf-\bsf}{\bsf} \right\rfloor ; \\
 & p_{\alpha +1}=\cdots= p_{\bsf}=\left\lceil \frac{\Nsf-\bsf}{\bsf} \right\rceil.
\end{align}
       \end{subequations}
       
We then choose datasets $$
E_1, \ E_{2+p_1}, \  E_{3+p_1+p_2}, \ldots,\  E_{\bsf+p_1+\cdots+p_{\bsf-1}}$$
 as the real datasets. 
It can be seen that  between each two real datasets, there are at least $\left\lfloor \frac{\Nsf-\bsf}{\bsf} \right\rfloor$ virtual datasets. Hence, in each adjacent $\Nsf-\Nsf_{\rm r}+1$ datasets, there are at most  
$$
\left\lceil  \frac{ \Nsf-\Nsf_{\rm r}+1 }{\left\lfloor  \frac{\Nsf-\bsf}{\bsf}+1  \right\rfloor }  \right\rceil=\left\lceil  \frac{ \Nsf-\Nsf_{\rm r}+1 }{\left\lfloor  \frac{\Nsf}{\bsf} \right\rfloor }  \right\rceil
$$
 real datasets. Hence,  we   prove that by the above choice, the number of real datasets assigned to each worker is no more than $\left\lceil  \frac{ \Nsf-\Nsf_{\rm r}+1 }{\left\lfloor  \frac{\Nsf}{\bsf}  \right\rfloor }  \right\rceil.$

\bibliographystyle{IEEEtran}
\bibliography{IEEEabrv,IEEEexample}

\begin{IEEEbiographynophoto}
						{Kai Wan} (S '15 -- M '18)
						received  the B.E. degree in    Optoelectronics from  Huazhong University of Science and Technology, China, in 2012, the   M.Sc. and Ph.D. degrees in Communications from Universit{\'e}  Paris-Saclay, France, in 2014 and 2018.  He is currently a post-doctoral    researcher with the Communications and Information Theory Chair   (CommIT) at Technische Universit\"at Berlin, Berlin, Germany. His   research interests include information theory, coding techniques, and   their applications on coded caching,  index coding, distributed storage,  distributed computing, wireless communications,   privacy and security. He has served as an Associate Editor of IEEE Communications Letters from Aug. 2021.
					\end{IEEEbiographynophoto}
					
					\begin{IEEEbiographynophoto}{Hua Sun} (S '12 -- M '17) received the B.E. degree in Communications Engineering from Beijing University of Posts and Telecommunications, China, in 2011, and the M.S. degree in Electrical and Computer Engineering and the Ph.D. degree in Electrical Engineering from University of California Irvine, USA, in 2013 and 2017, respectively. He is an Assistant Professor in the Department of Electrical Engineering at the University of North Texas, USA. His research interests include information theory and its applications to communications, privacy, security, and storage.

Dr. Sun is a recipient of the NSF CAREER award in 2021, and the UNT College of Engineering Distinguished Faculty Fellowship in 2021. His co-authored papers received the IEEE Jack Keil Wolf ISIT Student Paper Award in 2016, and an IEEE GLOBECOM Best Paper Award in 2016.
					\end{IEEEbiographynophoto}
					
										\begin{IEEEbiographynophoto}{Mingyue Ji}
(S '09 -- M '15) received the B.E. in Communication Engineering from Beijing University of Posts and Telecommunications (China), in 2006, the M.Sc. degrees in Electrical Engineering from Royal Institute of Technology (Sweden) and from University of California, Santa Cruz, in 2008 and 2010, respectively, and the PhD from the Ming Hsieh Department of Electrical Engineering at University of Southern California in 2015. He subsequently was a Staff II System Design Scientist with Broadcom Corporation (Broadcom Limited) in 2015-2016. He is now an Assistant Professor of Electrical and Computer Engineering Department and an Adjunct Assistant Professor of School of Computing at the University of Utah. He received the IEEE Communications Society Leonard G. Abraham Prize for the best IEEE JSAC paper in 2019, the best paper award in IEEE ICC 2015 conference, the best student paper award in IEEE European Wireless 2010 Conference and USC Annenberg Fellowship from 2010 to 2014.  He has served as an Associate Editor of IEEE Transactions on Communications from 2020. He is interested the broad area of information theory, coding theory, concentration of measure and statistics with the applications of caching networks, wireless communications, distributed storage and computing systems, distributed machine learning, and (statistical) signal processing.
					\end{IEEEbiographynophoto}

										\begin{IEEEbiographynophoto}{Giuseppe Caire}
 (S '92 -- M '94 -- SM '03 -- F '05) 
was born in Torino in 1965. He received the B.Sc. in Electrical Engineering  from Politecnico di Torino in 1990, 
the M.Sc. in Electrical Engineering from Princeton University in 1992, and the Ph.D. from Politecnico di Torino in 1994. 
He has been a post-doctoral research fellow with the European Space Agency (ESTEC, Noordwijk, The Netherlands) in 1994-1995,
Assistant Professor in Telecommunications at the Politecnico di Torino, Associate Professor at the University of Parma, Italy, 
Professor with the Department of Mobile Communications at the Eurecom Institute,  Sophia-Antipolis, France,
a Professor of Electrical Engineering with the Viterbi School of Engineering, University of Southern California, Los Angeles,
and he is currently an Alexander von Humboldt Professor with the Faculty of Electrical Engineering and Computer Science at the
Technical University of Berlin, Germany.

He received the Jack Neubauer Best System Paper Award from the IEEE Vehicular Technology Society in 2003,  the
IEEE Communications Society and Information Theory Society Joint Paper Award in 2004 and in 2011, 
the Okawa Research Award in 2006,   
the Alexander von Humboldt Professorship in 2014, the Vodafone Innovation Prize in 2015, an ERC Advanced Grant in 2018, 
the Leonard G. Abraham Prize for best IEEE JSAC paper in 2019, the IEEE Communications Society Edwin Howard Armstrong Achievement Award in 2020, 
and he is a recipient of the 2021 Leibinz Prize  of the German National Science Foundation (DFG). 
Giuseppe Caire is a Fellow of IEEE since 2005.  He has served in the Board of Governors of the IEEE Information Theory Society from 2004 to 2007,
and as officer from 2008 to 2013. He was President of the IEEE Information Theory Society in 2011. 
His main research interests are in the field of communications theory, information theory, channel and source coding
with particular focus on wireless communications.   
					\end{IEEEbiographynophoto}
					
\end{document}